\documentclass[fleqn,usenatbib]{mnras}

\usepackage{newtxtext,newtxmath}

\usepackage[T1]{fontenc}

\DeclareRobustCommand{\VAN}[3]{#2}
\let\VANthebibliography\thebibliography
\def\thebibliography{\DeclareRobustCommand{\VAN}[3]{##3}\VANthebibliography}

\usepackage{graphicx}	
\usepackage{amsmath}	
\usepackage{gensymb}
\usepackage{booktabs}
\usepackage{tabularx}
\usepackage{makecell}
\usepackage{enumerate}

\newcommand{\ctpo}{\texttt{C2PO-Torus}}
\newcommand{\myt}{\texttt{MYTorus}}

\title[X-Ray Analysis of 12MGS AGN]{\textit{A New Hope} for AGN SED Fitting: X-Ray Spectral Analysis of 12MGS AGN with the C2PO-Torus Model}

\author[C. J. E. Gilbert et al.]{
Carys J. E. Gilbert,$^{1}$\thanks{E-mail: carysjegilbert@gmail.com},
Lucia Marchetti,$^{1,2,3}$
Luigi Barchiesi,$^{1,2,3}$
Mattia Vaccari,$^{1,2,3}$
Cristian Vignali,$^{4,5}$
\newauthor
Stefano Marchesi,$^{4,5,6}$
Francesca Pozzi,$^{4,5}$
Francesco Salvestrini,$^{7,8}$
Anna Feltre,$^{9}$
Malebo Moloko,$^{1}$
\newauthor
and
Carlotta Gruppioni $^{4}$
\\
$^{1}$Department of Astronomy, University of Cape Town, Private Bag X3, Rondebosch 7701, South Africa\\
$^{2}$Inter-University Institute for Data Intensive Astronomy, University of Cape Town, Private Bag X3, Rondebosch 7701, South Africa\\
$^{3}$INAF–Istituto di Radioastronomia, Via Piero Gobetti 101, I-40129 Bologna, Italy\\
$^{4}$INAF–Osservatorio di Astrofisica e Scienza dello Spazio (OAS), Via Piero Gobetti 93/3, I-40129 Bologna, Italy\\
$^{5}$Dipartimento di Fisica e Astronomia (DIFA) Augusto Righi, Università di Bologna, via Gobetti 93/2, I-40129 Bologna, Italy \\
$^{6}$Department of Physics and Astronomy, Clemson University, Kinard Lab of Physics, Clemson, SC 29634, USA \\
$^{7}$INAF, Osservatorio Astronomico di Trieste, via Tiepolo 11, I-34131, Trieste, Italy\\
$^{8}$IFPU, Institute for Fundamental Physics of the Universe, Via Beirut 2, 34014 Trieste, Italy \\
$^{9}$INAF/Osservatorio Astrofisico di Arcetri, Largo E. Fermi 5, I-50125 Firenze, Italy \\
}

\date{Accepted XXX. Received YYY; in original form ZZZ}

\pubyear{\the\year{}}

\begin{document}
\label{firstpage}
\pagerange{\pageref{firstpage}--\pageref{lastpage}}
\maketitle

\begin{abstract}

In this paper, we present the results of comprehensive, broadband X-ray spectral analysis of a subsample of 43 mid-Infrared (IR) selected star-forming active galactic nuclei (AGN). We analysed archival \textit{NuSTAR}, \textit{Chandra}, and \textit{XMM-Newton} data, and present the first pointed X-ray observation of IRASF03450+0055. We introduce a new physically-based X-ray spectral model, \ctpo, based on a two-phase clumpy torus geometry. It is designed to provide constraints on physical parameters of the torus that can be directly linked to the parameters of the AGN model SKIRTOR. This model will allow us to exploit the synergies between the X-ray and IR regimes, place stronger constraints on the torus properties, and improve spectral energy distribution (SED) fitting by providing priors for the SKIRTOR parameters. We compare the results from the \ctpo \ fitting to those obtained from traditional models, finding that it is an extremely effective tool for fitting a wide variety of X-ray spectra. We observe that our sample closely follows the established  $L_{ \rm [OIII]}- L_{\rm X}$ AGN relation, while presenting an IR and Radio excess with respect to the $L_{\rm X} - L_{\rm 12 \ \mu m}$ and $ L_{ \rm X}- L_{\rm 1.4 \ GHz}$ AGN relations at low luminosities ($L_{\rm 2-10 \ keV} < 10^{41.5} \rm \ erg \ s^{-1} $). We believe that this excess is due to contamination from star formation related emission. We anticipate that the \ctpo \ model will aid substantially in completing full-spectrum SED fitting of this sample, allowing us to properly disentangle the different emission components of these objects, which are local analogues of AGN at Cosmic Noon.

\end{abstract}




\section{Introduction} \label{sec:Introduction}

    The comoving rates of central supermassive black hole accretion and star formation (SF) follow a similar trend of rising and falling as a function of redshift. Most notably, they peak at the same point in time \citep[at z $\sim 1-3$ or `Cosmic Noon', e.g.][]{Burgarella2013}. Additionally, there have been multiple scaling relations found between the mass of the central SMBH and its host galaxy's properties, such as the bulge mass and luminosity. Together, this indicates possible co-evolution of supermassive black holes (SMBHs) and their host galaxies \citep[e.g.][]{Shankar2009, Kormendy2013}, and potentially a positive feedback cycle \citep[see][]{King2005}. However, the exact nature of this feedback cycle, or if there even is any co-evolution at all, is still up for debate. One of the challenges in studying these cycles is that the galaxies at Cosmic Noon contain significant quantities of dust that extinguish most of the optical and ultraviolet (UV) radiation, making it incredibly difficult to study their inner workings. Overall, less than 10\% of the galaxies' emission is visible at optical or UV wavelengths \citep[see][]{Madau2014}. As the optical and UV photons are absorbed by dust grains, the grains are heated up, and re-emit this emission in the mid- and far-IR, causing the galaxies to appear incredibly IR bright. Telescopes such as \textit{Herschel} \citep{Herschel} have allowed for detailed photometric studies of these galaxies. These studies enabled the discovery of the dominance of this epoch by objects known as `SF-AGN' \citep[per][]{Gruppioni2013}. These galaxies have evidence for both high SF rates (SFR) and active galactic nuclei (AGN) through the study of their UV to far-IR spectral energy distributions (SEDs). These dusty SF-AGN hold the key to understanding the whole story of galaxy and AGN evolution during that obscured era, but given that so much of the emission is heavily obscured, the next best option is to instead study local analogues. In this case, even for incredibly dusty galaxies, their proximity makes it much easier to study them in detail, with both ground- and space-based telescopes where possible. 
    
    One local sample of galaxies that can be used to study SF-AGN objects is the 12 Micron Galaxy Sample \citep[12MGS,][]{Rush1993}, which appear to have similar SEDs to the SF-AGN population at Cosmic Noon. The 12MGS consists of 893 sources selected as having $12 \ \mu \rm m$ flux $\geq 0.22 \rm \ Jy$, from the \textit{Infrared Astronomical Satellite} (\textit{IRAS}) Faint Source Catalogue Version 2 \citep{Moshir1990}. 118 of these galaxies are classified as AGN \citep[see][]{Spinoglio1989, Hewitt1991, Veron1991}. The 12MGS is an incredibly well-studied sample with a plethora of multiwavelength observations to draw from, such as IR from \textit{Spitzer} \citep{Spitzer} and \textit{Herschel}, UV from \textit{GALEX} \citep{Galex}, X-ray from \textit{NuSTAR} \citep{NuSTAR}, \textit{XMM-Newton} \citep{XMM} and \textit{Chandra} \citep{Chandra}, and more. Thus, the 12MGS galaxies are the perfect laboratories for studying the interactions within galaxies with both high SFR and AGN, which will help with understanding their high-z analogues. \citet{Gruppioni2016} collected multiwavelength data for 76 of the 118 AGN that had available high-resolution \textit{Spitzer}-IRS spectra. Using this UV to far-IR coverage, \citet{Gruppioni2016} conducted a comprehensive study of this 12MGS subsample using SED fitting, constraining physical quantities such as SFR, intrinsic AGN luminosity, stellar mass, and AGN fraction. In order to properly do so, it is crucial that the contributions to the overall SED by the AGN and SF components are properly separated, which is a non-trivial task in the case of low-luminosity or heavily obscured AGN. Following this, \citet{Feltre2023} obtained homogeneous optical spectra using the \textit{Southern African Large Telescope} (\textit{SALT}) for the 43 members of this subsample that are within the Southern Sky. They tested how different optical line ratio diagrams can be used to identify AGN, and looked at using them to unravel the relative contributions of AGN and SF in these IR galaxies. However, these optical/IR line tracers are not perfect, especially when the entire galaxy's emission is captured within the spectral slit, and the AGN emission is diluted. Thus, in order to properly separate out the AGN and SF components within these galaxies, and better understand their interplay as proxies for the dusty IR-galaxies at Cosmic Noon, we need to explore wavelengths that exclusively trace the AGN or SF activity without contamination. The X-ray regime is ideal in this case, as the intrinsic X-ray AGN luminosity traces the innermost region of the AGN. The X-ray emission from AGN is believed to originate from the hot corona \citep[a hot and low-density region located `above' the SMBH accretion disk, see][for more details]{Haardt1991, Haardt1993} and is directly linked to the AGN power, making it one of the best proxies for the AGN bolometric luminosity. 
    
    In this paper, we aim to complete the first homogeneous X-ray spectral analysis of the 43 12MGS galaxies with SALT spectra from \citet{Feltre2023}. Our goal is to obtain the intrinsic AGN X-ray luminosity and to further constrain the physical geometry of the AGN, for the purpose of improving our understanding of the AGN emission of these galaxies and disentangling it from the galaxies' SF emission. There are a number of physically-based X-ray spectral models for different AGN torus geometries that can be used to characterise their properties (e.g. \texttt{MYTorus} from \citealt{MYTorus}, \texttt{borus02} from \citealt{borus1,borus2}, \texttt{XCLUMPY} from \citealt{Tanimoto2019}, and \texttt{UXCLUMPY} from \citealt{Buchner2019}). However, the geometric parameters constrained by these models are not necessarily the same as those that are used when modelling the SED of the AGN emission. \citet{EsparzaArrendondo2025} tested if X-ray and mid-IR models with similar geometries could be used to simultaneously fit mid-IR to X-ray SEDs, and found that most of their sample favoured a combination of the X-ray \texttt{UXCLUMPY} torus model with clumpy and two-phased IR torus models, that have similar (but not identical) geometries. They also found that linking parameters from X-ray and mid-IR models helps to constrain other parameters and break degeneracies. As such, we have created and are presenting here the Clumpy 2-Phase Obscuring Torus (\texttt{C2PO-Torus}) X-ray spectral model, as a counterpart to the SKIRTOR IR AGN model \citep[][]{Skirtor2012, Skirtor2016}, which was created using the state-of-the-art radiative transfer software SKIRT \citep{SKIRT2015, SKIRT2020}. Since 2023, SKIRT has been expanded to cover the X-ray regime, allowing for the modelling of X-ray spectra \citep[see][]{SKIRT2023}. Using the updated SKIRT, we recreated the `two-phase' clumpy torus geometry of the SKIRTOR model, ensuring that we sampled identical parameter spaces, and generated a suite of physically-based, self-consistent X-ray models. Thus, with \texttt{C2PO-Torus}, the X-ray spectra of the galaxies we are studying can be better utilised to understand the AGN emission, which would lead to improvements in the SED fitting and gaining a better understanding of the SF-AGN interactions within these galaxies. This paper is the first in a series of two, the second of which will cover the full-spectrum SED analysis of these sources, using the results presented in this paper. Our goal is to exploit our extensive multiwavelength coverage to properly disentangle the star formation and AGN emission components, and better understand the SF-AGN interplay within these local proxies of the IR-galaxies at Cosmic Noon.

    The paper is organised as follows: in Section \ref{sec:Observations and Data Reduction}, we describe the sample and the data reduction processes. In Section \ref{sec:C2PO-Torus Model}, we discuss the \ctpo \  model geometry and details of the simulation. Section \ref{sec:X-Ray Spectral Analysis} describes the spectral analysis of the sources using simple and other physically-based models, and \ctpo. In Section \ref{sec:Results and Discussion}, we present and compare the results from the traditional fitting methods with those from the physical models and \ctpo, and test our results against multiwavelength scaling relations. Finally, in Section \ref{sec:Conclusions} we present our conclusions. Throughout the paper we adopt a $\Lambda$CDM cosmology with $(\Omega_\text{M}, \Omega_\Lambda, \text{H}_0) = (0.315, 0.685, 67.4 \text{ km s}^{-1} \text{ Mpc}^{-1})$ from \citet{Planck2020}. All spectral fitting results are presented with uncertainties at a 90\% confidence level.

\section{X-ray Observations and Data Reduction} \label{sec:Observations and Data Reduction}

    \subsection{Sample} \label{sec:Sample}

        We analysed the broadband X-ray spectra of all 43 sources, introduced in Section \ref{sec:Introduction}, that were observed with the \textit{NuSTAR}, \textit{XMM-Newton} and/or \textit{Chandra} telescopes. The details of the observations are summarised in Table \ref{tab:sources}. The observations were chosen as follows: if there were simultaneous \textit{NuSTAR} and \textit{XMM-Newton}/\textit{Chandra} observations, we chose the longest pair of them. This was done to try to ensure there is no variability between the observations, and as longer observations generally have more counts and thus a higher spectra signal-to-noise ratio (SNR). Where no simultaneous observations are available, we chose the longest \textit{NuSTAR} observation, alongside the closest \textit{XMM-Newton}/\textit{Chandra} observation. In certain cases, for heavily-obscured sources where there was no significant variability, we preferred a longer \textit{XMM-Newton}/\textit{Chandra} over a temporally closer one to ensure the best available spectra were used. When more than one observation was available for a given instrument, and they were all taken within a short enough time span, we combined the spectra to improve the SNR after ensuring that there was no significant variability. We rebinned all final spectra into bins of 20 counts to improve the SNR and enable the usage of the standard goodness-of-fit test for Gaussian data (the $\chi^2$ test). 

    \subsection{\textit{NuSTAR}} \label{sec:NuSTAR}

        We reduced the \textit{NuSTAR} data for both Focal Plane Modules A and B (FPMA/B) using the \textit{NuSTAR} Data Analysis Software (NuSTARDAS; v2.1.4) commands \texttt{nupipeline} and \texttt{nuproducts}. The spectra were extracted from circular regions with a $50''$ radius centred on the source, which corresponds to at least $50 \%$ of the encircled energy fraction (EEF) on the $3-79 \rm \ keV$ energy band \citep{An2014}. The background was extracted from a nearby, source-free region with a $50''-75''$ radius. The FPMA/B spectra were kept separate during fitting.

    \subsection{\textit{XMM-Newton}} \label{sec:XMMNewton}

        We conducted the data reduction for all three \textit{XMM-Newton} cameras (EPIC-pn and EPIC-MOS1/2) using the Science Analysis Software (SAS; v21.0.0). The \texttt{epproc} and \texttt{emproc} commands were used to reprocess the data, and we generated the light-curves using \texttt{evselect}. Periods of background flaring were removed by generating Good-Time Interval (GTI) files using \texttt{tabgtigen}. We extracted the source counts from circular regions centred on the source of $15''-30''$ radius, which corresponds to an EEF of at least $70 \%$ for all three instruments \citep{XMMEFF}. We extracted the background counts from a nearby source-free circular region of radius $5''-40''$, on the same CCD chip. In general, we tried to ensure that the regions were as large as possible and that the background region radius was equal to or larger than the source radius, but in cases where there were bright patterns from the spider support of the mirrors, we chose a smaller region to ensure that the source contributions within the background spectrum were minimised. We checked the regions for pileup using \texttt{epatplot}, and observations with significant pileup $(\geq 20 \%)$ were discarded. Finally, we extracted the source/background spectra with \texttt{evselect}, and the ancillary response file (ARF) and response matrix file (RMF) were generated using \texttt{arfgen} and \texttt{rmfgen}, respectively. Once again, the spectra of each instrument were kept separate during fitting.
    
    \subsection{\textit{Chandra}} \label{sec:Chandra}
    
        We reduced the \textit{Chandra} data using the \textit{Chandra} Interactive Analysis of Observations software \citep[CIAO; v4.17.0 and CALDB 4.12.0,][]{CIAO}. We used \texttt{chandra\_repro} to reprocess the data, and then used a circular region of radius $3''-7''$ to extract the source spectrum, which corresponds to an EEF of at least $90 \%$ \citep{ChandraEFF}. We extracted the background from a $5''-15''$ radius source-free region on the same CCD. We checked the regions for pileup using \texttt{WebPIMMS}\footnote{\url{https://cxc.harvard.edu/toolkit/pimms.jsp}} and for observations with significant pileup $(\geq 20 \%)$, we instead extracted the source spectrum using an annulus centred at the source, to exclude the region suffering the most from pileup. To check for variability or flaring during the observation, we used \texttt{dmextract} to produce a light curve. We then extracted the spectra with \texttt{specextract}, and in cases where the spectra were combined, we used \texttt{combine\_spectra} to generate combined source and background spectra, with corresponding ARF and RMF files.

    \subsection{The First Pointed X-ray Observation of IRASF03450+0055}
    \label{sec:First Pointed Obs of IRAS}
    
            IRASF03450+0055 is a galaxy that has been classified as both a Narrow-Line Seyfert 1 \citep[NLS1,][]{Giannuzzo1996} and a Seyfert 1.5 \citep{Veron2006} from its optical spectra, but had previously not been studied in the X-ray. We submitted a \textit{NuSTAR} Cycle 11 proposal (Proposal ID 11077, PI: C. Gilbert) to collect data for IRASF03450+0055. We did so for two main reasons: firstly, to characterise and better understand the central AGN through the fitting of the X-ray spectra. In particular, we looked for the presence of features such as Fe K lines, reflection, scattering, and the underlying power law continuum that allowed us to measure the intrinsic luminosity of the AGN. We also aimed to better constrain physical attributes such as the AGN opening angle and inclination. Additionally, the ambiguity of its classification can be resolved using \textit{NuSTAR} data, as NLS1s have very steep photon indices ($\Gamma \sim 2.6$) and notable reflection in their X-ray spectra, which peaks at $E > 10 \ \rm{keV}$ \citep[see][]{Gallo2018, Fabian2002, Zoghbi2008}. Secondly, the \textit{NuSTAR} coverage would allow us to complete the X-ray catalogue of all of the sources in our sample.

\begin{figure}
                \centering
                \includegraphics[width=0.98\linewidth]{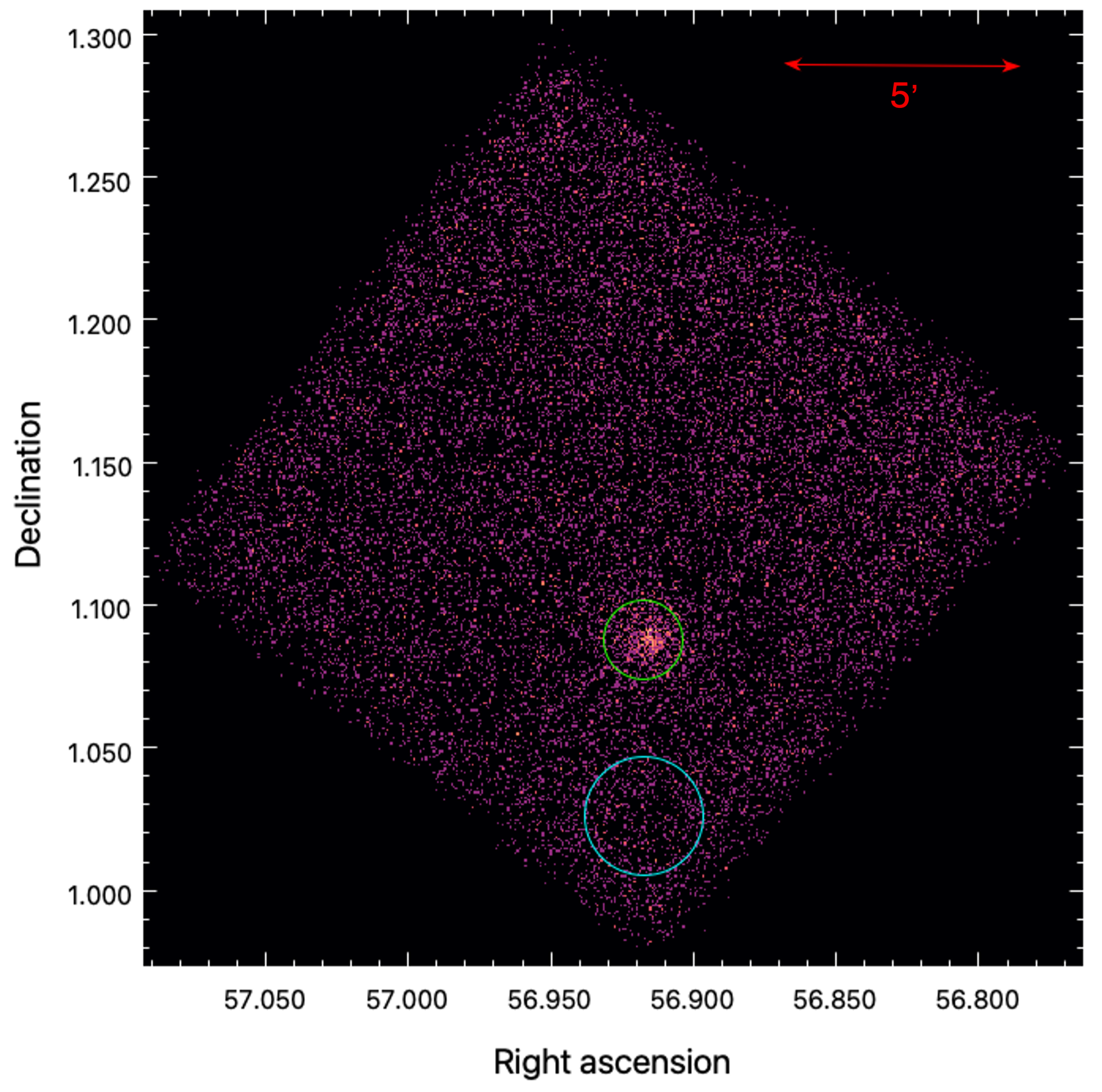}
                \caption{The FPMA Level 2 event file of the IRASF03450+0055 observation. The source and background extraction regions are plotted atop as the small green circle ($50''$ radius) and large blue circle ($75''$ radius) respectively.}
                \label{fig:ds9_regions}
            \end{figure}

            IRASF03450+0055 was observed by \textit{NuSTAR} starting on the 16th of August 2025, for a total of $\sim 22$ ks (OBSID 61101016002). The observation was offset by $3.2'$, and there was no noticeable flaring or variability during the observation. The data was processed following the methods outlined in Section \ref{sec:NuSTAR}. The background-corrected count rates for the FPMA and FPMB observations are $2.41 \times 10^{-2} \rm  \ counts \ s^{-1}$ and $2.28 \times 10^{-2} \rm  \ counts \ s^{-1}$ respectively. The FPMA event file is plotted in Figure \ref{fig:ds9_regions}, with the source and background count extraction regions plotted atop.

\section{C2PO-Torus Model} \label{sec:C2PO-Torus Model}

    Our goal in developing the Clumpy 2-Phase Obscuring Torus model was to create a self-consistent, physically-based model that can be used as a part of the SED fitting process for galaxies containing AGN. Specifically, we wanted a counterpart to the SKIRTOR SED models \citep{Skirtor2012, Skirtor2016}, which have been implemented inside of the CIGALE SED fitting software \citep{Boquien2019, Yang2020, Yang2022}. Using \ctpo, the results from X-ray spectral analysis can be used as priors for the SED fitting and vice-versa, which will help to separate out and put stronger constraints on the AGN and host galaxy components, improving our understanding of the AGN emission within the SED and the galaxy as a whole. This section describes the creation of \ctpo, and Section \ref{sec:C2PO-Torus Modelling} details how it may be used for spectral fitting.

    \subsection{Simulation with SKIRT} \label{sec:Simulation with SKIRT}
    
    To generate \ctpo, we used the SKIRT code \citep{SKIRT2015, SKIRT2020}, an open-source software for simulating continuum radiation transfer in astrophysical systems. Based on the Monte Carlo algorithm, it can emulate the physical processes present within the AGN torus, such as anisotropic scattering, absorption and (re-)emission by dust. It is able to handle any 3D geometry, making it ideal for simulating complex clumpy torus structures in great detail. The SKIRT model has recently been updated to include X-ray physical processes in gas and dust \citep{SKIRT2023}, and the authors verified its capability by testing it against other popular physical models, such as \texttt{MYTorus}. In \citet{XSKIRTORpol}, the authors released the \texttt{XSKIRTOR\_SMOOTH} model, which simultaneously describes X-ray spectra and spectro-polarimetry, and is ideal for the X-ray spectra obtained with the \textit{XRISM} telescope \citep{XRISM}. However, this model assumes a smooth, homogeneous dust distribution within the torus and is therefore not an ideal counterpart to SKIRTOR. Thus, we decided to utilise the proven capabilities of SKIRT to generate a new two-phase clumpy torus model that can be used to fit the X-ray spectra of our sample, and works as a direct counterpart for the SED model SKIRTOR.

    \subsection{Model Geometry and Description} \label{sec:Model Geometry and Description}

\begin{figure}
            \centering
            \includegraphics[width=\linewidth]{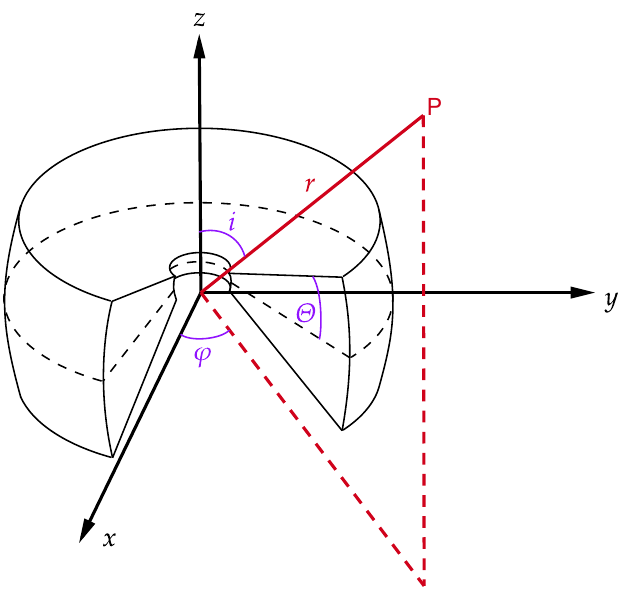}
            \caption{Schematic of the torus geometry and coordinate system. In this plot, $i$ is the inclination, $\Theta$ is the torus opening angle, and $\varphi$ is the azimuthal angle. Adapted from \citet{Skirtor2012}.}
            \label{fig:TorusGeometry}
        \end{figure}

         Following \citet{Skirtor2012, Skirtor2016}, our model consists of a `two-phase' clumpy wedge-shaped torus, in which the dust is distributed between high-density clumps and low-density interclump regions. There are many pieces of observational evidence that point towards AGN tori being clumpy \citep[see][and references within]{Tanimoto2019}, and clumpy torus models have been successfully used to reproduce AGN spectral features at both IR and X-ray wavelengths \citep[e.g.][]{Nenkova2002, Nenkova2008a, Nenkova2008b, Honig2007, Honig2010a, Honig2010b, Tanimoto2019, Tanimoto2020, Buchner2019, Buchner2021, Marchesi2022}. Thus, we expect that the torus geometry used for the SKIRTOR models should accurately reproduce the X-ray spectra observed from dusty AGN. The geometry of the torus and coordinate system used is shown in Figure \ref{fig:TorusGeometry}.
        
         The AGN primary emission is approximated as a point-like energy source at the centre of our coordinate system, anisotropically emitting according to the \citet{Netzer1987} emission profile, given in Equation \ref{eq:netzer}.
        \begin{equation}
            L(i) \propto \cos i(2\cos i +1)
        \label{eq:netzer}
        \end{equation}
    
        The shape and size of the torus are defined by the following parameters: the inner and outer radii, $R_{\rm in}$ and $R_{\rm out}$, and the half-opening angle $\Theta$, which defines the maximum vertical extent of the torus. The SKIRTOR models account for different $R_{\rm ratio}$ values, where $R_{\rm ratio} = R_{\rm out}/R_{\rm in}$. Due to computing constraints, we decided to work with a fixed $R_{\rm ratio} =30$. We chose this value as it is the default value used when fitting SEDs with SKIRTOR, and it was the original value used in \citet{Skirtor2012}. The anisotropic primary emission reshapes the inner radius of the torus according to the same polar angle relation, as can be seen in Equation \ref{eq:reshapewalls}.
        \begin{equation}
            { R_{\rm in} } \propto { R_{\rm iso}}\sqrt{\cos i(2\cos i +1)}
        \label{eq:reshapewalls}
        \end{equation}
        Where $R_{\rm iso}$ is the inner radius in the case of isotropic emission. The relation between the half-opening angle and the AGN covering factor (CF) is given by Equation \ref{eq:CF}.
        \begin{equation}
            \text{CF} = \sin\Theta
        \label{eq:CF}
        \end{equation}
        The dust and gas are initially distributed within the torus according to the density gradient described by Equation \ref{eq:dust_gradient}.
        \begin{equation}
            \rho(r,i) \propto r^{-p}e^{-q| \cos i|} 
        \label{eq:dust_gradient}
        \end{equation}
        where $r$ is the radius of the torus, and $i$ is the inclination, as shown in Figure \ref{fig:TorusGeometry}. Once the dust has been distributed throughout the torus according to Equation \ref{eq:dust_gradient}, the clumps are randomly generated within the torus.
        
        The parameters defining the ``clumpiness'' of the torus are the total number of clumps ($\rm N_{clumps}$), the radius of the clumps ($R_{\rm clump}$), the fraction of the total torus that is occupied by the clumps (the `filling factor'), and the fraction of total mass of the torus locked up inside clumps ($f_{\rm clumps}$), with the remainder of the mass being the interclump medium. The relation between the filling factor and the other parameters is given by Equation \ref{eq:fillingfactor}, which assumes spherical clumps and a wedge-shaped torus with no inner radius reshaping (i.e. a classical torus, ignoring the reshaping of Equation \ref{eq:reshapewalls}). While this is not the exact case for our model, the filling factor does not change significantly for our clump/torus configuration.
        \begin{equation}
            \text{Filling Factor} = \frac{V_{\text{clumps}}}{V_{\text{torus}}} = \frac{\frac{4}{3}\pi N_{\text{clumps}}R_{\text{clump}}^3}{\frac{4}{3}\pi(R^3_{\text{out}}-R^3_{\text{in}})\cos(90- \Theta)}
        \label{eq:fillingfactor}
        \end{equation}
         
\begin{figure}
            \centering
            \includegraphics[width=\linewidth]{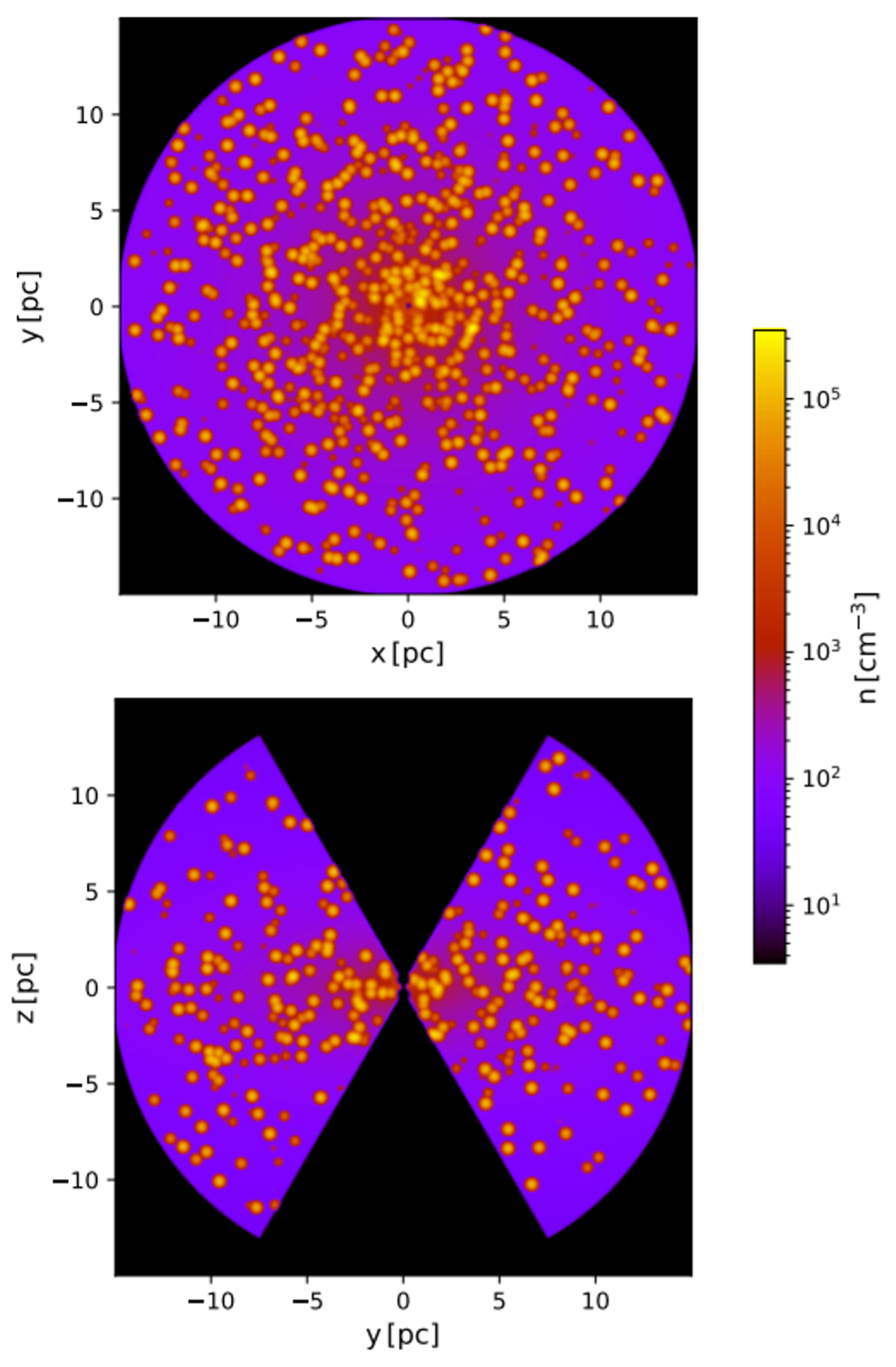}
            \caption{Density map of the xy plane (top) and xz plane (bottom) slices, showing the distribution of clumps within the torus. This particular torus has the following parameters: $\Theta = 60 \degree$, $p=1$, $q=1$, and $\mathrm{N_{H, eq}} = 5 \times 10^{23} \rm \ cm^{-2}$. Higher density clumps are shown in yellow, while lower density interclump regions are plotted in purple. The reshaping of the inner wall due to the anisotropic emission can be clearly seen in the bottom plot (see Equation \ref{eq:reshapewalls}).}
            \label{fig:Density}
        \end{figure}

        For the density profile of the individual clumps, we assumed the standard cubic spline density smoothing kernel\footnote{ \url{https://skirt.ugent.be/skirt9/class_cubic_spline_smoothing_kernel.html}}, as implemented in SKIRT. The total amount of obscuring matter in the torus was defined through the normalisation of the equatorial number column density ($\mathrm{N_{H, eq}}$), which is along the x-axis in Figure \ref{fig:TorusGeometry}. 
        Figure \ref{fig:Density} shows an example of a torus, and the matter distribution within it.

    \subsection{Parameter Grid} \label{sec:Parameter Grid}

        Here we present the range of values of parameters used to generate our models. In general, we tried to replicate the exact values used to generate the SKIRTOR models, but for the sake of computational and time efficiency, some values were omitted. For example, we chose not to sample the full range of the radial dust distribution gradient parameter ($p$), as it only results in small spectral differences that are not differentiable at the resolution of most current X-ray data. For explanations of the choice of the torus parameters, see \citet{Skirtor2012, Skirtor2016}. Table \ref{tab:ParameterGrid} lists the parameter ranges sampled when generating the models.
        
        \begin{table}
            \centering
            \begin{tabularx}{0.8\linewidth}{XlX}
                \toprule
                Parameter & Values & Units \\
                \midrule
                $L$ & 1 $\times 10^{43}$ & ergs s$^{-1}$  \\
                $R_{\mathrm{in}}$ & 0.5 & pc  \\
                $R_{\mathrm{out}}$ & 15 & pc  \\
                $R_{\mathrm{iso}}$ & 0.21 & pc  \\
                $R_{\mathrm{ratio}}$ & 30 &   \\
                $R_{\mathrm{clump}}$ & 0.4 & pc  \\
                Filling Factor & 0.25 &   \\
                $f_{\mathrm{clumps}}$ & 0.97 &   \\
                $p$ & 0, 1 &   \\
                $q$& 1 &  \\
                $\Theta$ & $10 -80$ & degrees  \\
                $i$ & $0 - 90$ & degrees  \\
                $\mathrm{N_{H, eq}}$  & $1 \times 10^{21} - 5\times 10^{25}$ & cm$^{-2}$  \\
                $\Gamma$ & $1.4-2.6$ &  \\
                \hline
            \end{tabularx}
            \caption{The grid of parameters used to compute the models. $L$ is the luminosity of the anisotropic primary emitter. $R_{\mathrm{in}}$ and $R_{\mathrm{out}}$ are the inner and outer radii of the torus. $R_{\mathrm{iso}}$ is the minimum radius of the torus once reshaped by the anisotropic emission. $R_{\mathrm{clump}}$ is the radius of the clumps. The filling factor is the fraction of the torus volume that is occupied by the clumps (see Equation \ref{eq:fillingfactor}). $f_{\mathrm{clumps}}$ is the fraction of total dust mass of the torus locked up inside clumps. $p$ and $q$ are the dust distribution gradient parameters, as per Equation \ref{eq:dust_gradient}. $\Theta$ is the opening angle of the torus, $i$ is the inclination of the torus relative to our line of sight, and both parameters are sampled in steps of $10^{\circ}$. $\mathrm{N_{H, eq}}$ is the equatorial column density of the torus, and it is sampled in steps of 0.1 dex. $\Gamma$ is the power law slope of the primary emission, sampled in steps of 0.1.}
            \label{tab:ParameterGrid}
        \end{table}
        In our notation, $i = 0\degree$ represents a Seyfert 1 (face-on) AGN and $i = 90\degree$ a Seyfert type 2 (edge-on) AGN. We note that the equatorial column density $\mathrm{N_{H, eq}}$ is the normalised value for the initial smooth model before the clumps are generated. In practice, the actual equatorial column density would be different depending on the random clump placement. Thus, it should be treated as an average equatorial column density that can be used to estimate the average line-of-sight (LOS) column density with Equation \ref{eq:nh_conversion}.
        \begin{equation}
            \mathrm{N_{H, LOS} = N_{H, eq}} \times e^{-|\cos i|}
        \label{eq:nh_conversion}
        \end{equation}
        We emphasize that Equation \ref{eq:nh_conversion} is derived from the smooth dust distribution given by Equation \ref{eq:dust_gradient}. It is used only to provide an approximate line-of-sight value for comparison with literature values, and plays no part in the generation or the fitting of the models. In reality, the random clump placement within the torii may lead to LOS values that differ from the estimates found using Equation \ref{eq:nh_conversion}, so caution should be exercised when utilising these values. Additionally, we stress throughout that $\mathrm{N_{H,eq}}$, and not $\mathrm{N_{H,LOS}}$, is the quantity constrained by our fits, and thus is a reliable estimate.
        
        All models were calculated on a spherical grid, with 150 linearly distributed bins along each axis. We found that this grid configuration provided the best balance between spectral detail and required compute time. For each model, we also calculated four different azimuthal angles ($\varphi = 0 \degree, 90\degree, 180\degree, 270\degree$), and then averaged the spectra across them. This was done to ensure that each spectrum represented a better approximation of the intrinsic equatorial column density, as the random clump generation may result in a given line-of-sight being under-/over-obscured. Additionally, each simulation run generates a new, random clump distribution, possibly leading to discontinuities between different runs. By averaging across multiple azimuthal angles, we solve this problem.
        
        Finally, we present the directly absorbed X-ray emission and reprocessed X-ray emission (which includes the scattered and reflected X-ray emission) as two separate components, \texttt{C2POTorusD} and \texttt{C2POTorusR}, respectively, which should be linked during fitting. Details of model usage are given in Section \ref{sec:C2PO-Torus Modelling}.

\section{X-Ray Spectral Analysis} \label{sec:X-Ray Spectral Analysis}

    We performed the X-ray spectral analysis using the X-Ray Spectral Fitting Package \citep[XSPEC; v12.14.1,][]{XSPEC}. We considered the standard energy ranges of $3-79 \ \rm keV$, $0.1-10 \ \rm keV$ and $0.3-7 \ \rm keV$ for \textit{NuSTAR}, \textit{XMM-Newton} and \textit{Chandra}, respectively. In cases of low SNR, we also excluded the energy ranges where the signal was below the background noise level. We fitted the spectra simultaneously for each source, using a cross-calibration constant to account for instrumental differences and source variability. This constant was fixed at unity for the FPMA spectrum, or in cases where \textit{NuSTAR} was not used, the \textit{XMM-Newton} EPIC-pn or \textit{Chandra} spectrum. The Galactic absorption was modelled using the \texttt{phabs} model, with the equivalent hydrogen column density ($\rm N_{H}$) fixed at the value given by the 2D HI4PI map \citep{HI4PI}. We initially modelled all the sources with a ``phenomenological'' model consisting of multiple components that account for the different spectral features, as detailed in Section \ref{sec:Phenomenological Modelling}. For sources with complex spectra that  could not be successfully fit with the phenomenological models, or those identified in the literature as being Compton-thick AGN, we instead used the \texttt{MYTorus} physically-based model, as explained in Section \ref{sec:Modelling with MYTorus}. We concluded the fitting once all major spectrum features had been modelled successfully, with a $\chi^2 \rm /d.o.f. \leq 1.5$ for the entire model. Once we were satisfied with our fits and results, we then refitted the sources with \ctpo \ (Section \ref{sec:C2PO-Torus Modelling}), in order to estimate physical torus parameters that can be used as priors during SED fitting. In all cases, we used the \texttt{error} function to calculate the errors of the fitted components with a $90\%$ confidence interval.

    \subsection{Phenomenological Modelling} \label{sec:Phenomenological Modelling}

        We started each fit with a base model of \texttt{const}$\times$\texttt{phabs}$\times$\texttt{po}, where \texttt{const} is the cross-calibration between \textit{NuSTAR} and the soft-X-ray observatories (\textit{XMM-Newton} and \textit{Chandra}), \texttt{phabs} models the Galactic absorption from our Galaxy, and \texttt{po} represents the intrinsic powerlaw emission of the AGN. After fitting with this initial model, we added additional components as necessary. In order to ensure that any added components made a statistically significant difference to the fits, we performed an ``F-test'' and discarded the new component if its significance was below 3-$\sigma$. We generally left the photon index of the primary power law to vary, but in cases where it could not be constrained, we fixed it at 1.8, which is a typical value for unobscured AGN \citep[e.g.][]{DelMoro2017}. If there was evidence for absorption within the soft part of the spectrum, we added the \texttt{zpcfabs} component to model any intrinsic absorption within the AGN caused by the obscuring torus. We first fixed the covering factor (CF) at 1, but we allowed it to vary if there were significant residuals at $ E \leq 3$ keV. We used the \texttt{zxipcf} model either instead of or in addition to \texttt{zpcfabs} to model partially ionised absorbing material where necessary. If there was remaining soft excess, we added the \texttt{mekal} component to account for diffuse hot plasma emission from the host galaxy. If this did not model the excess in its entirety, we added a scattering component consisting of \texttt{const}$\times$\texttt{po}, with the normalisation and slope of this power law linked to that of the primary, and the constant allowed to vary up to $10\%$. We accounted for any emission/absorption lines with the \texttt{zgauss} model, fixing the width of the line at $\sigma = 0.01 \ \rm keV$ unless otherwise stated. Finally, if there was excess in the hard part of the spectrum $( E \geq 10 \ \rm keV)$, we added either \texttt{pexrav} or \texttt{pexriv} to model reflection from cold neutral or warm ionised material close to the accretion disk. In the cases where there was evidence for relativistic reflection, such as residuals from a broadened Fe K emission line and/or excess reflection that could not be fit with the other models, we used \texttt{relxill}. These reflection models can reproduce many different spectral features, however we choose to use them in a ``reflection only'' mode by linking the photon index and normalisation to that of the main power law, and setting the reflection fraction $R_{\text{frac}} \leq 0$. We left all other values fixed at the default values unless otherwise stated. Once the fit procedure was optimised, we used \texttt{clumin} to calculate the intrinsic X-ray luminosity in the $2-10 \rm \ keV$ band.
        
    \subsection{Modelling with \texttt{MYTorus}} \label{sec:Modelling with MYTorus}

    For two of our sources (ESO033-G002 and MCG-03-34-064), we found that the spectra were too complex to fit with just phenomenological components, as we found residuals $> 3 \sigma$ at $ E \geq 3 \rm \ keV$ and $\chi^2 \rm /d.o.f. > 1.5$. Similarly, for all of the sources that were identified in literature as ``Compton-thick'' (11 sources with $\mathrm{N_{H, LOS}} \gtrsim 10^{23.5}-10^{24} \ \mathrm{cm}^{-2}$), the \texttt{zpcfabs} and \texttt{zxipcf} components are not ideal for modelling the obscuration as they do not account for the effects of Compton scattering (for less obscured sources this is not an issue, as the effects of Compton scattering are negligible). In these cases, we redid the fitting using the \texttt{MYTorus} model. The \texttt{MYTorus} model consists of three components: \texttt{MYTZ}, a multiplicative component that is applied to the main power law; \texttt{MYTS}, an additive component that represents the reprocessed photons; and \texttt{MYTL}, an additive component that accounts for the neutral Fe K$\alpha$ and K$\beta$ lines. We used \texttt{MYTorus} in its ``decoupled'' configuration, to mimic a clumpy torus by fitting a different $\rm N_{H, LOS}$ to the $\rm N_{H, eq}$. Practically, this is implemented by fixing the inclination angle of the continuum absorption at $90 \degree$ and having two sets of the additive components, one fixed at $90\degree$ and the other fixed at $0\degree$, with their own normalisation constants A$_{90}$ and A$_0$. The column densities of the two additive components are tied together, and represent the average $\rm N_{H}$, while the column density of the \texttt{MYTZ} component is fitted separately, and represents the $\rm N_{H, LOS}$. An example of a model in this configuration would appear in XSPEC as:
    \begin{equation*}
        \text{A}_{90}\times(\texttt{MYTS}_{90} + \texttt{MYTL}_{90}) + \text{A}_{0} \times (\texttt{MYTS}_{0} + \texttt{MYTL}_{0}) + (\texttt{MYTZ}_{90}\times \texttt{po})
    \end{equation*}
    Additional model components can be used alongside the \texttt{MYTorus} components, such as \texttt{mekal} for the soft excess. Once we had concluded the fitting of the spectra, we used \texttt{clumin} on the power law component (\texttt{po}) to calculate the intrinsic $2-10 \rm \ keV$ luminosity.

    \subsection{C2PO-Torus Modelling} \label{sec:C2PO-Torus Modelling}

\begin{figure}
        \centering
        \includegraphics[width=\linewidth]{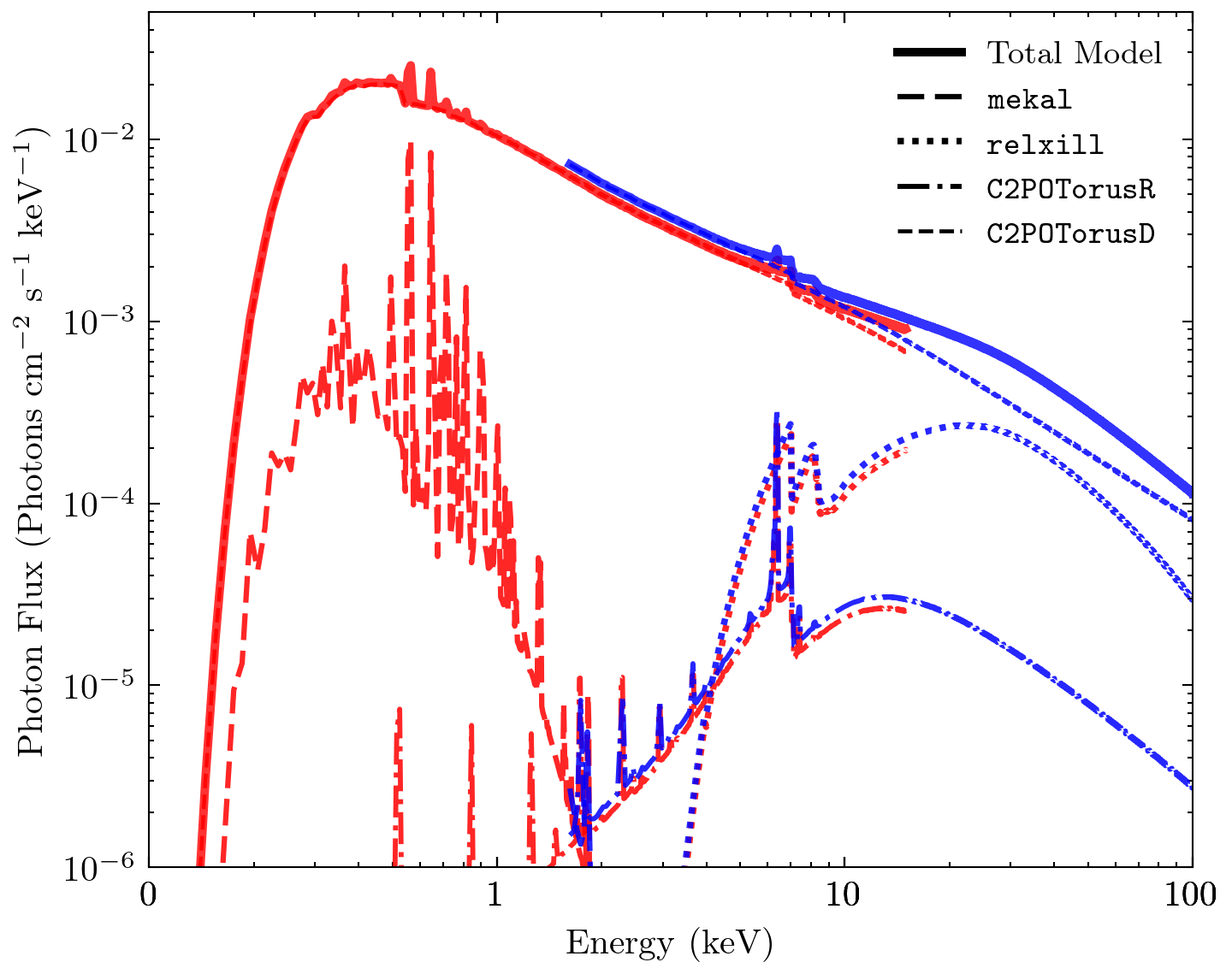}
        \caption{The ``unfolded'' best-fit model of ESO141-G055. The model consists of a \texttt{mekal} component (dashed line), a \texttt{relxill} component (dotted line) and \ctpo. The two \ctpo \ components, \texttt{C2POTorusR} and \texttt{C2POTorusD}, are plotted as the dot-dash and shorter dashed lines, respectively. The total model is plotted as the solid line, and the different colours represent the \textit{XMM-Newton} (blue) and \textit{NuSTAR} (red) models. We measured the best-fit values of $\Gamma = 2.23 \pm 0.01$, $\rm N_{H, eq} = 90.0^{+4.7}_{-5.4} \times 10^{22} \ cm^{-2}$, $\Theta = 64.74^{+0.15}_{-0.20}$ degrees, $i = 26.32^{+0.24}_{-0.36}$ degrees, and $p=1$, as given in Table \ref{tab:carrot_results}.}
        \label{fig:c2po_exampleplot}
    \end{figure}

    As described in Section \ref{sec:C2PO-Torus Model}, the \ctpo \ model consists of two additive components: \texttt{C2POTorusD}, which represents the emission that is directly absorbed by the torus; and \texttt{C2POTorusR}, which represents the reprocessed emission, such as emission lines, reflection, and scattering. During fitting, all of the parameters of the two components should be linked. There can also be an additional scaling constant in front of the reprocessed emission component, $\mathrm{A_R}$, which can be fixed at 1 or left free to vary. To fit a spectrum using \ctpo, we started with the following configuration in XSPEC:
    \begin{equation*}
        \texttt{const} \times \texttt{phabs} \times ( \mathrm{A_R} \times \texttt{C2POTorusR} + \texttt{C2POTorusD} )
    \end{equation*}
    As with \texttt{MYTorus}, additional components can be added alongside \ctpo. An example of this is plotted in Figure \ref{fig:c2po_exampleplot}, which shows the model fitted to ESO141-G055 that consists of \ctpo, a \texttt{mekal} component for the soft excess, and \texttt{relxill} to model relativistic reflection. When adding in \texttt{relxill}, it is important to ensure that the photon index and normalisation are linked to that of \ctpo \ (as is standard when working with reflection models). The \texttt{relxill} component should also be obscured, and this can be achieved by using \texttt{zphabs} (and \texttt{cabs} for sources that have significant obscuration), with the $\rm N_H$ of these components linked to the \ctpo \ $\rm N_{H, eq}$ via Equation \ref{eq:nh_conversion}.
    
    To calculate the intrinsic $2-10 \rm \ keV$ luminosity, we freeze all parameters at their best-fit values, delete all model components except for \texttt{C2POTorusD}, set the $\mathrm{N_H} = 0.1$ (the lowest possible value, at which the effect of the obscuration is negligible), and then use the \texttt{lum} command. The uncertainties on the intrinsic luminosity are found using the relative uncertainties of the normalisation. This complicated process is required as \texttt{C2POTorusD} represents the absorbed emission, and thus using \texttt{clumin} would produce the absorbed luminosity, rather than the intrinsic luminosity. Similarly, as \texttt{clumin} is meant to be used without changing any parameters from their best-fit values, simply setting $\mathrm{N_H} = 0.1$ while using \texttt{clumin} results in an unsuccessful fit and inaccurate luminosity estimate.

\section{Results and Discussion} \label{sec:Results and Discussion} 

    \subsection{Spectral Fitting Results} \label{sec:Spectral Fitting Results}

        We present the results of the spectral fitting with phenomenological models in Table \ref{tab:phenomenological_results}, physically-based models in Table \ref{tab:physical_results}, and the \ctpo \ model in Table \ref{tab:carrot_results}. More details on the individual fits and literature comparisons for each source can be found in Appendix \ref{appen:Notes}. We, however, present the details of the modelling of IRASF03450+0055 in Section \ref{sec:Modelling IRASF03450+0055} as an example, and as this is the first X-ray spectral analysis of this source.
     
        \begin{table*}
            \centering
            \caption{Results from the phenomenological fitting of the X-ray spectra. All parameters that were fixed are marked with $*$. A $-$ indicates that the model component was not used, and a $u$ indicates that the uncertainty could not be constrained. $\log (L_{2-10 \rm \ keV})$ is the intrinsic luminosity in the rest-frame $2-10 \rm \ keV$ energy band. $\Gamma$ is the photon index. $\mathrm{N_H}$ is the line-of-sight obscuring column density, with $\mathrm{N_{H, N}}$ referring to a neutral absorber and $\mathrm{N_{H, I}}$ an ionised absorber. $f_{\mathrm{cov}}$ is the covering factor of the line-of-sight obscuration, with the same subscript notation as the $\mathrm{N_H}$.}
            \begin{tabular}{l l l l l l l }
            \toprule
            Source & $\log (L_{2-10 \rm \ keV})$ & $\Gamma$ & $\mathrm{N_{H, N}}$ & $f_{\mathrm{cov, N}}$ & $\mathrm{N_{H, I}}$ & $f_{\mathrm{cov, I}}$ \\[2pt]
            & $\log(\rm erg \ s^{-1})$ & & $10^{22} \ \mathrm{cm}^{-2}$ & & $10^{22} \ \mathrm{cm}^{-2}$ & \\
            \midrule
            CGCG381-051 & $40.63_{-0.15}^{+0.13}$ & $1.73_{-0.39}^{+0.52}$ & $0.06_{-u}^{+0.12}$ & $1^{*}$ & - & - \\[2pt]
            ESO141-G055 & $43.17 \pm 0.02$ & $1.67 \pm 0.02$ & - & - & - & - \\[2pt]
            ESO362-G018 & $42.60 \pm 0.01$ & $1.66 \pm 0.01$ & - & - & $30.9_{-2.8}^{+3.3}$ & $0.37 \pm 0.02$ \\[2pt]
            IC4329A & $43.74 \pm 0.01$ & $1.77 \pm 0.01$ & $0.50 \pm 0.01$ & $1^{*}$ & $28.1_{-2.7}^{+2.5}$ & $0.25 \pm 0.02$ \\[2pt]
            IC5063 & $42.86 \pm 0.09$ & $1.97_{-0.02}^{+0.01}$ & $34.66 \pm 0.69$ & $1.00 \pm 0.01$ & - & - \\[2pt]
            IRASF01475-0740 & $41.73_{-0.05}^{+0.04}$ & $2.06 \pm 0.09$ & $0.43 \pm 0.04$ & $1^{*}$ & - & - \\[2pt]
            IRASF03450+0055 & $41.86_{-0.31}^{+0.18}$ & $3.14_{-0.13}^{+0.15}$ & - & - & - & - \\[2pt]
            IRASF05189-2524 & $43.44 \pm 0.01$ & $2.13 \pm 0.03$ & $7.64_{-0.17}^{+0.18}$ & $1^{*}$ & - & - \\[2pt]
            MCG+00-29-023 & $40.78_{-0.38}^{+0.36}$ & $1.83_{-0.51}^{+0.50}$ & - & - & - & - \\[2pt]
            MCG-02-33-034 & $42.79_{-0.06}^{+0.05}$ & $2.38 \pm 0.02$ & - & - & $60.8_{-5.9}^{+11.2}$ & $0.44_{-0.09}^{+0.06}$ \\[2pt]
            MCG-03-58-007 & $42.98_{-0.02}^{+0.03}$ & $2.47_{-0.09}^{+0.10}$ & $15.59_{-0.68}^{+0.69}$ & $1^{*}$ & $230_{-81}^{+72}$ & $0.58_{-0.06}^{+0.05}$ \\[2pt]
            MCG-06-30-015 & $42.95 \pm 0.02$ & $2.72_{-0.05}^{+0.06}$ & - & - & $4.45_{-0.19}^{+0.28}$ & $0.69 \pm 0.02$ \\[2pt]
            MRK0509 & $44.13 \pm 0.01$ & $1.85_{-0.04}^{+0.03}$ & $8.82_{-0.72}^{+0.78}$ & $0.37 \pm 0.02$ & - & - \\[2pt]
            MRK0897 & $41.68_{-0.15}^{+0.13}$ & $1.8^{*}$ & $10.1_{-3.8}^{+8.0}$ & $1^{*}$ & - & - \\[2pt]
            MRK1239 & $41.01_{-0.02}^{+0.04}$ & $2.78_{-0.04}^{+0.09}$ & $15.84_{-0.47}^{+0.63}$ & $1^{*}$ & - & - \\[2pt]
            NGC0034 & $41.85_{-0.13}^{+0.12}$ & $1.8^{*}$ & $49_{-12}^{+14}$ & $1^{*}$ & - & - \\[2pt]
            NGC0526A & $43.29 \pm 0.01$ & $1.47 \pm 0.01$ & $1.05 \pm 0.02$ & $1^{*}$ & - & - \\[2pt]
            NGC1365 & $41.59_{-0.17}^{+0.09}$ & $1.47_{-0.30}^{+0.03}$ & $20.78_{-0.37}^{+0.55}$ & $0.96 \pm 0.01$ & $25.3_{-3.1}^{+2.9}$ & $0.66 \pm 0.02$ \\[2pt]
            NGC1566 & $41.84 \pm 0.01$ & $1.75 \pm 0.02$ & - & - & - & - \\[2pt]
            NGC2992 & $43.09 \pm 0.01$ & $1.68 \pm 0.01$ & $0.79 \pm 0.01$ & $1^{*}$ & - & - \\[2pt]
            NGC4593 & $42.58 \pm 0.01$ & $1.92 \pm 0.01$ & - & - & $3.58_{-0.19}^{+0.14}$ & $0.43 \pm 0.01$ \\[2pt]
            NGC4602 & $39.70_{-0.11}^{+0.19}$ & $1.66_{-0.17}^{+0.26}$ & - & - & - & - \\[2pt]
            NGC5506 & $42.71 \pm 0.01$ & $1.83 \pm 0.02$ & $3.15 \pm 0.03$ & $1^{*}$ & - & - \\[2pt]
            NGC5995 & $43.43_{-0.07}^{+0.08}$ & $1.94_{-0.08}^{+0.09}$ & $1.06_{-0.10}^{+0.11}$ & $1^{*}$ & - & - \\[2pt]
            NGC6810 & $39.88_{-0.08}^{+0.07}$ & $1.94_{-0.18}^{+0.16}$ & - & - & - & - \\[2pt]
            NGC6860 & $42.31_{-0.22}^{+0.21}$ & $1.75_{-0.03}^{+0.04}$ & $0.21 \pm 0.01$ & $1^{*}$ & $2.99_{-0.22}^{+0.09}$ & $0.98_{-0.03}^{+u}$ \\[2pt]
            NGC7213 & $42.03 \pm 0.09$ & $1.88 \pm 0.01$ & - & - & $2.75_{-0.47}^{+0.39}$ & $0.22_{-0.02}^{+0.01}$ \\[2pt]
            NGC7469 & $43.32 \pm 0.01$ & $1.88 \pm 0.01$ & - & - & - & - \\[2pt]
            NGC7496 & $39.32_{-0.09}^{+0.08}$ & $1.98_{-0.40}^{+0.27}$ & - & - & - & - \\[2pt]
            NGC7603 & $43.44 \pm 0.01$ & $2.26 \pm 0.02$ & - & - & - & - \\[2pt]
            \bottomrule
            \end{tabular}
            \label{tab:phenomenological_results}
        \end{table*}

        \begin{table*}
            \centering
            \caption{Results from the \myt \ model fitting of the X-ray spectra. All parameters that were fixed are marked with $*$, and a $u$ indicates that the uncertainty could not be constrained. $\log (L_{2-10 \rm \ keV})$ is the intrinsic luminosity in the $2-10 \rm \ keV$ energy band. $\Gamma$ is the photon index. $\mathrm{N_{H, LOS}}$ is the line-of-sight obscuring column density, and $\mathrm{N_{H, Eq}}$ is the average equatorial column density. $\mathrm{A_{90}}$ and $\mathrm{A_0}$ are the normalisation constants required when using \myt \ in its decoupled configuration, as explained in Section \ref{sec:Modelling with MYTorus}.}
            
            \begin{tabular}{l l l l l l l }
            \toprule
            Source & $\log (L_{2-10 \rm \ keV})$ & $\Gamma$ & $\mathrm{N_{H, LOS}}$ & $\mathrm{N_{H, Eq}}$ & $ \mathrm{A_{90}}$ & $\mathrm{A_0}$ \\[2pt]
            & $\log(\rm erg \ s^{-1})$ & & $10^{24} \ \mathrm{cm}^{-2}$ & $10^{24} \ \mathrm{cm}^{-2}$ & & \\
            \midrule
            ESO033-G002 & $42.63 \pm 0.01$ & $1.80 \pm 0.01$ & $0.08 \pm 0.01$ & $0.31_{-0.04}^{+0.08}$ & $1^{*}$ & $1^{*}$ \\[2pt]
            IRASF04385-0828 & $42.72_{-0.04}^{+u}$ & $2.21_{-0.15}^{+0.16}$ & $10.0_{-7.2}^{+u}$ & $4.6_{-1.9}^{+u}$ & $1^{*}$ & $1^{*}$ \\[2pt]
            IRASF15480-0344 & $44.27_{-0.24}^{+0.17}$ & $2.24_{-0.19}^{+0.16}$ & $3.33_{-0.87}^{+u}$ & $1.82_{-0.37}^{+0.93}$ & $1^{*}$ & $1^{*}$ \\[2pt]
            MCG-03-34-064 & $42.59_{-0.25}^{+0.23}$ & $2.16_{-0.06}^{+0.08}$ & $0.51 \pm 0.04$ & $0.31_{-0.03}^{+0.06}$ & $77 \pm 10$ & $1^{*}$ \\[2pt]
            NGC0424 & $43.64_{-0.37}^{+0.35}$ & $1.57_{-u}^{+0.19}$ & $2.04_{-0.29}^{+0.34}$ & $0.21_{-0.05}^{+0.08}$ & $1^{*}$ & $1^{*}$ \\[2pt]
            NGC1125 & $42.84_{-1.42}^{+0.61}$ & $1.65_{-u}^{+0.45}$ & $0.93_{-0.25}^{+0.49}$ & $4.0_{-1.7}^{+u}$ & $1^{*}$ & $1.53_{-0.83}^{+1.98}$ \\[2pt]
            NGC1194 & $42.60_{-0.22}^{+0.07}$ & $1.89 \pm 0.07$ & $0.80_{-0.16}^{+0.11}$ & $9.0_{-4.2}^{+u}$ & $1^{*}$ & $2.92_{-0.62}^{+0.80}$ \\[2pt]
            NGC1320 & $42.59_{-0.03}^{+u}$ & $2.21_{-0.12}^{+0.19}$ & $5.4_{-2.3}^{+u}$ & $8.0_{-3.4}^{+u}$ & $1^{*}$ & $1^{*}$ \\[2pt]
            NGC5135 & $42.85_{-0.27}^{+0.17}$ & $2.43_{-0.18}^{+0.13}$ & $2.51_{-0.55}^{+1.49}$ & $5_{-2}^{+4}$ & $1^{*}$ & $1^{*}$ \\[2pt]
            NGC6890 & $41.19_{-0.79}^{+0.70}$ & $1.86_{-0.12}^{+0.13}$ & $0.22_{-0.05}^{+0.08}$ & $4.0_{-1.4}^{+4.1}$ & $1^{*}$ & $1^{*}$ \\[2pt]
            NGC7130 & $42.62_{-0.13}^{+0.10}$ & $1.97_{-0.14}^{+0.17}$ & $1.86_{-0.28}^{+0.39}$ & $5.0_{-1.6}^{+2.7}$ & $1^{*}$ & $1^{*}$ \\[2pt]
            NGC7674 & $42.71_{-0.05}^{+0.04}$ & $2.60_{-0.16}^{+u}$ & $0.34 \pm 0.09$ & $9.0_{-4.2}^{+u}$ & $84_{-58}^{+73}$ & $10_{-4}^{+6}$ \\[2pt]
            TOLOLO1238-364 & $42.26_{-0.13}^{+u}$ & $2.33_{-0.28}^{+0.26}$ & $4.6_{-1.5}^{+u}$ & $8.0_{-3.4}^{+u}$ & $1^{*}$ & $1^{*}$ \\[2pt]
            \bottomrule
            \end{tabular}
            \label{tab:physical_results}
        \end{table*}

        \begin{table*}
            \centering
            \caption{Results from the fitting of the spectra with the \ctpo \ model. All parameters that were fixed are marked with $*$, and a $u$ indicates that the uncertainty could not be constrained. $\log (L_{2-10 \rm \ keV} )$ is the intrinsic luminosity in the $2-10 \rm \ keV$ energy band. $\Gamma$ is the photon index. $\mathrm{N_{H, eq}}$ is the equatorial obscuring column density. $\Theta$ is the opening angle of the torus, and $i$ is the inclination of the torus relative to our line-of-sight. $p$ is the radial dust distribution gradient parameter, $\mathrm{A_R}$ is the scaling constant between the reprocessed and directly absorbed components of \ctpo, and $\mathrm{N_{H, LOS}}$ is the line-of-sight column density found using Equation \ref{eq:nh_conversion}.}
            \label{tab:carrot_results}
            \begin{tabular}{l l l l l l l l l }
            \toprule
            Source & $\log (L_{2-10 \rm \ keV})$ & $\Gamma$ & $\mathrm{N_{H,eq}}$ & $\Theta$ & $i$ & $p$ & $\mathrm{A_R}$ & $\mathrm{N_{H, LOS}}$ \\[2pt]
             & $\log(\rm erg \ s^{-1})$ & & $10^{22} \ \mathrm{cm}^{-2}$ & $^{\circ}$ & $^{\circ}$ & & & $10^{22} \ \mathrm{cm}^{-2}$ \\
            \midrule
            CGCG381-051 & $40.62_{-0.07}^{+0.06}$ & $1.8^{*}$ & $0.10_{-u}^{+0.09}$ & $60^{*}$ & $45^{*}$ & $0^{*}$ & $1^{*}$ & $0.05_{-u}^{+0.04}$ \\[2pt]
            ESO033-G002 & $42.64 \pm 0.02$ & $2.05_{-0.03}^{+0.02}$ & $12.01_{-0.91}^{+0.69}$ & $78.98 \pm 0.07$ & $18.42_{-0.24}^{+0.14}$ & $1^{*}$ & $7.00_{-0.83}^{+1.09}$ & $4.65_{-0.35}^{+0.27}$ \\[2pt]
            ESO141-G055 & $43.32_{-0.02}^{+0.01}$ & $2.23 \pm 0.01$ & $90.0_{-5.4}^{+4.7}$ & $64.74_{-0.20}^{+0.15}$ & $26.32_{-0.36}^{+0.24}$ & $1^{*}$ & $0.26_{-0.10}^{+0.13}$ & $36.7_{-2.2}^{+1.9}$ \\[2pt]
            ESO362-G018 & $42.65 \pm 0.01$ & $1.82_{-0.01}^{+0.02}$ & $78.7_{-6.0}^{+5.0}$ & $72.98_{-0.27}^{+0.23}$ & $12.05_{-0.32}^{+0.34}$ & $0^{*}$ & $3.19_{-0.27}^{+0.30}$ & $0.01$ \\[2pt]
            IC4329A & $43.73 \pm 0.01$ & $1.76 \pm 0.01$ & $14.12_{-0.32}^{+0.31}$ & $71.23 \pm 0.15$ & $19.82_{-0.09}^{+0.09}$ & $0^{*}$ & $2.05 \pm 0.14$ & $5.51 \pm 0.12$ \\[2pt]
            IC5063 & $42.69 \pm 0.36$ & $1.84 \pm 0.04$ & $39.52_{-0.89}^{+1.24}$ & $59.85 \pm 0.04$ & $39.97 \pm 0.01$ & $0^{*}$ & $1.30_{-0.18}^{+0.19}$ & $18.36_{-0.41}^{+0.58}$ \\[2pt]
            IRASF01475-0740 & $41.74_{-0.05}^{+0.04}$ & $2.04 \pm 0.09$ & $0.63 \pm 0.06$ & $60^{*}$ & $45^{*}$ & $1^{*}$ & $1^{*}$ & $0.31 \pm 0.03$ \\[2pt]
            IRASF03450+0055 & $42.42_{-0.04}^{+0.09}$ & $2.60_{-0.04}^{+u}$ & $0.15_{-u}^{+2.35}$ & $60^{*}$ & $45^{*}$ & $0^{*}$ & $1^{*}$ & $0.07_{-u}^{+1.16}$ \\[2pt]
            IRASF04385-0828 & $43.54_{-0.17}^{+0.11}$ & $2.60_{-0.10}^{+u}$ & $570_{-200}^{+220}$ & $60^{*}$ & $40.00 \pm 0.01$ & $0^{*}$ & $1^{*}$ & $260_{-90}^{+100}$ \\[2pt]
            IRASF05189-2524 & $43.47 \pm 0.02$ & $2.19 \pm 0.04$ & $11.4_{-1.0}^{+1.4}$ & $69.36_{-5.52}^{+0.37}$ & $38.8_{-1.5}^{+1.2}$ & $1^{*}$ & $1.18_{-0.56}^{+0.63}$ & $5.22_{-0.46}^{+0.66}$ \\[2pt]
            IRASF15480-0344 & $43.98_{-0.09}^{+0.11}$ & $2.54_{-0.10}^{+u}$ & $447_{-73}^{+138}$ & $59.68_{-0.24}^{+0.16}$ & $39.99 \pm 0.01$ & $0^{*}$ & $1^{*}$ & $208_{-34}^{+64}$ \\[2pt]
            MCG+00-29-023 & $40.89 \pm 0.10$ & $1.8^{*}$ & $0.10_{-u}^{+0.10}$ & $60^{*}$ & $45^{*}$ & $1^{*}$ & $1^{*}$ & $0.05_{-u}^{+0.05}$ \\[2pt]
            MCG-02-33-034 & $42.99 \pm 0.05$ & $2.59_{-0.02}^{+u}$ & $254_{-35}^{+29}$ & $74.99_{-0.56}^{+0.49}$ & $14.33_{-0.37}^{+0.34}$ & $0^{*}$ & $2.53_{-0.50}^{+0.57}$ & $0.01$ \\[2pt]
            MCG-03-34-064 & $42.74_{-0.02}^{+0.06}$ & $1.88_{-0.05}^{+0.06}$ & $84.9_{-3.3}^{+6.6}$ & $80.00_{-0.01}^{+u}$ & $19.99_{-0.12}^{+0.51}$ & $0^{*}$ & $2.88 \pm 0.32$ & $33.2_{-1.3}^{+2.6}$ \\[2pt]
            MCG-03-58-007 & $43.00_{-0.02}^{+0.03}$ & $2.28_{-0.04}^{+0.05}$ & $28.85_{-0.78}^{+0.52}$ & $69.94 \pm 0.01$ & $30.00_{-0.01}^{+1.57}$ & $0^{*}$ & $1.88_{-0.50}^{+0.52}$ & $12.13_{-0.33}^{+0.22}$ \\[2pt]
            MCG-06-30-015 & $42.93_{-0.02}^{+0.01}$ & $2.41 \pm 0.02$ & $139_{-20}^{+17}$ & $70.00_{-4.47}^{+0.45}$ & $15.59_{-0.13}^{+3.41}$ & $0.38_{-0.09}^{+0.08}$ & $0.10_{-u}^{+0.18}$ & $0.01$ \\[2pt]
            MRK0509 & $44.16 \pm 0.01$ & $1.87 \pm 0.01$ & $65.7_{-7.5}^{+7.2}$ & $41.83_{-1.08}^{+0.96}$ & $43.67_{-1.07}^{+0.68}$ & $0.00_{-u}^{+0.33}$ & $1.26_{-0.23}^{+0.25}$ & $0.01$ \\[2pt]
            MRK0897 & $41.69_{-0.15}^{+0.12}$ & $1.8^{*}$ & $7.7_{-1.9}^{+5.2}$ & $60^{*}$ & $45^{*}$ & $0^{*}$ & $1^{*}$ & $3.80_{-0.91}^{+2.54}$ \\[2pt]
            MRK1239 & $42.42 \pm 0.02$ & $2.60_{-0.04}^{+u}$ & $26.28_{-0.55}^{+1.05}$ & $30.04_{-0.78}^{+9.46}$ & $69.87_{-0.17}^{+0.01}$ & $0^{*}$ & $1^{*}$ & $18.62_{-0.39}^{+0.74}$ \\[2pt]
            NGC0034 & $42.00^{+0.21}_{-0.15}$ & $1.8^{*}$ & $113_{-35}^{+60}$ & $69.51_{-0.25}^{+0.21}$ & $29.71_{-0.29}^{+0.22}$ & $0^{*}$ & $1^{*}$ & $48_{-15}^{+25}$ \\[2pt]
            NGC0424 & $43.34_{-0.41}^{+0.35}$ & $2.56_{-0.11}^{+u}$ & $377_{-51}^{+119}$ & $39.98_{-0.03}^{+0.02}$ & $59.56_{-0.76}^{+0.09}$ & $0^{*}$ & $1.78_{-0.80}^{+3.18}$ & $227_{-31}^{+72}$ \\[2pt]
            NGC0526A & $43.30 \pm 0.01$ & $1.60_{-0.03}^{+0.04}$ & $6.24_{-0.16}^{+0.21}$ & $11.14_{-0.62}^{+0.50}$ & $79.92 \pm 0.01$ & $0^{*}$ & $10.7_{-2.9}^{+2.8}$ & $5.24_{-0.13}^{+0.18}$ \\[2pt]
            NGC1125 & $42.75_{-0.12}^{+0.23}$ & $2.52_{-0.22}^{+u}$ & $383_{-32}^{+42}$ & $79.83_{-0.27}^{+0.09}$ & $19.13_{-1.77}^{+0.32}$ & $0.95_{-0.18}^{+u}$ & $2.3_{-1.1}^{+3.7}$ & $149_{-12}^{+16}$ \\[2pt]
            NGC1194 & $42.63_{-0.26}^{+0.19}$ & $1.81_{-0.14}^{+0.15}$ & $432_{-111}^{+58}$ & $52.5_{-4.5}^{+4.5}$ & $39.91_{-0.09}^{+7.09}$ & $0^{*}$ & $4.1_{-1.3}^{+1.8}$ & $201_{-52}^{+27}$ \\[2pt]
            NGC1320 & $43.74_{-0.07}^{+0.19}$ & $2.60_{-0.11}^{+u}$ & $2390_{-580}^{+1870}$ & $60^{*}$ & $45^{*}$ & $0^{*}$ & $1^{*}$ & $1180_{-280}^{+920}$ \\[2pt]
            NGC1365 & $42.17 \pm 0.01$ & $1.96 \pm 0.01$ & $45.31_{-0.55}^{+0.50}$ & $69.81_{-0.01}^{+0.02}$ & $30.21_{-0.19}^{+0.37}$ & $0^{*}$ & $0.66 \pm 0.16$ & $19.09_{-0.23}^{+0.21}$ \\[2pt]
            NGC1566 & $41.82_{-0.01}^{+0.13}$ & $1.77 \pm 0.02$ & $80_{-20}^{+31}$ & $60^{*}$ & $0.02_{-u}^{+30.98}$ & $1^{*}$ & $1.79_{-0.34}^{+0.46}$ & $0.01$ \\[2pt]
            NGC2992 & $43.07 \pm 0.01$ & $1.69 \pm 0.01$ & $1.48 \pm 0.04$ & $71.12_{-0.52}^{+0.58}$ & $30.00_{-0.01}^{+0.43}$ & $0^{*}$ & $14.88_{-0.92}^{+0.96}$ & $0.62 \pm 0.02$ \\[2pt]
            NGC4593 & $42.56 \pm 0.01$ & $1.70 \pm 0.01$ & $100.0_{-1.9}^{+3.6}$ & $69.97_{-0.65}^{+0.40}$ & $0.04_{-u}^{+3.04}$ & $1^{*}$ & $1^{*}$ & $0.01$ \\[2pt]
            NGC4602 & $39.84 \pm 0.09$ & $1.91 \pm 0.11$ & $0.10_{-u}^{+0.02}$ & $60^{*}$ & $45^{*}$ & $1^{*}$ & $1^{*}$ & $0.05_{-u}^{+0.01}$ \\[2pt]
            NGC5135 & $43.72_{-0.16}^{+0.11}$ & $2.33_{-0.06}^{+0.07}$ & $5000_{-4020}^{+u}$ & $62.5_{-2.5}^{+3.2}$ & $39.99 \pm 0.01$ & $0^{*}$ & $1^{*}$ & $2300_{-1900}^{+u}$ \\[2pt]
            NGC5506 & $42.88 \pm 0.03$ & $1.95 \pm 0.01$ & $123.1_{-1.8}^{+1.7}$ & $72.42_{-0.63}^{+0.52}$ & $19.98 \pm 0.01$ & $0^{*}$ & $1.50 \pm 0.01$ & $48.08 \pm 0.68$ \\[2pt]
            NGC5995 & $43.39_{-0.10}^{+0.08}$ & $1.83 \pm 0.03$ & $1.31_{-0.21}^{+0.24}$ & $18.65_{-0.75}^{+0.55}$ & $89.80_{-0.12}^{+0.06}$ & $0^{*}$ & $1^{*}$ & $1.31_{-0.21}^{+0.24}$ \\[2pt]
            NGC6810 & $39.78 \pm 0.06$ & $2.41 \pm 0.18$ & $0.19_{-u}^{+0.10}$ & $60^{*}$ & $45^{*}$ & $0^{*}$ & $1^{*}$ & $0.09_{-u}^{+0.05}$ \\[2pt]
            NGC6860 & $42.39_{-0.28}^{+0.27}$ & $1.63 \pm 0.03$ & $4.22_{-0.56}^{+0.17}$ & $38.73_{-0.16}^{+0.13}$ & $51.48_{-0.79}^{+0.67}$ & $0^{*}$ & $8.9_{-1.8}^{+1.9}$ & $2.27_{-0.30}^{+0.09}$ \\[2pt]
            NGC6890 & $42.78_{-0.93}^{+0.80}$ & $2.43 \pm 0.10$ & $770_{-170}^{+130}$ & $79.46_{-1.16}^{+0.25}$ & $19.29_{-0.32}^{+0.23}$ & $0^{*}$ & $0.45_{-0.17}^{+0.20}$ & $300_{-67}^{+50}$ \\[2pt]
            NGC7130 & $43.14_{-0.15}^{+0.16}$ & $2.13_{-0.09}^{+0.08}$ & $450_{-120}^{+160}$ & $59.63_{-0.38}^{+5.00}$ & $39.97_{-0.03}^{+0.01}$ & $1^{*}$ & $1^{*}$ & $208_{-56}^{+76}$ \\[2pt]
            NGC7213 & $42.04_{-0.10}^{+0.09}$ & $1.95 \pm 0.02$ & $15.0_{-2.1}^{+2.0}$ & $62.93_{-0.54}^{+0.45}$ & $27.07_{-0.40}^{+0.53}$ & $0^{*}$ & $4.65_{-0.70}^{+0.88}$ & $0.01$ \\[2pt]
            NGC7469 & $43.37 \pm 0.01$ & $1.97 \pm 0.01$ & $100.2_{-3.0}^{+15.8}$ & $79.2_{-1.3}^{+u}$ & $1.69_{-0.21}^{+2.88}$ & $0^{*}$ & $1.77_{-0.13}^{+0.14}$ & $0.01$ \\[2pt]
            NGC7496 & $39.25_{-0.07}^{+0.04}$ & $2.60_{-0.10}^{+u}$ & $0.10_{-u}^{+0.01}$ & $60^{*}$ & $45^{*}$ & $0^{*}$ & $1^{*}$ & $0.05_{-u}^{+0.01}$ \\[2pt]
            NGC7603 & $43.70_{-0.05}^{+0.06}$ & $2.24_{-0.01}^{+0.02}$ & $26.9_{-5.6}^{+6.9}$ & $60^{*}$ & $34.73_{-0.67}^{+0.55}$ & $0^{*}$ & $0.91_{-0.75}^{+0.96}$ & $11.8_{-2.5}^{+3.0}$ \\[2pt]
            NGC7674 & $43.74_{-0.05}^{+0.06}$ & $2.60_{-0.03}^{+u}$ & $1630_{-510}^{+490}$ & $78.45_{-0.31}^{+0.20}$ & $19.86 \pm 0.02$ & $0^{*}$ & $4.74_{-0.68}^{+0.63}$ & $640_{-200}^{+190}$ \\[2pt]
            TOLOLO1238-364 & $43.30_{-0.20}^{+0.29}$ & $2.44_{-0.22}^{+0.13}$ & $4230_{-2350}^{+770}$ & $59.94_{-0.88}^{+2.45}$ & $39.99 \pm 0.01$ & $0^{*}$ & $1^{*}$ & $1970_{-1090}^{+360}$ \\[2pt]
            \bottomrule
            \end{tabular}
        \end{table*}

        \subsubsection{Spectral Analysis of IRASF03450+0055} \label{sec:Modelling IRASF03450+0055}

\begin{figure}
            \centering
            \includegraphics[width=0.95\linewidth]{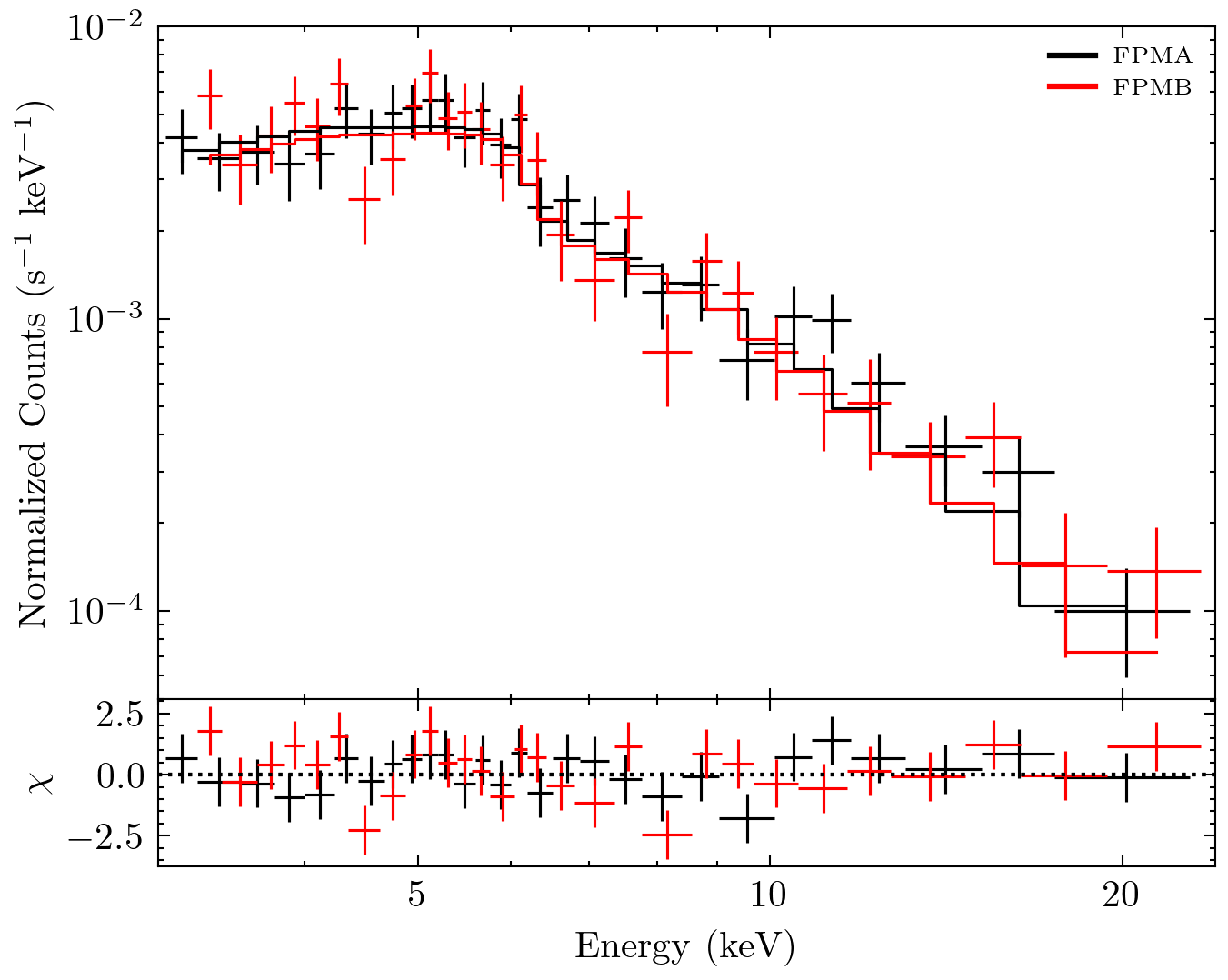}
            \caption{The folded and phenomenologically fitted spectrum (top panel) and residuals (bottom panel) of IRASF03450+0055. The counts have been grouped into bins of 20 counts per bin. The FPMA/FPMB points and models are plotted in black and red, respectively.}
            \label{fig:foldedspec}
        \end{figure}

            We analysed IRASF03450+0055 using \textit{NuSTAR} data from our proposal. As the spectrum is background-dominated at $E<3$ keV and $E>25$ keV, we focused on the $3-25$ keV range. The fitted spectrum is shown in Figure \ref{fig:foldedspec}. When analysing this galaxy, we not only wanted to find its intrinsic luminosity, but also to confirm if it may be a NLS1 galaxy. After fitting the spectrum with a basic phenomenological model, we found large residuals $(\sim 5 \sigma)$ around the Fe K$\alpha$ emission line, indicative of strong relativistic reflection. The addition of a relativistic reflection component (\texttt{relxill}) improved the fit significantly, although we had to fix the reflection fraction at $-1$ due to the data quality. We also froze the iron abundance at solar values ($\mathrm{A_{Fe}} = 1$), the inclination at its lowest possible value ($i=3^{\circ}$), and found the ionisation of the accretion disk to be $\log \xi = 2.70^{+0.34}_{-0.39}$. Due to the energy range covered by our observations, we were not able to put any constraints on the obscuration with a phenomenological model. We found that IRASF03450+0055 has an incredibly steep photon index with $\Gamma = 3.14^{+0.15}_{-0.13}$, and is completely reflection dominated. This is consistent with the typical X-ray spectral characteristics of a NLS1 (see Section \ref{sec:First Pointed Obs of IRAS}). We estimated the intrinsic luminosity of this source to be $\log(L_{ \rm 2-10 \ keV} / \rm \ erg \ s^{-1}) = 41.86^{+0.18}_{-0.31}$, but given how reflection-dominated this source is, it is difficult to properly constrain. 
            
            When using \ctpo \ to fit this spectrum, we added in \texttt{relxill} to account for the strong relativistic reflection. It was multiplied by \texttt{zphabs} to account for the LOS absorption, with the column density linked to the \ctpo \ $\rm N_{H, eq}$ via Equation \ref{eq:nh_conversion}. We once again kept all \texttt{relxill} values frozen at their defaults except for the ionisation parameter ($\log \xi = 2.31^{+0.60}_{-0.28}$), and fixed the reflection fraction at $-1$. We tied the photon index and normalisation to that of \ctpo. As the reflection completely dominates the spectrum, we were only able to find an upper limit for the average obscuration of $\rm N_{H, eq} \leq 2.5 \times 10^{22} \ cm^{-2}$. Our intrinsic luminosity estimate is  $\log (L_{ \rm 2-10 \ keV} / \rm \ erg \ s^{-1}) = 42.42^{+0.09}_{-0.04}$, which is higher than the original estimate, but we expect this given the higher obscuration. Overall, this is a very interesting and complex source, and we found compelling evidence that it is a NLS1.

        \subsubsection{Comparing Model Results}
        \label{sec:Comparing Models}

\begin{figure}
                \centering
                \includegraphics[width=0.95\linewidth]{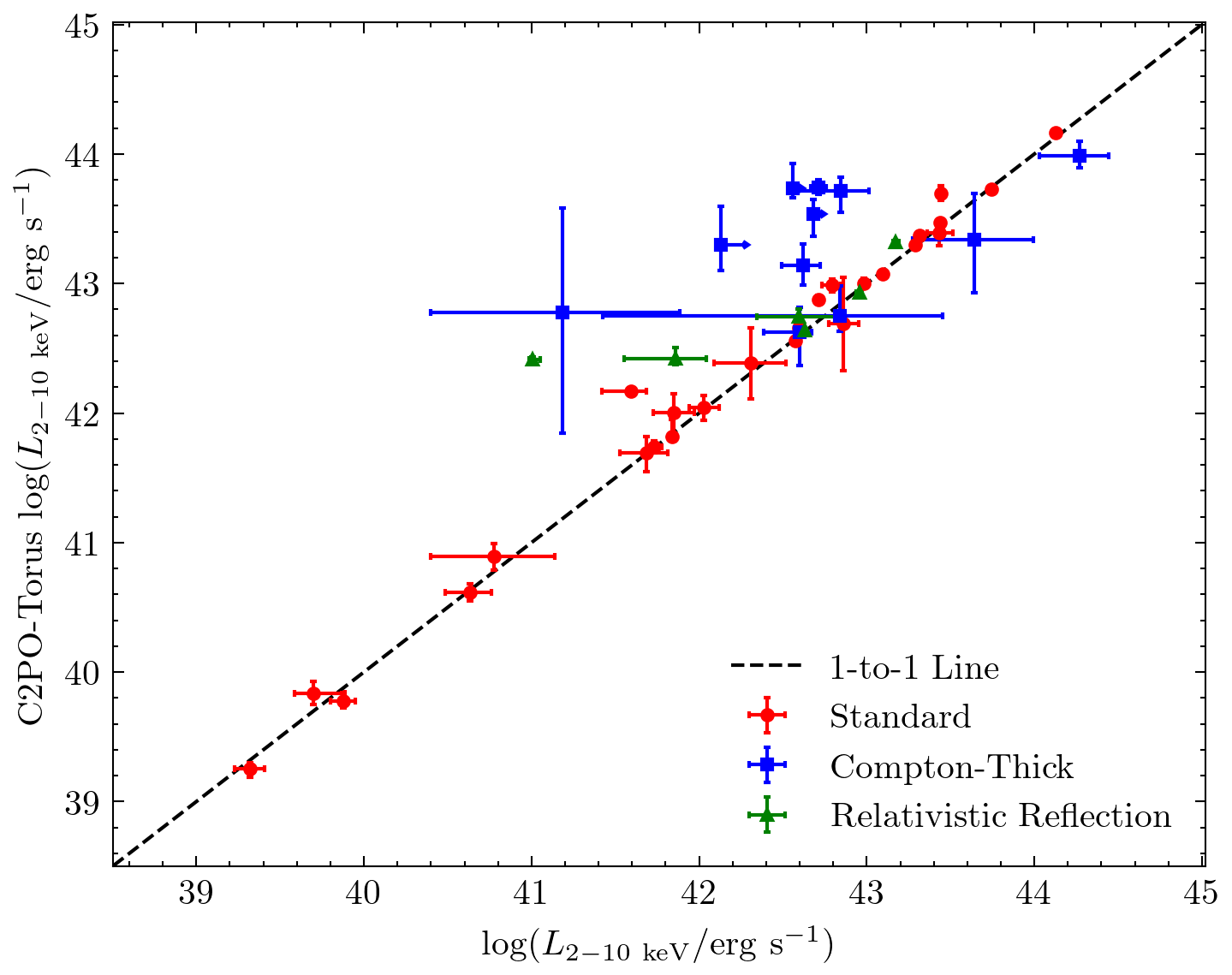}
                \caption{Comparison of the intrinsic $2-10 \rm \ keV$ luminosities derived from traditional phenomenological or physically-based models against those derived using the \ctpo \ model. We have highlighted the sources that are known from literature to be Compton-thick (blue squares) or contain relativistic reflection (green triangles). The remaining ``standard'' (non-CT and/or not featuring relativistic refection) sources are plotted as red circles, with the 1-to-1 line plotted as the dashed black line. It is clear that the luminosities of the majority of sources agree with each other within the uncertainties, with the Compton-thick/relativistically-reflected sources being harder to constrain.}
                \label{fig:c2po_comparison}
            \end{figure}

\begin{figure}
                \centering
                \includegraphics[width=0.95\linewidth]{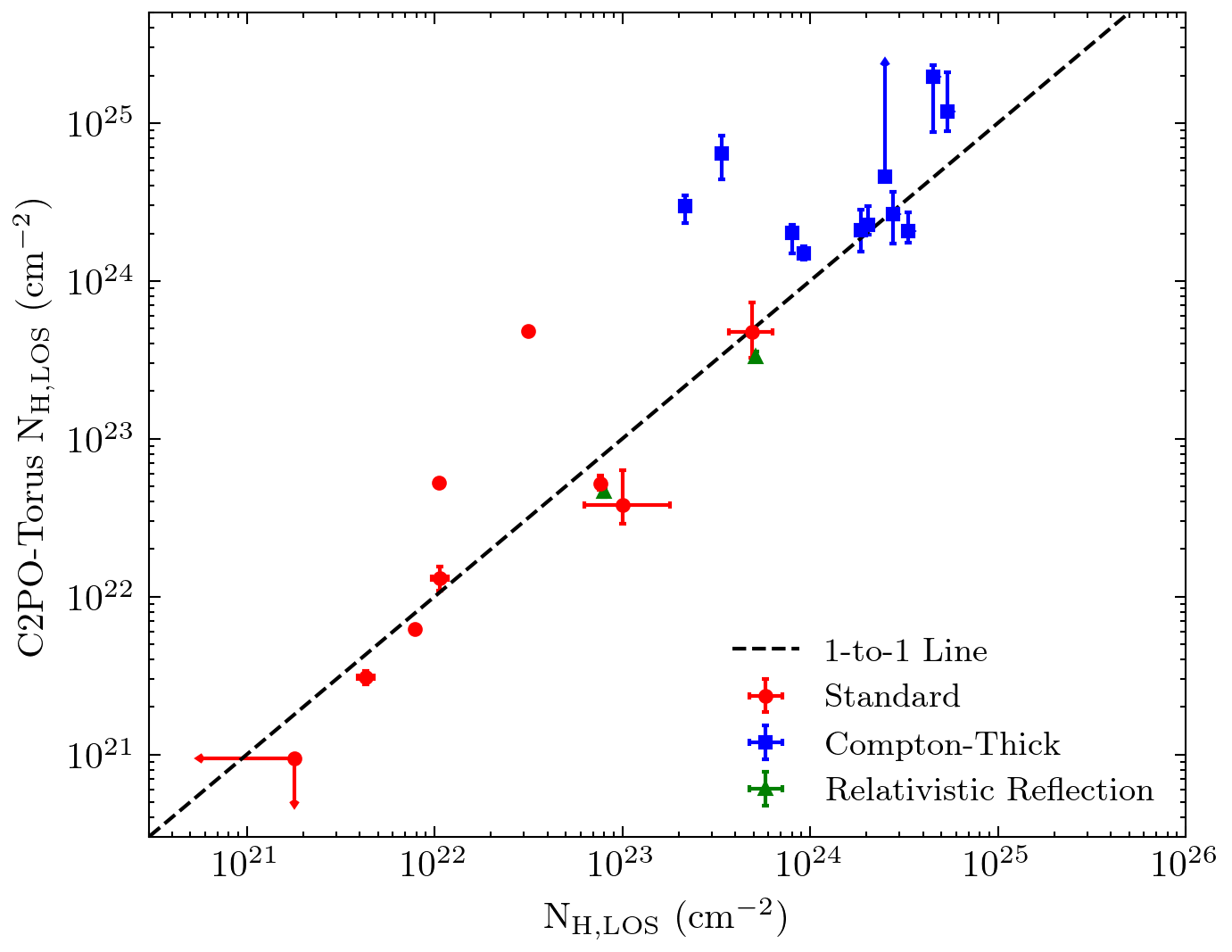}
                \caption{Comparison of the LOS column densities derived from traditional phenomenological or physically-based models against those derived using the \ctpo \ model. The markers are the same as in Figure \ref{fig:c2po_comparison}. The column densities of the standard sources generally agree with each other, while some of those that are Compton-thick have higher $\rm N_{H, LOS}$ when fitted with \ctpo, but this may be due to differences in model geometries and/or overestimations due to the implicit assumptions of Equation \ref{eq:nh_conversion}.}
                \label{fig:c2po_nh_comparison}
            \end{figure}

            To check whether the \ctpo \ model is consistent with traditional phenomenological and physically-based models, we compared the intrinsic $2-10 \rm \ keV$ luminosities and line-of-sight column densities found using the different models in Figures \ref{fig:c2po_comparison} and \ref{fig:c2po_nh_comparison}. Previous comparisons in the literature have found that clumpy models generally find column densities in agreement with their smooth counterparts \citep[see][]{Tanimoto2019, Buchner2021}. The differences between the geometries instead appear in the reflected/reprocessed emission; In a clumpy model, radiation that is reflected from the far side of the torus and from the inner wall can escape through the gaps in between the clumps, adding additional reprocessed flux at $\rm E \geq 10$ keV and changing the shape/intensity of the Fe~K$\alpha$ line. In a smooth torus with the same column density, this radiation would instead be absorbed. These differences are most prominent in Compton-thick sources, for which there are both high reflection fractions and levels of obscuration. As such, we do not expect significant differences between the obscuration values found using \ctpo \ and phenomenological/smooth models, especially given that we used \myt \ in its ``decoupled'' configuration, which mimics a clumpy torus. However,if we were to compare the reflection fractions and/or Fe K$\alpha$ equivalent widths, we would expect notable differences, especially for CT sources.
            
\begin{figure}
                \centering
                \includegraphics[width=0.95\linewidth]{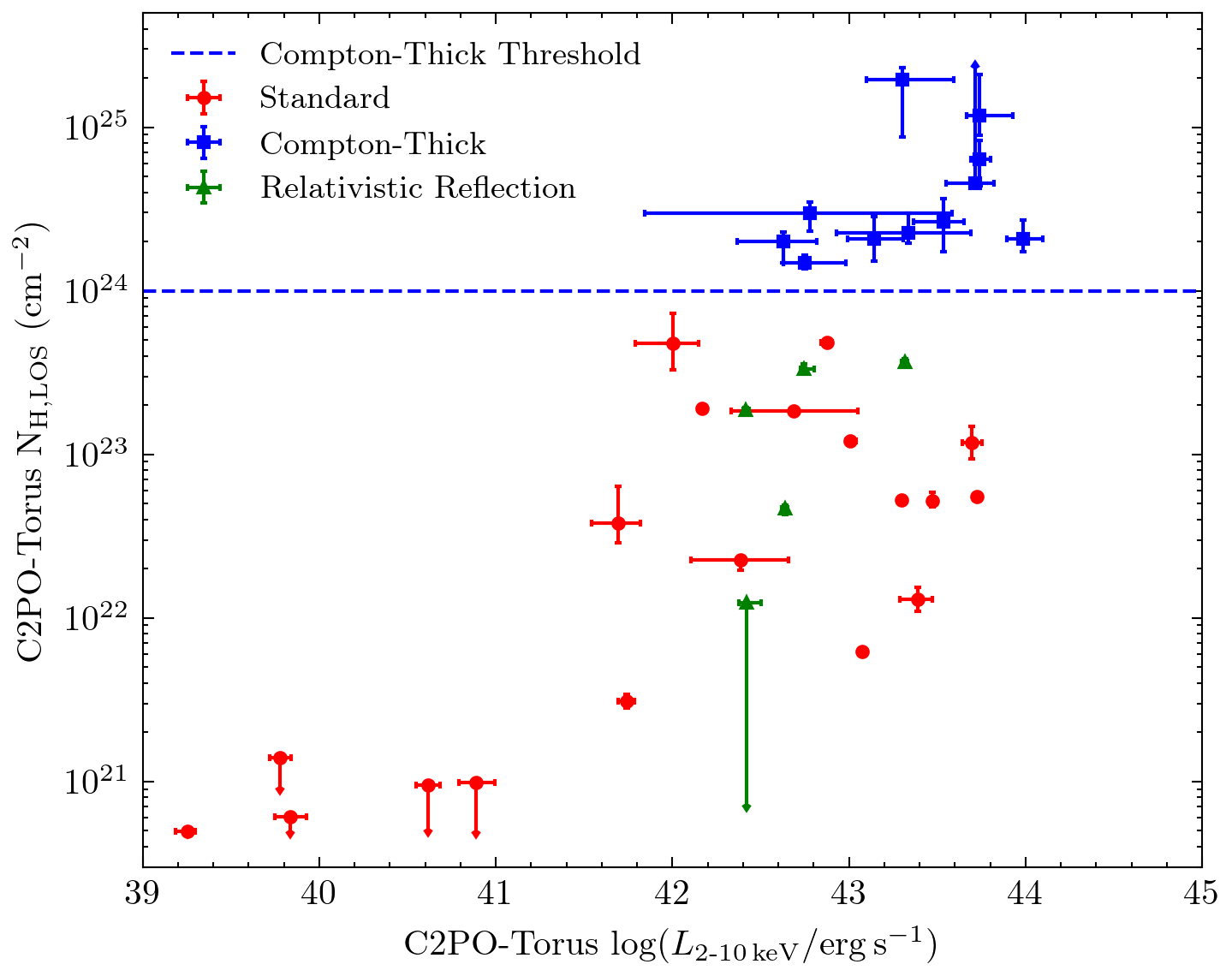}
                \caption{Comparison of the intrinsic rest-frame $2-10$ keV luminosities against the LOS column density, both derived using the \ctpo \ model. The markers are the same as in Figure \ref{fig:c2po_comparison}. The threshold for Compton-thick sources $(\rm N_{H} \geq 10^{24} \ cm^{-2})$ is plotted as the dashed blue line. \ctpo \ is able to correctly identify all literature-identified Compton-thick sources.}
                \label{fig:c2po_lumvsnh}
            \end{figure}

\begin{figure}
                \centering
                \includegraphics[width=0.95\linewidth]{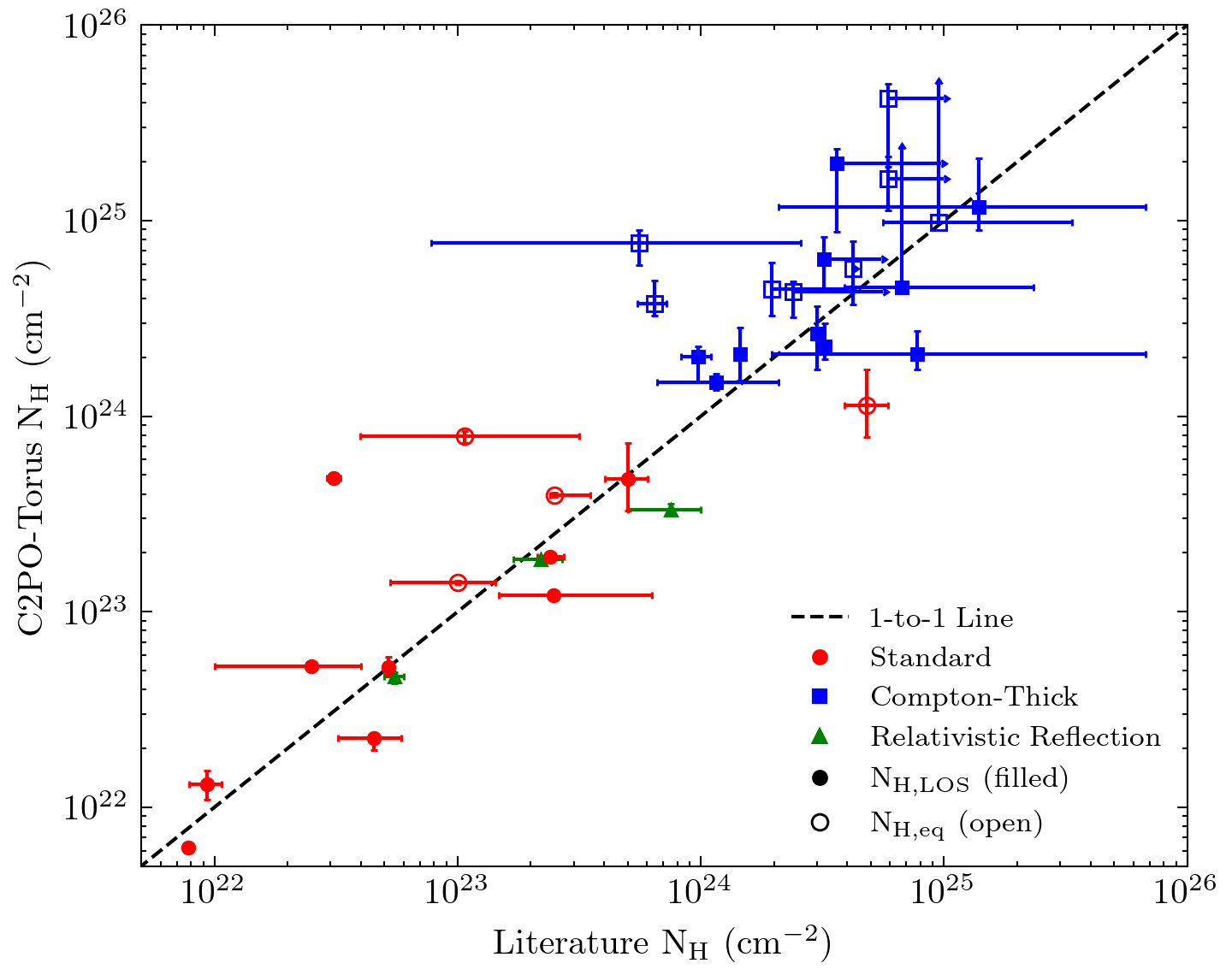}
                \caption{Comparison of the equatorial and LOS column densities found in literature against those found using \ctpo, for sources with significant obscuration ($\rm N_H > 10^{21} \ cm^{-2}$). The marker colours and shapes are the same as in Figure \ref{fig:c2po_comparison}, but the $\rm N_{H, LOS}$ are plotted as the filled markers, while the $\rm N_{H, eq}$ are plotted as the open markers. Most of the \ctpo \ estimates agree with the literature values, and those that differ significantly are usually due to differences in model geometries (e.g. phenomenological vs smooth torus vs clumpy torus). Refer to Appendix \ref{appen:Notes} for details on the comparison with literature column densities}.
                \label{fig:c2povslit}
            \end{figure}

            In order to fairly compare the LOS column densities estimated from the fitted equatorial values using Equation \ref{eq:nh_conversion}, we only compared sources that were either fit with \myt, or only required a single phenomenological \texttt{zpcfabs} component with $\mathrm{CF}=1$ to fit the obscuration. However, we would like to re-emphasise that the LOS column density values found using Equation \ref{eq:nh_conversion} are indicative estimates. \ctpo \ fits the equatorial column density only, and one can convert this value to a LOS estimate based on an assumption of a smooth torus. In many cases, the LOS will pass through clouds ``as expected'', and this approximation will be more accurate. However, there are cases where the LOS passes through more clouds than expected, or ``peeks through'' a gap in the clouds, for which the values found using Equation \ref{eq:nh_conversion} will be under-/overestimates of the true LOS value. We also wanted to confirm that \ctpo \ would successfully fit sources that are defined as Compton-thick in literature, i.e. find $\mathrm{N_{H, LOS}} \geq 10^{24} \ \mathrm{cm}^{-2}$ for those sources, and this is shown in Figure \ref{fig:c2po_lumvsnh}. Finally, we compared both the LOS and Equatorial column densities found using \ctpo \ with literature values in Figure \ref{fig:c2povslit}. It should also be re-emphasised that the values derived using Equation \ref{eq:nh_conversion} are only indicative estimates, as this conversion assumes a smooth, clump-free torus. 

            In general, \ctpo \ performs well, with values in agreement with those from other phenomenological/physical models while also providing insights into the physical properties of the AGN that can be used to assist with SED fitting. We find that it is able to successfully fit sources that are reported in the literature as being Compton-thick (see Appendix \ref{appen:Notes} for more details), for which the high levels of obscuration make spectral fitting difficult, even for standard models. For these CT sources, the large $\rm N_{H, eq}$ values available when using \ctpo \ made it much more suitable for fitting the absorbed and reprocessed emission of the spectra than models such as \myt, even in its decoupled configuration. $11/43$ or $\sim 30 \%$ of our sources have been classified in the literature as Compton-thick. This fraction agrees with what is expected for the local AGN population \citep{Ricci2017, Yamada2021, Annuar2025}, and \ctpo \ can be used to find LOS column densities that agree with values from various models in literature (as seen in Figure \ref{fig:c2povslit}). We found that \ctpo \ struggles to properly fit sources that feature strong relativistic reflection (those that required \texttt{relxill} when fitting). This is expected, as relativistic reflection effects are not accounted for within these simulations. Much like when working with \myt, the best course of action to fit these sources is to add in an additional component to properly reproduce relativistic reflection, such as \texttt{relxill}, making sure that the physical parameters are linked between the models where possible. Overall, \ctpo \ is able to remain consistent with the results from well-established models while successfully fitting a wide range of X-ray spectra and providing physical information on the torus properties (opening angle, inclination, etc.). In Figure \ref{fig:c2po_nh_comparison}, there are a few outliers for which the phenomenological/\myt \ LOS column density does not agree with the \ctpo \ estimate (e.g. for NGC5506 and NGC7674), despite the general agreement with literature values seen in Figure \ref{fig:c2povslit}. These discrepancies may arise for a number of reasons, including the aforementioned caveats regarding Equation \ref{eq:nh_conversion}. The detailed comparisons between model results for the individual sources are presented in Appendix \ref{appen:Notes}.
        
    \subsection{Multi-Wavelength AGN Tracers} \label{sec:Multi-Wavelength AGN Tracers}

        Using the results from the spectral modelling with \ctpo, we compare our sample of galaxies to well-known multiwavelength AGN relations. We aim to check whether our sample follows the established relations, or if new correlations can be found for these mid-IR-bright SF-AGN. As our fluxes are extracted from entire galaxies, as opposed to just the central ``nucleus'', we made sure to use relations found for ``total fluxes'' from entire galaxies. We fitted all lines following the methods outlined in \citet{Hogg2010} for data with both x- and y-axis uncertainties, and intrinsic scatter. The fitting was completed using the \texttt{emcee} \citep[][]{ForemanMackey2013} Python package to implement Markov chain Monte Carlo (MCMC) methods. The $90\%$ confidence intervals of the fitted lines are also plotted.

        \subsubsection{[OIII]$\lambda$5007 Line}
        
\begin{figure}
                \centering
                \includegraphics[width=0.95\linewidth]{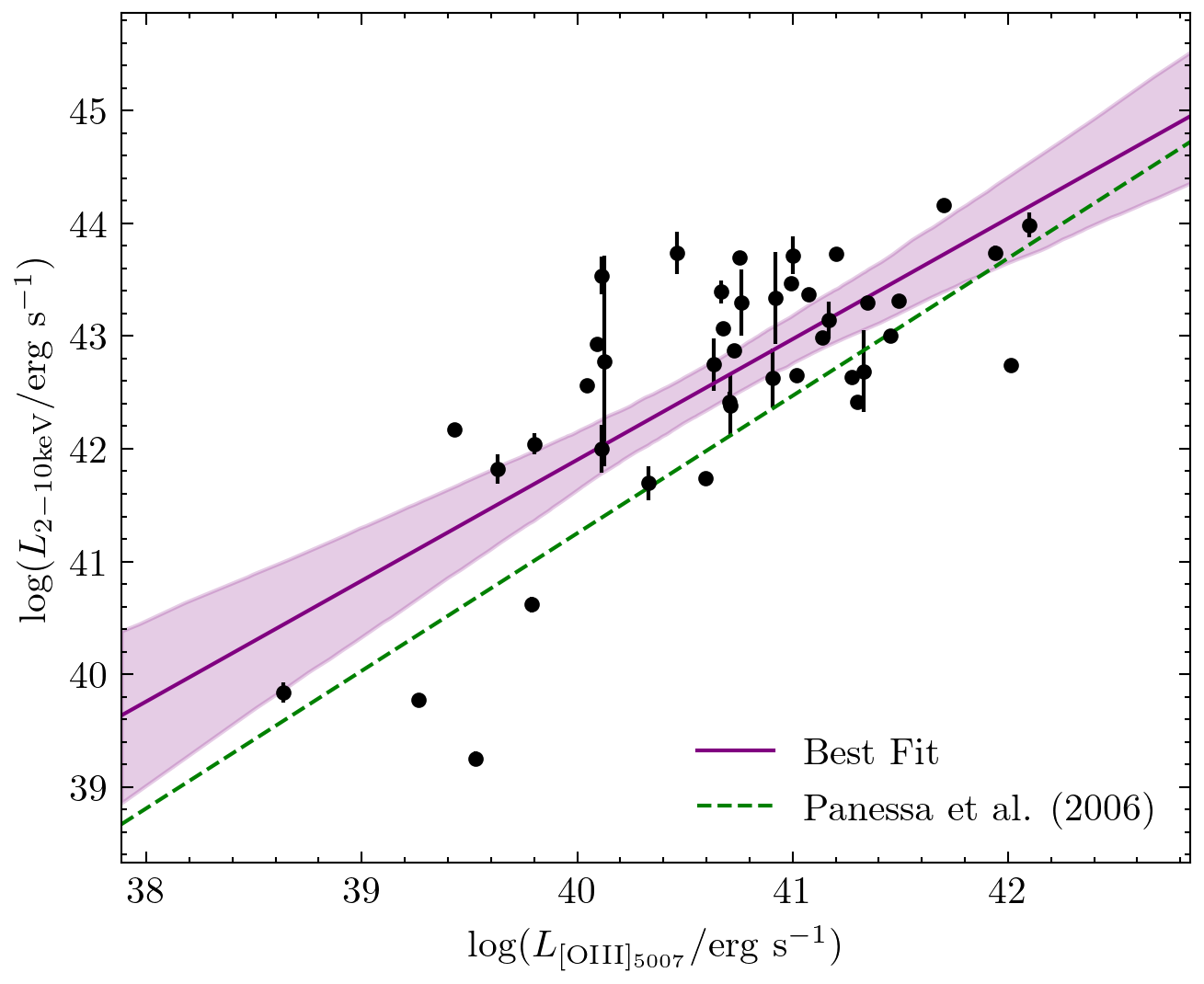}
                \caption{Log of the $2-10$ keV luminosity vs log of the [OIII]$_{5007}$ line luminosity. We find there is a strong linear correlation between the luminosities, with a Pearson's correlation coefficient of $r=0.73$. The solid, purple line shows our best fit to the data, which is given by $\log (L_{ \rm 2-10 \ keV}) = (1.07 \pm 0.16) \log (L_{ \rm [OIII]}) + (0.95 \pm 6.32)$. The $90\%$ confidence intervals of the fit are plotted as the light purple shading. Comparing our fitted line to the $ L_{ \rm [OIII]}- L_{\text{X}}$ relation from \citet{Panessa2006} (plotted as the green, dashed line), we find that our fit agrees with this relation within the uncertainties.}
                \label{fig:LX_OIII_comparison}
            \end{figure}

            The [OIII]$\lambda5007$ line is one of the most prominent optical lines found in AGN spectra. It is generally not strongly contaminated by emission from star formation due to its high ionization potential \citep[$35.11 \ \rm eV$, see][]{Feltre2023}, and thus is commonly used as an indicator of the intrinsic AGN power \citep[see][]{LaMassa2010}. Its luminosity has been found to strongly correlate with the $2-10$ keV X-ray luminosity for different types of AGN \citep[e.g.][]{Netzer2006,Lamastra2009,Georgantopoulos2010}. We used the [OIII]$\lambda5007$ line luminosities from \citet{Feltre2023} to compare with our estimates of the intrinsic $2-10$ keV luminosity, as shown in Figure \ref{fig:LX_OIII_comparison}.
            
            As we can see from Figure \ref{fig:LX_OIII_comparison}, our sample does not deviate from the $ L_{\rm [OIII]}- L_{\text{X}}$ relation derived by \citet{Panessa2006}, indicating that the AGN in our sample are not significantly different from the local population.

        \subsubsection{$12 \ \mu$m Luminosity}

\begin{figure}
                \centering
                \includegraphics[width=0.95\linewidth]{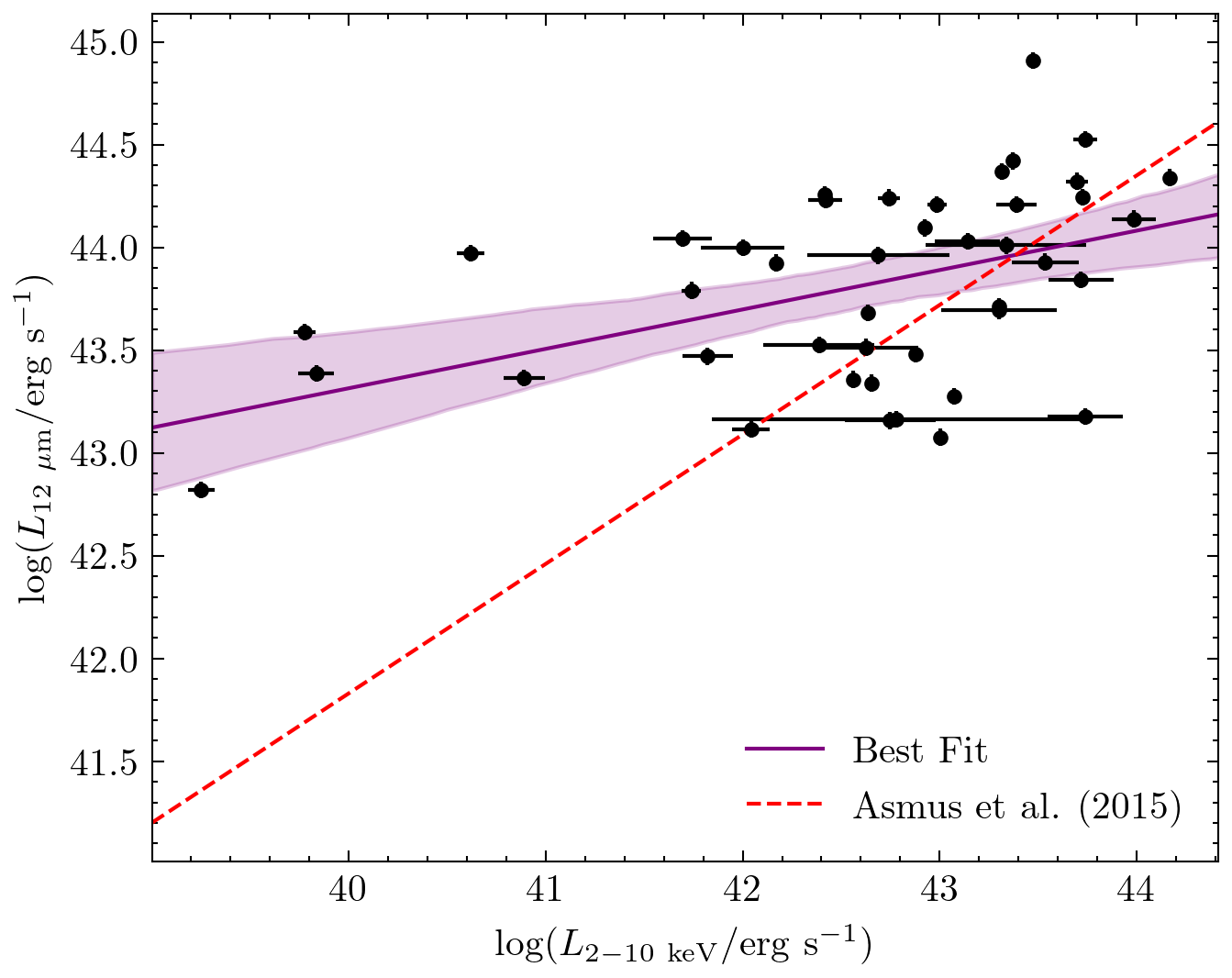}
                \caption{Log of the $2-10$ keV luminosity vs log of the $12 \ \mu \rm m$ luminosity. The solid, purple line shows our best fit to the data, which is given by $\log(L_{ \rm 12 \ \mu m}) = (0.19 \pm 0.06) \log(L_{ \rm 2-10 \ keV}) + (35.7 \pm 2.5)$. The $90\%$ confidence intervals of the fit are plotted as the light purple shading. Comparing our fitted line to the $L_{ \rm 12 \ \mu \rm m, \ tot}-L_{\rm X}$ relation from \citet{Asmus2015} (plotted as the red, dashed line), we find that our sample deviates from this relation, at $\log (L_{\rm 2-10 \ keV} / \rm \ erg \ s^{-1}) \lesssim 41.5$. This is likely due to high SFRs, see Section \ref{sec:radiolum} for more details.}
                \label{fig:LX_12mum_comparison}
            \end{figure}

            In SF-AGN galaxies, the total mid-IR emission is generally a combination of central AGN emission and SF-region emission. However, as the origin of the X-ray and mid-IR AGN emission is thought to be from a common source (the UV emission from the accretion disk, either Compton up-scattered to the X-ray or reprocessed by dust into the mid-IR), it is expected that there should be a correlation between these wavelengths. Both \citet{Gandhi2009} and \citet{Asmus2015} found a strong correlation between the $12 \ \mu \rm m$ and $2-10 \rm \ keV$ luminosities for local AGN for $ \rm 10^{39} \ erg \ s^{-1} \leq L_{2-10 \ keV} \leq 10^{46} \ erg \ s^{-1}$. The correlation was particularly strong for the $12 \ \mu \rm m$ luminosities extracted from the central ``nuclear'' region of the galaxies, which were separated out using sub-arcsecond resolution mid-IR measurements. However, \citet{Asmus2015} still found a clear relation between the two luminosities when using the total $12 \ \mu \rm m$ measurements from \textit{IRAS}. For our sample, we used the $12 \ \mu \rm m$ fluxes from \citet{Gruppioni2016}, which are extracted from the entire galaxy using the \textit{Spitzer} IRS instrument, to compare these galaxies to the \citet{Asmus2015} relation for the total $12 \ \mu \rm m$ luminosity, shown in Figure \ref{fig:LX_12mum_comparison}. To ensure consistency, we compared our \textit{Spitzer} IRS fluxes with the \textit{IRAS} fluxes, and found them to be in agreement.

            From Figure \ref{fig:LX_12mum_comparison} it is clear that our sample deviates from the \citet{Asmus2015} $L_{ \rm 12 \ \mu m, \ tot}-L_{\rm X}$ relation at $\log (L_{\rm 2-10 \ keV} / \rm \ erg \ s^{-1}) \lesssim 41.5$. We do expect that all of our sources would have high $L_{ \rm 12 \ \mu m, \ tot}$, as this sample was specifically selected as being bright at $12 \rm \ \mu m$. This excess at low X-ray luminosities is likely caused by the high SFR of the host galaxies contributing to the $12 \ \mu \rm m$ luminosity. The sources with $\log (L_{\rm 2-10 \ keV} / \rm \ erg \ s^{-1}) >41.5$ may suffer from the same effect, but their higher intrinsic AGN luminosities dominate over the contributions from the star formation.

        \subsubsection{$1.4$ GHz Radio Luminosity}
        \label{sec:radiolum}

\begin{figure}
                \centering
                \includegraphics[width=0.95\linewidth]{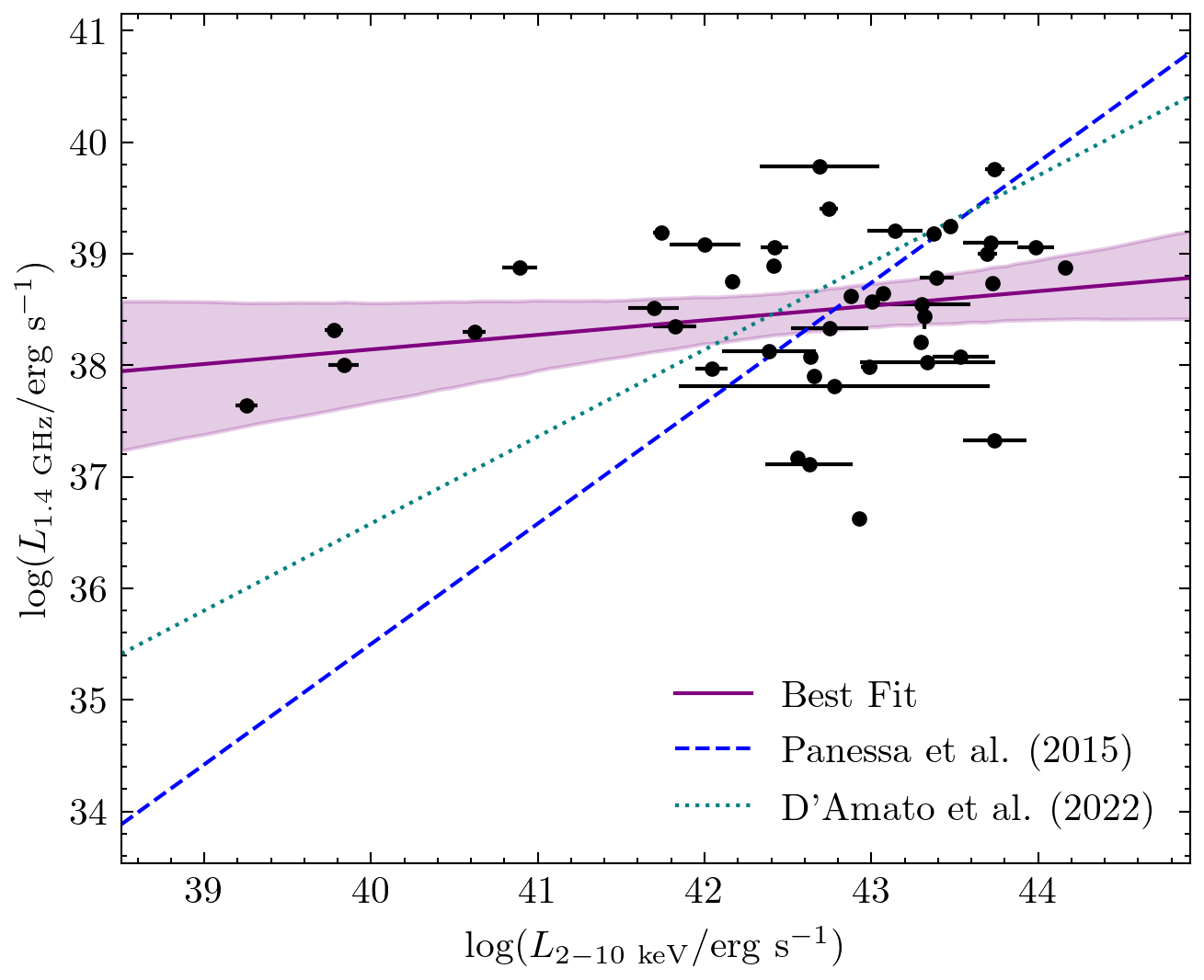}
                \caption{Log of the $2-10$ keV luminosity vs log of the $1.4 \rm \ GHz$ luminosity. The solid, purple line shows our best fit to the data, which is given by $\log ( L_{1.4 \rm \ GHz}) = (0.13 \pm 0.10) \log(L_{ \rm 2-10 \ keV}) + (32.9 \pm 4.1)$. The $90\%$ confidence intervals of the fit are plotted as the light purple shading. Comparing our fitted line to the $ L_{1.4 \rm \ GHz}-L_{\rm X}$ relations from \citet{Panessa2015} and \citet{DAmato2022} (plotted as the blue, dashed and teal, dotted lines respectively), we find that our sample does not follow these relations at $\log (L_{\rm 2-10 \ keV } / \ \rm erg \ s^{-1}) \lesssim 41.5$. This is likely due to high SFRs, see Section \ref{sec:radiolum} for more details.}
                \label{fig:LX_radio_comparison}
            \end{figure}

             We decided to test another SF-contaminated AGN luminosity relation to see if our sample of galaxies may have higher than normal emission from SF regions and due to high SFRs. Specifically, we decided to look at $ L_{\rm Radio}-L_{ \rm X}$ relations, as Radio emission from SF-AGN galaxies comes both from the jets of the central engines, and from SF-related processes present throughout the galaxy (thermal bremsstrahlung in $\rm HII$ regions, non-thermal emission from pulsars, etc.). We used \textit{MeerKAT} $1.28 \ \rm GHz$ total flux densities acquired from \citet{Condon2021} for 14 of the sources, and the remaining 29 sources were observed as part of \textit{MeerKAT} Proposal ID: MKT-22054, PI: L. Marchetti, for which we received the fluxes through private communication (to be released in Moloko et al., in prep.). We converted these fluxes to $1.4 \rm \ GHz$ luminosities assuming a constant spectral index of $\alpha = 0.7$ \citep[the canonical value from][]{Condon2002}. We then compared our sample with the $ L_{ \rm 1.4 \ GHz}-L_{\rm X}$ relations from \citet{Panessa2015} and \citet{DAmato2022}, as seen in Figure \ref{fig:LX_radio_comparison}.

            Much like for the $ L_{12 \ \mu \rm m, \ tot}-L_{\rm X}$ relation, it can be clearly seen in Figure \ref{fig:LX_radio_comparison} that our sample deviates from the \citet{Panessa2015} and \citet{DAmato2022} $ L_{\rm 1.4 \ GHz}-L_{\rm X}$ relations at $\log (L_{\rm 2-10 \ keV } / \ \rm erg \ s^{-1}) \lesssim 41.5$. This seems to confirm that our sample could have particularly high SFR, which contribute to the measured $12 \mu \rm m$ and Radio luminosities. This could also be the cause of the mid-IR brightness that led to them being selected as part of the 12MGS in the first place. 
            
            To test this theory, we calculated the required SFRs to produce the excess $12 \ \mu m$ and $1.4 \ \rm GHz$ fluxes observed, using the \citet{Donoso2012} and \citet{Davies2017} relations respectively. We found that the galaxies need SFRs $\leq 126 \rm \ M_{\odot} \ yr^{-1}$ to account for the $12 \ \mu \rm m$ excess, or $\leq 55 \rm \ M_{\odot} \ yr^{-1}$ to account for the $1.4 \ \rm GHz$ excess. Having these SFRs would classify them as star-forming or even starburst galaxies. However, it is only possible to confirm this theory by performing comprehensive SED fitting to properly disentangle the AGN and SF emission components and determine the SFRs of the galaxies. We intend to release the results of the X-ray-Radio SED fitting completed with the assistance of the \ctpo \ results in a follow-up paper.

\section{Conclusions} \label{sec:Conclusions}

    In this paper, we studied the X-ray spectra of a sample of 43 $\rm 12 \ \mu m$-selected SF-AGN galaxies within the Southern Sky. We analysed archival \textit{NuSTAR}, \textit{Chandra}, and \textit{XMM-Newton} data for 42 of the galaxies, and present the first pointed X-ray observation of the remaining source, IRASF03450+0055, taken by \textit{NuSTAR} in 2025. We performed comprehensive X-ray spectral fitting of all sources, using phenomenological and physically-based models. We also presented a new physically-based X-ray spectral model, \ctpo, designed to provide constraints on physical parameters of the torus that can be directly linked to the parameters of the IR AGN model SKIRTOR (e.g. the torus opening angle, line-of-sight inclination, and radial dust distribution gradient). This model will allow us to exploit the synergies between the X-ray and IR regimes to place stronger constraints on the torus properties, that we are otherwise unable to do so without significant degeneracies, especially in the case of sparse photometric coverage. We tested the \ctpo \ model on our entire sample, and compared its results to those obtained from the phenomenological and physically-based modelling. We also compared our results with AGN scaling relations, specifically the $ L_{ \rm [OIII]}- L_{\text{X}}$, $L_{\rm X} - L_{\rm 12 \ \mu m, \ tot}$ and $ L_{ \rm 1.4 \ GHz}- L_{\text{X}}$ relations, to compare this sample to the general AGN population. The main results of this paper can be summarised as follows:

    \begin{enumerate}[i)]
    
        \item The fitting of the \textit{NuSTAR} spectrum of IRASF03450+0055 revealed an incredibly steep photon index $(\Gamma \sim 3)$, and clear evidence of relativistic reflection, which strongly suggests that this source is a Narrow-Line Seyfert 1. 
        
        \item We showed that \ctpo \ can be extremely effective when fitting a wide variety of X-ray spectra, successfully replicating the results of well-established phenomenological and physically-based models, while providing information on the physical structure of the torus that can be immediately linked to those obtained in the IR from the SKIRTOR AGN model. As our simulations do not account for soft excess or relativistic reflection from the accretion disk, additional model components such as \texttt{mekal} and \texttt{relxill} should be added to account for these features, as is the case for other physically-based models.

        \item When we compared \ctpo \ with the \myt \ smooth torus model for the Compton-thick sources, and with phenomenological absorbers for sources with lower levels of obscuration, we find that the LOS column density is generally consistent, irrespective of the assumed geometry. The differences between clumpy and smooth model geometries appear instead in the reprocessed and reflected emission, with an advantage of our clumpy model being that one can decouple the LOS and equatorial column densities self-consistently rather than through arbitrary normalisation constants as are required by a decoupled smooth model such as \myt. However, we highlight that the LOS column densities we present for \ctpo \ are indicative estimates, as they are not fitted quantities, as opposed to the equatorial column densities, which are.

        \item Our sample does not deviate from the \citet{Panessa2006} $ L_{\rm [OIII]}- L_{\text{X}}$ relation, which indicates that our sources are not significantly different from the whole AGN population. However, we find that the sources with $L_{\rm 2-10 \ keV}<10^{41.5} \rm \ erg \ s^{-1}$ have significantly higher $L_{\rm 12 \ \mu m}$ and $L_{\rm 1.4 \ GHz}$ luminosity than what is expected from the \citet{Asmus2015} $L_{ \rm 12 \ \mu m, \ tot}-L_{\rm X}$ and \citet{Panessa2015} $ L_{\rm 1.4 \ GHz}-L_{\rm X}$ relations at low intrinsic X-ray luminosities. We suggest that this is due to SF-related emission outshining the low-luminosity AGN in the $12 \ \mu \rm m$ and $1.4 \rm \ GHz$ bands.
        
    \end{enumerate}

    We plan to update the \ctpo \ model to incorporate the full parameter space explored by the SKIRTOR model. We will present and show the improvements provided by the use of \ctpo \ in a following paper (Gilbert et al., in prep), where we will perform full spectrum (X-ray to Radio) SED fitting of our sample. Once completed, the SED fitting will help to fully disentangle the AGN and SF emission components, and will reveal if the excess mid-IR and Radio emission in the low-luminosity AGN is a result of high SFRs. As these sources may act as local proxies of the dusty SF-AGN found at Cosmic Noon, by studying their AGN and SF emission, we will continue to gain a better understanding of the interplay and possible co-evolution of these galactic components across Cosmic time.

\section*{Acknowledgements}

CJEG acknowledges the financial assistance of the South African Radio Astronomy Observatory (SARAO) (\url{www.sarao.ac.za}).
CJEG, LM, LB, MV acknowledge financial support from the Inter-University Institute for Data Intensive Astronomy (IDIA). IDIA is a partnership of the University of Cape Town, the University of Pretoria and the University of the Western Cape. IDIA is registered on the Research Organization Registry with ROR ID 01edhwb26, and on Open Funder Registry with funder ID 100031500.
CJEG, LM, LB, MV acknowledge financial support from the South African Department of Science and Innovation’s National Research Foundation under the ISARP RADIOMAP Joint Research Scheme (DSI-NRF Grant Number 150551) and the CPRR Projects (DSI-NRF Grant Numbers SRUG22031677 and SRUG2204254729).
FS acknowledges financial support from the PRIN MUR 2022 2022TKPB2P - BIG-z, Ricerca Fondamentale INAF 2023 Data Analysis grant 1.05.23.03.04 ``ARCHIE ARchive Cosmic HI \& ISM  Evolution'', Ricerca Fondamentale INAF 2024 under project 1.05.24.07.01 MINI-GRANTS RSN1 "ECHOS", Bando Finanziamento ASI CI-UCO-DSR-2022-43 CUP:C93C25004260005 project ``IBISCO: feedback and obscuration in local AGN''.

This research has made use of the NASA/IPAC Extragalactic Database, which is funded by the National Aeronautics and Space Administration and operated by the California Institute of Technology, and of the SIMBAD database and the VizieR catalogue access tool (both operated at CDS, Strasbourg, France). This work has made use of data from the \textit{NuSTAR} mission, a project led by the California Institute of Technology, managed by the Jet Propulsion Laboratory, and funded by the National Aeronautics and Space Administration. We thank the \textit{NuSTAR} Operations, Software and Calibration teams for support with the execution and analysis of these observations. This research has made use of the (NUSTARDAS) jointly developed by the ASI Science Data Center (ASDC, Italy) and the California Institute of Technology (USA). The work is also based on observations obtained with \textit{XMM-Newton}, an ESA science mission with instruments and contributions directly funded by ESA Member States and the USA (NASA). This research has made use of data obtained from the Chandra Data Archive and the Chandra Source Catalogue, both provided by the Chandra X-ray Center (CXC). This research has made use of software provided by the CXC in the application packages CIAO and WebPIMMS. For data reduction, we used software provided by the High Energy Astrophysics Science Archive Research Center (HEASARC) at NASA/Goddard Space Flight Center. Part of this work is based on archival data, software or online services provided by the Space Science Data Center - ASI. We acknowledge the use of the ilifu cloud computing facility – \url{www.ilifu.ac.za}, a partnership between the University of Cape Town, the University of the Western Cape, Stellenbosch University, Sol Plaatje University and the Cape Peninsula University of Technology. The ilifu facility is supported by contributions from the Inter-University Institute for Data Intensive Astronomy (IDIA – a partnership between the University of Cape Town, the University of Pretoria and the University of the Western Cape), the Computational Biology division at UCT and the Data Intensive Research Initiative of South Africa (DIRISA).

\section*{Data Availability}
 
The X-ray data used in this paper is all publicly accessible through the \textit{Chandra}, \textit{NuSTAR} and \textit{XMM-Newton} data archives. The \texttt{C2PO-Torus} model is available to download from Zenodo at \url{https://doi.org/10.5281/zenodo.21888565}.



\bibliographystyle{mnras}
\bibliography{ref} 



\appendix

\section{Observation IDs} \label{appen:Observation IDs}

The observation IDs of all of the observations used in this paper are presented in Table \ref{tab:obsIDs}.

\begin{table*}
\centering
\caption{Details of our sample and the X-ray observations used. The observation durations for \textit{NuSTAR} are given as FPMA/FPMB. In cases where more than one observation is listed for a given telescope, the spectra were combined per instrument (e.g. all FPMA spectra were combined).}
\label{tab:sources}
\begin{tabular}{@{}llllllll@{}}
\toprule
Source & RA & Dec & z & Instrument & ObsID & Date & Duration (ks) \\ \midrule
CGCG381-051 & 357.173788 & 2.239784 & 0.030921 & \textit{XMM-Newton} & 0655380201 & 2010-12-02 & 16.4 \\
 &  &  &  &  & 0912590101 & 2023-01-05 & 28.0 \\
ESO033-G002 & 73.995459 & -75.54118 & 0.019197 & \textit{NuSTAR} & 60601002002 & 2020-06-09 & 174.7/173.2 \\
 &  &  &  & \textit{XMM-Newton} & 0863050201 & 2020-06-09 & 136.9 \\
ESO141-G055 & 290.308978 & -58.670275 & 0.037109 & \textit{NuSTAR} & 60801011002 & 2022-10-01 & 124.1/122.9 \\
 &  &  &  & \textit{XMM-Newton} & 0913190101 & 2022-10-01 & 124.5 \\
ESO362-G018 & 79.899235 & -32.657752 & 0.012445 & \textit{NuSTAR} & 60201046002 & 2016-09-24 & 101.9/101.6 \\
 &  &  &  & \textit{XMM-Newton} & 0790810101 & 2016-09-24 & 120.8 \\
IC4329A & 207.330257 & -30.309506 & 0.016054 & \textit{NuSTAR} & 60701015002 & 2023-01-11 & 82.4/81.5 \\
 &  &  &  & \textit{XMM-Newton} & 0890670201 & 2023-01-11 & 90.0 \\
IC5063 & 313.009818 & -57.068756 & 0.011348 & \textit{NuSTAR} & 60061302002 & 2013-07-08 & 18.4/18.4 \\
 &  &  &  & \textit{Chandra} & 21466 & 2019-07-23 & 95.4 \\
 &  &  &  &  & 21467 & 2018-12-11 & 27.6 \\
 &  &  &  &  & 21999 & 2018-12-12 & 35.0 \\
 &  &  &  &  & 22000 & 2018-12-13 & 16.5 \\
 &  &  &  &  & 22001 & 2018-12-15 & 30.0 \\
 &  &  &  &  & 22002 & 2018-12-16 & 45.0 \\
IRASF01475-0740 & 27.511238 & -7.430136 & 0.017666 & \textit{NuSTAR} & 60360005002 & 2019-06-16 & 30.7/30.5 \\
 &  &  &  & \textit{XMM-Newton} & 0200431101 & 2004-01-12 & 11.9 \\
IRASF03450+0055 & 56.9175 & 1.08722 & 0.03149 & \textit{NuSTAR} & 61101016002 & 2025-08-16 & 22.4/22.2 \\
IRASF04385-0828 & 70.228999 & -8.372797 & 0.0151 & \textit{NuSTAR} & 60701043002 & 2022-02-14 & 30.7/30.4 \\
 &  &  &  & \textit{XMM-Newton} & 0890690401 & 2022-02-04 & 18.0 \\
IRASF05189-2524 & 80.255821 & -25.362564 & 0.044064 & \textit{NuSTAR} & 60201022002 & 2016-09-05 & 155.1/154.6 \\
 &  &  &  & \textit{XMM-Newton} & 0790580101 & 2016-09-06 & 98.95 \\
IRASF15480-0344 & 237.672908 & -3.888341 & 0.0303 & \textit{NuSTAR} & 90601603002 & 2020-03-23 & 47.95/47.5 \\
 &  &  &  & \textit{XMM-Newton} & 0600690201 & 2010-01-30 & 51.9 \\
MCG+00-29-023 & 170.300993 & -2.984116 & 0.024294 & \textit{XMM-Newton} & 0890690201 & 2022-05-24 & 19.0 \\
 &  &  &  &  & 0890690901 & 2022-05-24 & 9.9 \\
MCG-02-33-034 & 193.051964 & -13.414765 & 0.01463 & \textit{NuSTAR} & 60663002002 & 2021-01-24 & 80.6/79.9 \\
 &  &  &  & \textit{XMM-Newton} & 0723100401 & 2014-01-14 & 68.6 \\
MCG-03-34-064 & 200.601925 & -16.728445 & 0.017179 & \textit{NuSTAR} & \multicolumn{1}{c}{60101020002} & \multicolumn{1}{c}{2016-01-17} & 78.5/78.3 \\
 &  &  &  & \textit{XMM-Newton} & 0763220201 & 2016-01-18 & 142.5 \\
MCG-03-58-007 & 342.404758 & -19.273973 & 0.031462 & \textit{NuSTAR} & 60101027002 & 2015-12-06 & 137.9/137.6 \\
 &  &  &  & \textit{XMM-Newton} & 0764010101 & 2015-12-08 & 134.3 \\
MCG-06-30-015 & 203.974045 & -34.295617 & 0.007749 & \textit{NuSTAR} & 60902004002 & 2024-02-06 & 130.6/129.4 \\
 &  &  &  & \textit{XMM-Newton} & 0921420101 & 2024-02-06 & 135.7 \\
MRK0509 & 311.040636 & -10.723539 & 0.034397 & \textit{NuSTAR} & 91001647004 & 2024-11-15 & 21.2/21.0 \\
 &  &  &  & \textit{XMM-Newton} & 0953791001 & 2024-11-14 & 56.4 \\
MRK0897 & 316.941054 & 3.877807 & 0.02634 & \textit{NuSTAR} & 60701042002 & 2022-04-14 & 20.8/20.6 \\
 &  &  &  & \textit{XMM-Newton} & 0890690301 & 2022-04-27 & 21.0 \\
MRK1239 & 148.07957 & -1.612079 & 0.019927 & \textit{NuSTAR} & 60701038002 & 2021-11-04 & 100.9/99.9 \\
 &  &  &  & \textit{XMM-Newton} & 0891070101 & 2021-11-04 & 105.0 \\
NGC0034 & 2.777255 & -12.107684 & 0.019617 & \textit{NuSTAR} & 60101068002 & 2015-07-31 & 21.4/21.4 \\
 &  &  &  & \textit{XMM-Newton} & 0150480501 & 2002-12-22 & 22.2 \\
 &  &  &  & \textit{Chandra} & 15061 & 2013-06-05 & 15.0 \\
NGC0424 & 17.865163 & -38.083459 & 0.011764 & \textit{NuSTAR} & 60662007002 & 2020-10-24 & 33.7/33.4 \\
 &  &  &  & \textit{Chandra} & 21417 & 2019-02-07 & 16.0 \\
NGC0526A & 20.976547 & -35.065445 & 0.01931 & \textit{XMM-Newton} & 0150940101 & 2003-6-21 & 47.9 \\
NGC1125 & 42.91856 & -16.65064 & 0.010931 & \textit{NuSTAR} & 60510001002 & 2019-06-10 & 31.7/31.5 \\
 &  &  &  & \textit{Chandra} & 21418 & 2018-10-24 & 58.0 \\
NGC1194 & 45.954598 & -1.103714 & 0.013631 & \textit{NuSTAR} & 60501011002 & 2020-01-17 & 58.3/57.9 \\
 &  &  &  & \textit{Chandra} & 22552 & 2019-10-22 & 45.0 \\
 &  &  &  &  & 22880 & 2019-10-25 & 22.0 \\
 &  &  &  &  & 22881 & 2019-10-27 & 23.0 \\
 &  &  &  &  & 23688 & 2020-11-05 & 7.66 \\
NGC1320 & 51.202866 & -3.042254 & 0.009283 & \textit{NuSTAR} & 60061036004 & 2013-02-10 & 27.998/27.96 \\
 &  &  &  & \textit{Chandra} & 23686 & 2020-11-16 & 10.62 \\
NGC1365 & 53.401696 & -36.140186 & 0.005457 & \textit{NuSTAR} & 60002046003 & 2012-07-26 & 40.6/40.5 \\
 &  &  &  & \textit{XMM-Newton} & 0692840201 & 2012-07-27 & 138.5 \\
NGC1566 & 65.001642 & -54.937944 & 0.005017 & \textit{NuSTAR} & 60501031006 & 2019-08-21 & 86.0/85.5 \\
 &  &  &  & \textit{XMM-Newton} & 0851980101 & 2019-08-11 & 17.999 \\
 \hline
 \label{tab:obsIDs}
\end{tabular}
\end{table*}

\begin{table*}
\centering
\contcaption{Details of the sources and the observations used. In cases where more than one observation is listed for a given source and telescope, the spectra were combined for each instrument (e.g. PN, MOS1/2 for each observation were combined).}
\label{tab:sources_continued}
\begin{tabular}{@{}llllllll@{}}
\toprule
\textbf{Source} & \textbf{RA} & \textbf{Dec} & \textbf{z} & \textbf{Instrument} & \textbf{ObsID} & \textbf{Date} & \textbf{Duration (ks)} \\ \midrule
NGC2992 & 146.425211 & -14.326382 & 0.00771 & \textit{NuSTAR} & 90501623002 & 2019-05-10 & 57.5/57.2 \\
 &  &  &  & \textit{XMM-Newton} & 0840920201 & 2019-05-07 & 134.3 \\
NGC4593 & 189.914348 & -5.344176 & 0.008312 & \textit{NuSTAR} & 60001149002 & 2014-12-29 & 23.3/23.3 \\
 &  &  &  & \textit{XMM-Newton} & 0784740101 & 2016-07-14 & 142.1 \\
NGC4602 & 190.153791 & -5.132826 & 0.008469 & \textit{XMM-Newton} & 0890690101 & 2021-12-24 & 20.5 \\
NGC5135 & 201.43345 & -29.83344 & 0.013693 & \textit{NuSTAR} & 60001153002 & 2015-01-14 & 33.4/33.3 \\
 &  &  &  & \textit{Chandra} & 2187 & 2001-09-04 & 30.0 \\
NGC5506 & 213.311982 & -3.207694 & 0.006084 & \textit{NuSTAR} & 60061323002 & 2014-04-01 & 56.6/56.4 \\
 &  &  &  & \textit{XMM-Newton} & 0761220101 & 2015-07-07 & 132.0 \\
NGC5995 & 237.103942 & -13.757563 & 0.025194 & \textit{NuSTAR} & 60061267002 & 2014-08-28 & 21.2/21.1 \\
 &  &  &  & \textit{Chandra} & 17123 & 2015-06-17 & 10.0 \\
NGC6810 & 295.893505 & -58.65559 & 0.006775 & \textit{XMM-Newton} & 0205220101 & 2004-04-25 & 48.7 \\
NGC6860 & 302.195379 & -61.099952 & 0.015054 & \textit{NuSTAR} & 60160726002 & 2024-03-31 & 41.7/41.3 \\
 &  &  &  & \textit{XMM-Newton} & 0903030101 & 2023-03-18 & 126.1 \\
NGC6890 & 304.575444 & -44.806895 & 0.008069 & \textit{NuSTAR} & 60375003002 & 2018-05-25 & 34.6/34.5 \\
 &  &  &  & \textit{XMM-Newton} & 0301151001 & 2005-09-29 & 12.8 \\
NGC7130 & 327.081366 & -34.951248 & 0.016151 & \textit{NuSTAR} & 60261006002 & 2016-12-15 & 42.1/42.0 \\
 &  &  &  & \textit{Chandra} & 2188 & 2001-10-23 & 40.0 \\
NGC7213 & 332.317708 & -47.16675 & 0.005839 & \textit{NuSTAR} & 60001031002 & 2014-10-05 & 53.4/52.9 \\
 &  &  &  & \textit{XMM-Newton} & 0605800301 & 2009-11-11 & 132.5 \\
NGC7469 & 345.815059 & 8.873917 & 0.016268 & \textit{NuSTAR} & 60101001014 & 2015-12-28 & 23.4/23.4 \\
 &  &  &  & \textit{XMM-Newton} & 0760350801 & 2015-12-28 & 101.6 \\
NGC7496 & 347.447043 & -43.42787 & 0.0055 & \textit{XMM-Newton} & 0912590201 & 2022-11-02 & 23.0 \\
NGC7603 & 349.73609 & 0.24398 & 0.028762 & \textit{XMM-Newton} & 0305600601 & 2006-06-14 & 16.8 \\
NGC7674 & 351.986263 & 8.778966 & 0.02903 & \textit{NuSTAR} & 60001151002 & 2014-09-30 & 51.99/51.9 \\
 &  &  &  & \textit{XMM-Newton} & 0200660101 & 2004-06-02 & 10.4 \\
TOLOLO1238-364 & 190.220159 & -36.755825 & 0.010924 & \textit{NuSTAR} & 60001164002 & 2015-01-09 & 58.7/58.6 \\
 &  &  &  & \textit{Chandra} & 4844 & 2004-03-07 & 10.0 \\ \bottomrule
\end{tabular}
\end{table*}


\section{Notes on Individual Sources} \label{appen:Notes}

\subsection*{CGCG381-051}

    Also known as Z 381-051, this nearby Seyfert-2 AGN was analysed using archival \textit{XMM-Newton} observations from 2010 and 2023. \textit{NuSTAR} coverage of this source was available, but we found no signal above the background. We fit the spectrum with an obscured power law model, with an additional \texttt{mekal} component (kT $= 0.82\pm 0.17$ keV) to account for the additional emission in the soft X-ray band. We were only able to find an upper limit for the intrinsic obscuration of $\rm N_{H}\leq 1.8 \times 10^{21} \ cm^{-2}$. We found an intrinsic luminosity of $\log(L_{\rm 0.1-2.4 \ keV} / \rm \ erg \ s^{-1}) = 40.67^{+0.38}_{-0.25}$, which is significantly lower than the estimate of $ \log(L_{\rm 0.1-2.4 \ keV} / \rm erg \ s^{-1}) = 42.24 \pm 0.16$ from \citet{Rush1996}. However, since their estimate was found by converting the count rate using an assumed slope of $\Gamma = 2.3$ (much steeper than our estimate of $\Gamma \sim 1.7 $), and does not account for any soft excess, we assumed that this was likely an overestimation. When using the \ctpo \ model, we had to fix the values of the slope at $\Gamma = 1.8$, opening angle at OA$ = 60\degree$, inclination at $i = 45\degree$, and dust density gradient at $p = 0$ due to the quality of the data. We again needed to add a \texttt{mekal} component with kT $= 0.81^{+0.10}_{-0.14}$ keV. The intrinsic luminosity found agrees with the phenomenological estimate. As the \citet{Rush1996} analysis did not include fitting the obscuration, we do not have any $\rm N_H$ values to compare our estimates against.
    
\subsection*{ESO033-G002}

    This bright Seyfert 2 galaxy was analysed using archival simultaneous \textit{XMM-Newton} and \textit{NuSTAR} observations from 2020. We originally fit the spectrum with a complicated phenomenological model, but we were not satisfied with this fit as there were significant residuals at $E \leq 10$ keV, and many of the fitted parameters were poorly constrained. We decided to redo the fitting using the \texttt{MYTorus} and \texttt{relxill} models, with additional \texttt{zgauss}, \texttt{mekal} and \texttt{zbbody} components to model an absorption line and the soft excess, respectively. The \texttt{mekal} component had a plasma temperature of $\text{kT}= 0.34^{+0.05}_{-0.04}$ keV, and the \texttt{zbbody} component had a temperature of $\text{kT}_\text{e}= 1.34 \pm 0.01$ keV. We found an absorption feature at an energy of $6.83^{+0.02}_{-0.03}$ keV, which is likely due to highly ionised iron \citep[$\rm Fe\, XXV$ and/or $\rm Fe\, XXVI$, as found by][]{Walton2021}. The \texttt{relxill} model, which we added to fit the relativistic reflection hump seen at energies greater than 20 keV, yielded a reflection fraction of $-0.08_{-0.03}^{+0.02}$, an accretion disk ionisation of $\log \xi=2.29^{+0.06}_{-0.11}$, and an iron abundance of $\text{A}_\text{Fe} =3.37^{+0.24}_{-0.21}$. From \myt, we estimated the obscuration to be $\rm N_{H,LOS} = 7.98_{-0.18}^{+0.16} \times 10^{22} \ cm^{-2}$, which is slightly higher than the \citet{Walton2021} estimate of $\rm N_{H,LOS}\sim (5-6) \times 10^{22} \ cm^{-2}$. We estimated the intrinsic luminosity of this model to be $\log (L_{2-10 \rm \ keV} / \rm erg \ s^{-1}) = 42.63 \pm 0.01$, which is slightly lower than the \citet{Walton2021} estimate of $\log (L_{2-10 \rm \ keV} / \rm erg \ s^{-1}) \sim 42.7$, but this may be due to the difference in emission models \citep[][used \texttt{relxill} for both emission and reflection]{Walton2021}. When fitting with \ctpo, we added an additional \texttt{zgauss} component to model the absorption line at $ 7.08 \pm 0.06$ keV. We also added \texttt{relxill} to account for the relativistic reflection that cannot be modelled by \ctpo. It was multiplied by \texttt{zphabs} and \texttt{cabs} components to account for the LOS absorption, with the column density of both components linked to the \ctpo \ $\rm N_{H, eq}$ via Equation \ref{eq:nh_conversion}. We again kept all \texttt{relxill} parameters frozen at the default values except for the ionisation parameter $(\log\xi = 1.70_{-0.15}^{+0.04})$, iron abundance $(\rm A_{Fe} =4.48_{-0.23}^{+0.19})$, and we fixed the reflection fraction $= -1$. We also tied the inclination, photon index and normalisation to that of the main \texttt{C2POTorusD} component. We estimated the LOS column density to be $\rm N_{H,LOS} = 4.65_{-0.35}^{+0.27} \times 10^{22} \ cm^{-2}$, which agrees with the \citet{Walton2021} estimate within the uncertainties. We found an intrinsic luminosity of $\log (L_{2-10 \rm \ keV} / \rm erg \ s^{-1}) = 42.64 \pm 0.02$, which is higher than the estimate found using \myt \ and \texttt{relxill}, but is much closer to the estimate from \citet{Walton2021}.

\subsection*{ESO141-G055}

    ESO141-G055 is a broad-line Seyfert 1 galaxy that has been classified as a `bare AGN' due to its lack of absorption along the line of sight \citet{Porquet2024}. We analysed this galaxy using simultaneous \textit{XMM-Newton} and \textit{NuSTAR} observations from 2022. Following the methods outlined in \citet{Porquet2024}, we used the \texttt{comptt} warm corona model for the soft excess, finding a corona plasma temperature of $\text{kT}_{\text{e}} = 0.65^{+0.21}_{-0.12}$ keV, and an optical depth of $\tau_p = 7.56^{+1.09}_{-1.26}$. These values do not agree with what was found by \citet{Porquet2024}; however, we let our input temperature vary (finding $T_0 = 0.07 \pm 0.01$), which would affect the previous values significantly. We used one \texttt{zgauss} component for an emission line at $6.40_{-0.01}^{+0.02}$ keV, which we assume is the Fe K$\alpha$ line. We used the \texttt{relxill} model to fit the relativistic reflection, keeping all parameters frozen at the default values except for the inclination ($i = 41.59^{+4.04}_{-4.25}$ degrees), ionisation parameter $(\log \xi = 3.71_{-0.06}^{+0.05})$, iron abundance $(\rm A_{Fe} =2.22_{-0.38}^{+0.35})$, and we fixed the reflection fraction $= -1$. We linked the photon index and normalisation to that of the primary power law. We did not use any obscuring component in this model. We found the intrinsic luminosity of this model to be $\log (L_{2-10 \rm \ keV} / \rm erg \ s^{-1}) = 43.17 \pm 0.02$, which is much lower than the value found by \citet{Porquet2024} of $\log (L_{2-10 \rm \ keV} / \rm erg \ s^{-1}) \sim 44$, however this is the total luminosity, not the intrinsic. When we fitted using \ctpo, we had to use an additional \texttt{mekal} component with kT $=0.19 \pm 0.01$ keV for the soft excess. We again used \texttt{relxill} to model the relativistic reflection, keeping all parameters frozen at the default values except for the inclination ($i = 68.93^{+2.12}_{-1.98}$ degrees), ionisation parameter $(\log \xi = 2.36_{-0.14}^{+0.15})$, iron abundance $(\rm A_{Fe} =3.42_{-0.58}^{+0.48})$, and we fixed the reflection fraction $= -1$. We linked the photon index and normalisation to that of the main \texttt{C2POTorusD} component. We did not link the inclination to that of \ctpo \ as the torus and accretion disk may not lie in the same plane, and thus used the \ctpo \ inclination to tie the \texttt{zphabs} and \texttt{cabs} $\rm N_{H, eq}$. Despite this being classified as a `bare AGN', we find an inclination that intersects with the torus, albeit right on the edge. As such, even though the LOS obscuration estimated is $\mathrm{N_{H, LOS}} =36.71_{-2.20}^{+1.93} \times 10^{22} \ \mathrm{cm}^{-2}$, we emphasise that this is likely to be inaccurate, as Equation \ref{eq:nh_conversion} relies on simplifications of the torus geometry. It might be such in this case that although the inclination does appear to intersect the torus, the clumpy geometry may be such that there is no direct LOS obscuration, as we may be `peeking through' the clumps. We estimate the intrinsic luminosity to be $\log (L_{2-10 \rm \ keV} / \rm erg \ s^{-1}) = 43.32_{-0.02}^{+0.01}$, which is lower than the total luminosity estimate from \citet{Porquet2024}, as is expected.

\subsection*{ESO362-G018}

    We analysed this Seyfert 1.5 galaxy using simultaneous \textit{XMM-Newton} and \textit{NuSTAR} observations from 2016. To model the intrinsic obscuration, we used the \texttt{zxipcf} model and found $\log \xi = 2.07^{+0.08}_{-0.07}$ with $37 \pm 2 \%$ coverage. We added a \texttt{zgauss} component to model the Fe K$\alpha$ emission line at $6.39\pm 0.01$ keV. The soft part of the spectrum was complex, requiring both a \texttt{mekal} component with kT $=0.14_{-0.03}^{+0.01}$ keV and a \texttt{zgauss} component at $0.90 \pm 0.01$ keV to model the soft excess. This is likely to be the $\rm Ne\, IX$ line, which can be detected in the soft X-ray spectra of AGN \citep[e.g.][]{Braito2007}. We estimated the intrinsic luminosity of this model to be $\log (L_{2-10 \rm \ keV} / \rm erg \ s^{-1}) = 42.60 \pm 0.01$, which is very close to the estimate of $\log (L_{2-10 \rm \ keV} / \rm erg \ s^{-1}) \sim 42.5$ from \citet{Xu2021}. Using \ctpo, we also needed an additional \texttt{mekal} component (kT $= 0.12 \pm 0.01$ keV). We estimate the average equatorial absorption to be $\rm N_{H, eq} = 78.7_{-6.0}^{+5.0} \times 10^{22} \ cm^{-2}$, which is higher than the estimate of $\rm N_{H, eq} = 10.7_{-6.7}^{+20.9} \times 10^{22} \ cm^{-2}$ found by \citet{Xu2021} using \texttt{Borus}, but they employed an additional warm absorber. Our estimate of the intrinsic luminosity from this model was $\log (L_{2-10 \rm \ keV} / \rm erg \ s^{-1}) = 42.65 \pm 0.01$, which is slightly higher than the \citet{Xu2021} estimate, but this difference can be explained by the difference in intrinsic absorption.

\subsection*{IC4329A}

    IC4329A is a Seyfert 1 galaxy that we analysed using simultaneous \textit{XMM-Newton} and \textit{NuSTAR} observations taken in 2023. We modelled the obscuration of this bright source using both a neutral absorber model (\texttt{zphabs}) and a partially-covering partially ionised absorber model (\texttt{zxipcf}). For the latter, we found $\log \xi = 2.21^{+0.06}_{-0.05}$ and $25^{+1}_{-2}\%$ coverage. To model the soft excess, we used both \texttt{mekal} (kT $= 0.27^{+0.01}_{-0.02}$ keV) and scattering (f$_{\text{scatt}} = 6.0_{-0.5}^{+0.4} \%$). We also found the Fe K$\alpha$ and K$\beta$ emission lines at $6.41\pm 0.01$ keV and $7.06^{+0.09}_{-0.13}$ keV, which we modelled using two \texttt{zgauss} components. We estimate the intrinsic luminosity to be $\log (L_{2-10 \rm \ keV} / \rm erg \ s^{-1}) = 43.74 \pm 0.01$. This is slightly lower than the average of the estimates from \citet{Ogawa2019} of $\log (L_{2-10 \rm \ keV} / \rm erg \ s^{-1}) \sim 43.86$,  but given that different instruments were used \citep[\textit{XMM-Newton} vs \textit{Suzaku},][]{Suzaku} and the observations were taken 11 years apart, this slight difference is likely due to the different observations analysed. Using \ctpo, we again utilised a \texttt{mekal} component to model the soft excess, with kT $= 0.22 \pm 0.01$ keV. We estimated the obscuration to be $\rm N_{H,eq} = 14.12^{+0.31}_{-0.32} \times 10^{22} \ cm^{-2}$, which agrees with the average of the estimates of $\rm N_{H,eq} = 10.1^{+4.3}_{-4.8} \times 10^{22} \ cm^{-2}$ from \citet{Ogawa2019}, found using \texttt{XCLUMPY} and additional partial absorbers. We found a luminosity of $\log (L_{2-10 \rm \ keV} / \rm erg \ s^{-1}) = 43.73 \pm 0.01$, which is again slightly lower than the average \citet{Ogawa2019} estimate, but agrees with our phenomenological estimate.

\subsection*{IC5063}

    We analysed this Seyfert 2 galaxy using a \textit{NuSTAR} observation from 2013 and \textit{Chandra} observations from 2018/2019. There was a significant variation in flux between the two observation epochs, and so we have reported the average intrinsic luminosity in Tables \ref{tab:phenomenological_results} and \ref{tab:carrot_results}. Using the neutral refection model \texttt{pexrav}, we found a relative reflection of $-0.81 \pm 0.06$. We also added a \texttt{zgauss} component at $6.39 \pm 0.01$ keV to account for the Fe K$\alpha$ emission line, and a \texttt{mekal} component with kT $= 0.66^{+0.14}_{-0.09}$ keV for the soft excess. We found a LOS column density of $\rm N_{H, LOS} = 33.66 \pm 0.69 \times 10^{22} \ cm^{-2}$, which agrees with the average of the \citet{EsparzaArredondo2019} estimates of $\rm N_{H, LOS} = 18_{-12}^{+25} \times 10^{22} \ cm^{-2}$, found by simultaneously fitting the X-ray and IR spectra using the \texttt{borus02} model and different IR models. We found an intrinsic luminosity of $\log (L_{2-10 \rm \ keV} / \rm erg \ s^{-1}) = 42.94 \pm 0.01$ for the \textit{NuSTAR} epoch, and $\log (L_{2-10 \rm \ keV} / \rm erg \ s^{-1}) = 42.79 \pm 0.01$ for the \textit{Chandra} epoch. \citet{Tazaki2011} estimated an intrinsic luminosity $ \log (L_{2-10 \rm \ keV} / \rm erg \ s^{-1}) \sim 42.8$ using observations from the \textit{Suzaku} telescope, which is slightly lower than our average estimate, but a difference is expected due to instrumental differences and the intrinsic long-term variability of the source. When fitting with the \ctpo \ model we also used a \texttt{mekal} component to model the soft excess, for which we found kT $= 0.69^{+0.22}_{-0.21}$ keV. We found an average equatorial column density of $\rm N_{H,eq} = 39.52^{+1.24}_{-0.87} \times 10^{22} \ cm^{-2}$, which is slightly higher than the \citet{Tazaki2011} estimate of $\rm N_{H,eq} = 25^{+10}_{-1} \times 10^{22} \ cm^{-2}$, found using the \texttt{Torus} model. This discrepancy is likely due to the difference in model geometries, the different observations used, and/or the intrinsic source variability. Comparing with the average of the \citet{EsparzaArredondo2019} estimates for the equatorial obscuration of $\rm N_{H,eq} = 93^{+24}_{-27} \times 10^{22} \ cm^{-2}$, we find that our value is lower by a factor of $\sim2-3$. This may also be due to a difference in model geometries, as \texttt{borus02} employs a smooth torus. We estimated an intrinsic luminosity of $\log (L_{2-10 \rm \ keV} / \rm erg \ s^{-1}) = 43.02\pm 0.02$ for the \textit{NuSTAR} epoch, and $\log (L_{2-10 \rm \ keV} / \rm erg \ s^{-1}) = 42.35 \pm 0.02$ for the \textit{Chandra} epoch with this model. This average luminosity agrees with the previous and \citet{Tazaki2011} estimates within the uncertainties.

\subsection*{IRASF01475-0740}

    We used \textit{XMM-Newton} and \textit{NuSTAR} observations from 2004 and 2019 respectively to analyse this Seyfert 2 galaxy. Despite the large temporal gap, we found no significant flux variation between the observations. We used a simple model for this source, only needing an extra \texttt{zgauss} component at $6.6^{+0.1}_{-0.3}$ keV to model the Fe K$\alpha$ line. We estimated the LOS column density to be $\rm N_{H,LOS} = 0.43 \pm 0.04 \times 10^{22} \ cm^{-2}$, which agrees with the \citet{Akylas2024} estimate of $\rm N_{H,LOS} = 0.09_{-0.33}^{+0.34} \times 10^{22} \ cm^{-2}$ within the uncertainties. Our estimated intrinsic luminosity also agrees with the value from \citet{Akylas2024} of $\log (L_{2-10 \rm \ keV} / \rm erg \ s^{-1}) \sim 41.74$. Using the \ctpo \ model, we did not need any extra model components. We found a LOS obscuration of $\rm N_{H,LOS} = 0.31 \pm 0.03 \times 10^{22} \ cm^{-2}$, and an intrinsic $2-10$ keV luminosity of $\log (L_{2-10 \rm \ keV} / \rm erg \ s^{-1}) = 41.74^{+0.04}_{-0.05}$, both of which agree with the previous and \citet{Akylas2024} estimates.

\subsection*{IRASF04385-0828}

    We analysed this Seyfert 1 galaxy using simultaneous 2022 \textit{XMM-Newton} and \textit{NuSTAR} observations. It is a known Compton-thick source \citep[see][]{Akylas2024}, with a LOS column density $\rm \sim 10^{24} \ cm^{-2}$. We started by fitting it using the \myt \ model. When using \myt, we used an additional \texttt{mekal} component with kT $= 0.36^{+0.24}_{-0.14}$ keV to model the soft excess. We were only able to find lower limits on the LOS and equatorial column densities of $\mathrm{N_{H, LOS}} \geq 2.76 \times 10^{24} \ \mathrm{cm}^{-2}$ and $\rm N_{H, eq} \geq 2.70 \times 10^{24} \ cm^{-2}$, much like \citet{Akylas2024} who found  $\mathrm{N_{H, LOS}} \geq 3 \times 10^{24} \ \mathrm{cm}^{-2}$ and $\rm N_{H, eq} \geq 4.21 \times 10^{24} \ cm^{-2}$ using the \texttt{RXTORUS} model. As such, we can only estimate a lower limit on the intrinsic luminosity, which we find by fixing the column densities at their lower limit values and freezing all other parameters at their best-fit values. This resulted in an intrinsic luminosity estimate of  $\log (L_{2-10 \rm \ keV} / \rm erg \ s^{-1}) \geq 42.68$, which agrees with the \citet{Akylas2024} estimate of $\log (L_{2-10 \rm \ keV} / \rm erg \ s^{-1}) \sim 42.88$. Using \ctpo, we again used a \texttt{mekal} component (kT $= 0.38^{+0.11}_{-0.08}$ keV) to model the soft excess. We were able to constrain the obscuration, finding $\rm N_{H,eq} = 5.7^{+2.2}_{-2.0} \times 10^{24} \ cm^{-2}$, which agrees with both the previous and \citet{Akylas2024} estimates. We then found an intrinsic luminosity of $\log (L_{2-10 \rm \ keV} / \rm erg \ s^{-1}) = 43.54^{+0.11}_{-0.17}$, which agrees with our original lower limit estimate. It is higher than the \citet{Akylas2024} estimate; however, this discrepancy could be explained by assuming that their estimate also may be a lower limit, as they were also unable to constrain the obscuration.
    
\subsection*{IRASF05189-2524}

    We used simultaneous \textit{XMM-Newton} and \textit{NuSTAR} observations from 2016 to study this bright Seyfert 1 galaxy. To model its soft excess, we included a scattering component with $f_{\text{scatt}} = 1.16 \pm 0.09 \%$, and two \texttt{mekal} components with kT $= 0.10 \pm 0.01$ keV and kT $= 0.68^{+0.08}_{-0.06}$ keV. We also found the Fe K$\alpha$ and K$\beta$ emission lines at $6.44 \pm 0.06$ keV and $6.72 \pm 0.05$ keV, which we modelled with two \texttt{zgauss} components. We estimated the LOS obscuration to be $\rm N_{H,LOS} = 7.64_{-0.17}^{+0.18} \times 10^{22} \ cm^{-2}$, which is slightly higher than the value found by \citet{Teng2015} for their main absorber of $\rm N_{H,LOS} = 5.19_{-0.18}^{+0.20} \times 10^{22} \ cm^{-2}$, however they also employed an additional partial absorber. \citet{Teng2015} found an intrinsic luminosity $\log (L_{2-10 \rm \ keV} / \rm erg \ s^{-1}) \sim 43.57$, which is slightly higher than our estimate of $\log (L_{2-10 \rm \ keV} / \rm erg \ s^{-1}) = 43.44 \pm 0.01$. The difference could be explained by the different observations used, and/or intrinsic variability of the source \citep[][reported that during a 2006 \textit{Suzaku} observation, the $2-10$ keV flux of the source dropped by a factor of $\sim 30$]{Teng2015}. When modelling with \ctpo, we again utilised two \texttt{mekal} components, with kT $= 0.10 \pm 0.01$ keV and kT $= 0.69^{+0.09}_{-0.06}$ keV respectively. We estimated the LOS column density to be $\rm N_{H,LOS} = 5.22_{-0.46}^{+0.66} \times 10^{22} \ cm^{-2}$, and the intrinsic luminosity to be $\log (L_{2-10 \rm \ keV} / \rm erg \ s^{-1}) = 43.47 \pm 0.02$. The $\mathrm{N_{H, LOS}}$ agrees with the \citet{Teng2015} estimate, while the luminosity is slightly higher, but does agree with our previous estimate within the uncertainties.
    
\subsection*{IRASF15480-0344}

    This Compton-thick Seyfert 2 galaxy \citep[see][]{Boorman2025} was analysed using \textit{XMM-Newton} observations from 2010 and \textit{NuSTAR} observations from 2020. Despite the large temporal gap, we found no significant flux differences. When fitting with \myt, we required a \texttt{mekal} component with kT $= 0.81\pm 0.04$ keV, and scattering with $f_{\text{scatt}} = 0.41_{-0.17}^{+0.24} \%$. With this model we were only able to find a lower limit for the LOS obscuration of $\mathrm{N_{H, LOS}} \geq 2.46 \times 10^{24} \ \mathrm{cm}^{-2}$, which agrees within the uncertainties with the estimates of \citet{Boorman2025}, that range from $\rm 1.95\times 10^{24} \ cm^{-2} \leq N_{H, LOS} \leq 67.61 \times 10^{24} \ cm^{-2}$ depending on what model is used. With \myt \ we estimate an intrinsic luminosity of $\log (L_{2-10 \rm \ keV} / \rm erg \ s^{-1}) = 44.27^{+0.17}_{-0.24}$, which is much higher than the \citet{Brightman2011} estimate of $\log (L_{2-10 \rm \ keV} / \rm erg \ s^{-1}) \sim 43.6$, however they do not account for the Compton-thick nature of this source, as they only find $\mathrm{N_{H, LOS}} \sim 10^{22} \ \mathrm{cm}^{-2}$. As such, it may be that they underestimate the intrinsic luminosity. When fitting with \ctpo, we used an extra \texttt{mekal} component with kT $= 0.83 \pm 0.04$ keV to model the soft excess. We were able to constrain the obscuration, finding $\rm N_{H,LOS} = 2.08_{-0.34}^{+0.64} \times 10^{24} \ cm^{-2}$, which also agrees with the \citet{Boorman2025} estimates. We found an intrinsic luminosity of $\log (L_{2-10 \rm \ keV} / \rm erg \ s^{-1}) = 43.98^{+0.11}_{-0.09}$, which agrees with our previous estimate within the uncertainties.

\subsection*{MCG+00-29-023}

    MCG+00-29-023 is a faint Seyfert 2 galaxy, which we were only able to detect within \textit{XMM-Newton} observations from 2022. \textit{NuSTAR} observations were available, but there was no signal detected above the background. Due to the low counts $(\sim 150)$, we utilised C-statistics during fitting. The initial model consisted only of an unobscured power law, which yielded an intrinsic luminosity of $\log (L_{2-10 \rm \ keV} / \rm erg \ s^{-1}) = 40.78^{+0.36}_{-0.38}$, which is in agreement with the upper limit estimate of $\log (L_{2-10 \rm \ keV} / \rm erg \ s^{-1}) \leq 41.64$ from \citet{Rush1996}. Working with \ctpo, we had to freeze the majority of parameters, and only left the column density free to vary. We estimated an intrinsic luminosity of $\log (L_{2-10 \rm \ keV} / \rm erg \ s^{-1}) = 40.89 \pm 0.10$, which agrees with our previous estimate and with the \citet{Rush1996} upper limit. As \citet{Rush1996} did not estimate a column density for this source, we cannot compare our results for the obscuration.

\subsection*{MCG-02-33-034}

    MCG-02-33-034, also known as NGC4748, is a NLS1 galaxy \citep[see][]{Vasylenko2018} that we analysed using \textit{NuSTAR} data from 2021 and \textit{XMM-Newton} data from 2014. Despite both the temporal difference between the observations and the expected inherent variability of NLS1 sources, there was no significant flux difference between the two epochs. To model the spectrum phenomenologically, we used \texttt{mekal} with kT $= 0.78_{-0.07}^{+0.06}$ keV to model the soft excess, \texttt{pexrav} with relative reflection $= -1.06_{-0.65}^{+0.48}$ for the reflection, and one \texttt{zgauss} component for the Fe K$\alpha$ emission line at $6.44_{-0.07}^{+0.11}$ keV. For the obscuration, we used \texttt{zxipcf} with $\log \xi = 2.42^{+0.22}_{-0.13}$. This resulted in an estimated intrinsic luminosity of $\log (L_{2-10 \rm \ keV} / \rm erg \ s^{-1}) = 42.79_{-0.06}^{+0.05}$, which agrees with the average of the estimates from \citet{Chen2025}. We needed an additional \texttt{mekal} component (kT $= 0.15 \pm 0.03$ keV) when fitting with \ctpo, and found an inclination angle that indicates that the source should be unobscured, despite a significant equatorial column density $(\rm N_{H, eq} = 2.54_{-0.35}^{+0.29} \times 10^{24} \ \mathrm{cm}^{-2})$. This result agrees with \citet{Ricci2017}, who also found this source to be unobscured $(\mathrm{N_{H, LOS}} \sim 10^{20} \ \mathrm{cm}^{-2})$. We estimated the intrinsic luminosity to be $\log (L_{2-10 \rm \ keV} / \rm erg \ s^{-1}) = 42.99 \pm 0.05$, which is slightly higher than our previous estimate, but still agrees with \citet{Chen2025}.

\subsection*{MCG-03-34-064}

    MCG-03-34-064, also known as IRAS13197-1627, is a Seyfert 1.8 galaxy that is known to have a complex X-ray spectrum \citep[e.g.][]{Walton2018}, and possibly may contain a dual AGN \citep{TrindadeFalcao2024, Lamperti2026}. To study this galaxy, we used simultaneous \textit{NuSTAR} and \textit{XMM-Newton} observations from January 2016. We originally modelled MCG-03-34-064 phenomenologically, using scattering ($f_{\text{scatt}} = 1.11 \pm 0.08 \%$) and a \texttt{mekal} component (kT $= 0.78 \pm 0.01$ keV) to model the soft excess, a \texttt{zgauss} component at $6.39 \pm 0.01$ keV to model the Fe K$\alpha$ emission line, and \texttt{pexrav} with relative reflection $= -0.06 \pm 0.01$ for the reflection. We found an intrinsic luminosity of $\log (L_{2-10 \rm \ keV} / \rm erg \ s^{-1}) = 43.04 \pm 0.02$, which is lower than the estimate of $\log (L_{2-10 \rm \ keV} / \rm erg \ s^{-1}) \sim 43.18$ from \citet{Walton2018}. Overall, we were unsatisfied with this fit, as we had excessive residuals across the whole spectrum. Thus, we decided to try to fit the spectrum using the \myt \ model. When doing this, we also required two additional \texttt{mekal} components with kT $= 0.66 \pm 0.01$ keV and kT $= 2.38 \pm 0.11$ keV. The second component has quite a high temperature, but as it only affects the spectrum at $E \lesssim 2$ keV, we are not too concerned with it affecting the rest of the model parameters. We also used \texttt{relxill}, as the spectrum showed clear evidence of relativistic reflection. For this component, we found an inclination of $i=22.94^{+1.71}_{-2.15}$ degrees, $\log \xi = 1.70^{+0.15}_{-0.29}$, and A$_{\text{Fe}} = 3.64_{-0.25}^{+0.18}$. We estimated the intrinsic luminosity of this model to be $\log (L_{2-10 \rm \ keV} / \rm erg \ s^{-1}) = 42.59_{-0.25}^{+0.23}$, which is significantly lower than the \citet{Walton2018} estimate. However, it is not clear if their value includes the contributions from the reflection, which we expect to be significant. When fitting with \ctpo, we again required two \texttt{mekal} components with kT $= 0.69 \pm 0.01$ keV and kT $= 1.79_{-0.10}^{+0.09}$ keV respectively, for the soft excess. We again used \texttt{relxill} to model the relativistic reflection, keeping all parameters frozen at the default values except for the ionisation parameter $(\log\xi = 0.3_{-0.3}^{+1.1})$ and iron abundance $(\rm A_{Fe} =4.49_{-0.19}^{+0.28})$, and we fixed the reflection fraction $= -1$. We linked the photon index, inclination and normalisation to that of the main \texttt{C2POTorusD} component. The estimated LOS obscuration is $\rm N_{H,LOS} = 33.2_{-1.3}^{+2.6} \times 10^{22} \ cm^{-2}$, which is lower than the \citet{Walton2018} estimate of $\rm N_{H,LOS} \sim 5-10 \times 10^{23} \ cm^{-2}$, however they were using a partial absorber. For this model, we estimated an intrinsic luminosity $\log (L_{2-10 \rm \ keV} / \rm erg \ s^{-1}) = 42.74_{-0.02}^{+0.06}$, which agrees with our previous estimate within the uncertainties, but is still lower than the \citet{Walton2018} estimate that may include the reflection.

\subsection*{MCG-03-58-007}

    We analysed this Seyfert 1 galaxy using simultaneous \textit{NuSTAR} and \textit{XMM-Newton} observations from December 2015. To model the spectrum phenomenologically, we used both \texttt{zphabs} and \texttt{zxipcf} with $\log \xi = 2.82^{+0.11}_{-0.10}$ for the absorption, scattering ($f_{\text{scatt}} = 0.24_{-0.06}^{+0.07}\%$) and \texttt{mekal} (kT $= 0.20 \pm 0.01$ keV) for the soft excess, and two \texttt{zgauss} components for emission lines at $6.47 \pm 0.05$ keV and $0.92 \pm 0.01$ keV. These emission lines are likely the Fe K$\alpha$ and Fe L emission lines, respectively, but more detailed spectroscopy would be needed to confirm this. For this model we found an intrinsic luminosity of $\log (L_{2-10 \rm \ keV} / \rm erg \ s^{-1}) = 42.98_{-0.02}^{+0.03}$, which is only slightly lower than the estimate of $\log (L_{2-10 \rm \ keV} / \rm erg \ s^{-1}) \sim 43.04$ from \citet{Braito2018}. Using \ctpo \ we again required a \texttt{zgauss} component at $0.92 \pm 0.01$ keV for the Fe L emission line, and \texttt{mekal} with kT $= 0.20 \pm 0.01$ keV for the soft excess. We found a LOS column density of $\rm N_{H, LOS} = 12.13_{-0.33}^{+0.22} \times 10^{22} \ cm^{-2}$, which is slightly lower than the average of the estimates found by \citet{Boorman2025} of $\rm N_{H, LOS} = 24.7_{-10.0}^{+38.4} \times 10^{22} \ cm^{-2}$, but this may be due to the different model geometries. We estimated the intrinsic luminosity of this model to be $\log (L_{2-10 \rm \ keV} / \rm erg \ s^{-1}) = 43.00_{-0.02}^{+0.03}$, which agrees with both the phenomenological and \citet{Braito2018} estimates.
    
\subsection*{MCG-06-30-015}

    MCG-06-30-015, also known as ESO383-G035, is a Seyfert 1 galaxy that was simultaneously observed with \textit{NuSTAR} and \textit{XMM-Newton} in February 2024. Due to pile-up, we did not use the MOS1/2 observations during fitting. It is an incredibly variable source, with flux variations occurring on timescales similar to the observation lengths \citep[see][]{Marinucci2014}. Despite this, we attempted to fit combined spectrum with one model. The soft part of this source's spectrum is complex, so we fitted our models at $E\geq 2$ keV. The spectrum also features strong reflection, seemingly from both a neutral and ionised reflector, and possibly also relativistic reflection \citep[see][for more details]{Pal2024}. Overall, this was a challenging spectrum to fit, both phenomenologically and with \ctpo. Our first model used \texttt{pexrav} for the reflection, with $\Gamma=2.42^{+0.05}_{-0.06}$, relative reflection $= -1.18_{-0.28}^{+0.20}$, and $ \rm \cos Incl=0.80^{+0.15}_{-0.21}$. We needed the \texttt{pexrav} photon index to be unlinked from the main power law to best fit the reflection. Additionally, we used two \texttt{zgauss} components for the Fe K$\alpha$ emission line at $6.43 \pm 0.03$ keV and an absorption line at $7.07_{-0.05}^{+0.04}$ keV. We think that this absorption line is due to an outflow of highly ionised $\rm Fe\,XXVI$, as was identified by \citet{Miniutti2007}. To model the absorption, we used \texttt{zxipcf} with $\log \xi= -0.55^{+0.18}_{-0.06}$. This model resulted in an intrinsic luminosity estimate of $\log (L_{2-10 \rm \ keV} / \rm erg \ s^{-1}) = 42.95 \pm 0.02$, which is higher than the average of the estimates from \citet{Marinucci2014}. However they did not use the same observations, and given that this source is known to be variable, we can expect a difference in average flux between observations. To fit the spectrum with \ctpo, we used \texttt{relxill} to model the additional relativistic reflection. We kept all parameters frozen at the default values except for the inclination ($i=25.49^{+2.36}_{-2.91}$ degrees), ionisation parameter $(\log \xi = 2.21_{-0.08}^{+0.14})$ and iron abundance $(\rm A_{Fe} =4.57 \pm 0.36)$, and we fixed the reflection fraction $= -1$. We linked the photon index and normalisation to that of the main \texttt{C2POTorusD} component. We did not link the inclinations, as the accretion disk may not lie in the exact same plane as the torus. The estimated torus inclination and opening angle result in no LOS obscuration, which agrees with what was found by \citet{Pal2024}. We estimate the intrinsic luminosity to be $\log (L_{2-10 \rm \ keV} / \rm erg \ s^{-1}) = 42.93_{-0.02}^{+0.01}$, which is nearly identical to the phenomenological estimate.

\subsection*{MRK0509}

    We studied this Seyfert 1 galaxy using simultaneous \textit{NuSTAR} and \textit{XMM-Newton} observations from November 2024. The soft part of the spectrum was incredibly complex, so we fitted the spectrum at $E\geq 1$ keV. Initially, we fit the spectrum using a \texttt{mekal} component (kT $= 0.11_{-0.11}^{+0.12}$ keV) for the soft excess, one \texttt{zgauss} component at $6.40 \pm 0.02$ keV for the Fe K$\alpha$ emission line, and \texttt{pexrav} with relative reflection $= -1.81_{-1.04}^{+1.06}$ for the reflection. This model resulted in an intrinsic luminosity estimate of $\log (L_{2-10 \rm \ keV} / \rm erg \ s^{-1}) = 44.13 \pm 0.01$, which agrees with the estimate from \citet{Palit2024}. When fitting with \ctpo, we didn't require any additional components, and the estimated inclination and opening angle indicates that this source is unobscured. \citet{Palit2024} found an average value of $\rm N_{H,LOS} = 0.25_{-0.06}^{+0.16} \times 10^{22} \ cm^{-2}$, but used an ionised warm absorber (\texttt{Cloudy}) to model the absorption, making it difficult to fairly compare the two estimates as they are fundamentally different models. We found an intrinsic luminosity of $\log (L_{2-10 \rm \ keV} / \rm erg \ s^{-1}) = 44.16 \pm 0.01$, which also agrees with the \citet{Palit2024} estimate.

\subsection*{MRK0897}

    We analysed this faint Seyfert 2 galaxy using joint 2022 \textit{NuSTAR} and \textit{XMM-Newton} observations. Due to low counts ($\sim 700$), we utilised C-statistics during fitting. To fit the spectrum initially, we used both scattering ($f_{\text{scatt}} = 7.96_{-3.27}^{+4.39}\%$) and a \texttt{mekal} component (kT $= 0.53_{-0.12}^{+0.13}$ keV) for the soft excess. We estimated the intrinsic luminosity of this source to be $\log (L_{2-10 \rm \ keV} / \rm erg \ s^{-1}) = 41.68_{-0.15}^{+0.13}$, which is higher than the value of $\log (L_{2-10 \rm \ keV} / \rm erg \ s^{-1}) \sim 41.17$ from \citet{Rosen2016}, however that is the observed, not intrinsic luminosity. To fit the spectrum with \ctpo, we again used a \texttt{mekal} component with kT $= 0.61_{-0.11}^{+0.07}$ keV, and found an intrinsic luminosity of $\log (L_{2-10 \rm \ keV} / \rm erg \ s^{-1}) = 41.69_{-0.15}^{+0.12}$, which agrees with our previous estimate. As we only have the observed flux/luminosity from \citet{Rosen2016}, we cannot compare our obscuration estimates.

\subsection*{MRK1239}

    MRK1239 is a NLS1 galaxy, with an incredibly complex spectrum \citep[described as ``a hot mess'' by][]{Buhariwalla2024}, that we analysed using simultaneous \textit{NuSTAR} and \textit{XMM-Newton} observations from November 2021. There was a significant flaring event during the second half of the \textit{NuSTAR} observation that was not recorded by \textit{XMM-Newton}, so to avoid contamination, we used only the first 100ks of the \textit{NuSTAR} observation \citep[see][for more details]{Buhariwalla2024}. We modelled this spectrum using a complicated phenomenological model, following \citet{Buhariwalla2020, Buhariwalla2024}. We used \texttt{mekal} for the soft excess (kT $= 0.71 \pm 0.02$ keV), and a \texttt{zgauss} component for an absorption line at $7.00 \pm 0.04$ keV, which we think is due to a high-velocity outflow of ionised $\rm Fe\, XXVI$ \citep[evidence for outflows was found in the soft part of the spectrum by ][]{Buhariwalla2023}. Following \citet{Buhariwalla2020, Buhariwalla2024}, we used both \texttt{relxill} and \texttt{xillver} for the reflection. We linked the inclinations between the two reflection components, finding $i = 25.91 ^{+1.21 }_{-1.03}$. We froze the iron abundance of the \texttt{xillver} model at the solar value while allowing the \texttt{relxill} value to vary, finding A$_{\text{Fe}} = 3.66^{+0.19}_{-0.24}$. We allowed the ionisation parameters of both reflection models to vary, finding $\log \xi = 3.48^{+0.13}_{-0.08}$ for \texttt{xillver} and $\log \xi = 2.00^{+0.11}_{-0.09}$ for \texttt{relxill}. We froze the \texttt{relxill} reflection fraction at $-1$, and found a reflection fraction $= -1.26_{-0.09}^{+0.14} \times 10^{-2}$ for \texttt{xillver}. All other parameters were frozen at the default values, with the photon index and normalisation linked to those of the primary power law. The LOS column density found, $\rm N_{H,LOS} = 15.84_{-0.47}^{+0.63} \times 10^{22} \ cm^{-2}$, which is close to the value of the \citet{Buhariwalla2020} ``blurred reflection'' model of $\rm N_{H,LOS} = 22 \pm 5 \times 10^{22} \ cm^{-2}$ for the \textit{Swift}/\textit{NuSTAR} epoch. We estimated an intrinsic luminosity of $\log (L_{2-10 \rm \ keV} / \rm erg \ s^{-1}) = 41.01_{-0.02}^{+0.04}$, which is much lower than the average estimate of $\log (L_{2-10 \rm \ keV} / \rm erg \ s^{-1}) = 42.90_{-0.05}^{+0.04}$ from \citet{Jiang2021}, however their estimate may include the reflection. As this source is completely reflection-dominated, it is very hard to properly constrain the underlying intrinsic emission. To fit the spectrum with \ctpo, we required additional \texttt{mekal} and \texttt{zbbody} components for the soft excess, with kT $= 0.71 \pm 0.02$ keV and kT $= 0.08 \pm 0.01$ keV, respectively. We again added a \texttt{zgauss} component for the absorption line at $7.01_{-0.04}^{+0.03}$ keV. We again used \texttt{relxill} to model the relativistic reflection that cannot be accounted for by \ctpo. This component was multiplied by \texttt{zphabs} and \texttt{cabs} to account for the absorption due to the torus, with the column density of both of these components linked to the \ctpo \ $\rm N_{H, eq}$ via Equation \ref{eq:nh_conversion}. We kept all \texttt{relxill} parameters frozen at the default values except for the inclination ($i=25.96_{-1.35}^{+1.05}$ degrees), ionisation parameter $(\log\xi = 0)$, and iron abundance $(\rm A_{Fe} =4.63_{-0.17}^{+0.19})$. We fixed the reflection fraction $= -1$, and tied the photon index and normalisation to that of the main \texttt{C2POTorusD} component. We did not link the \texttt{relxill} inclination to that of \ctpo \ as the torus and accretion disk may not lie in the same plane, and thus used the \ctpo \ inclination to tie \texttt{zphabs} and \texttt{cabs} to $\rm N_{H, eq}$. The estimated LOS obscuration is $\rm N_{H,LOS} = 18.62_{-0.39}^{+0.74} \times 10^{22} \ cm^{-2}$, which agrees with the \citet{Buhariwalla2020} estimate. The intrinsic luminosity we estimated with this model was $\log (L_{2-10 \rm \ keV} / \rm erg \ s^{-1}) = 42.42 \pm 0.02$, which is still lower than the \citet{Jiang2021} estimates, a result that could be explained assuming that their ``absorption-corrected'' estimates include contributions from the reflection and soft excess.

\subsection*{NGC0034}

    We analysed the Seyfert 2 galaxy NGC0034, also known as NGC0017, using a \textit{NuSTAR} observation from 2015, \textit{Chandra} observation from 2013, and \textit{XMM-Newton} observation from 2002. Despite the temporal gaps between the observations, there was no significant flux difference. We first fit the spectrum using both scattering ($f_{\text{scatt}} = 7.89_{-2.35}^{+1.23}\%$) and a \texttt{mekal} component (kT $= 0.67_{-0.13}^{+0.12}$ keV) for the soft excess. We found an intrinsic luminosity of $\log (L_{2-10 \rm \ keV} / \rm erg \ s^{-1}) = 41.85_{-0.13}^{+0.12}$, which agrees with the estimate from \citet{Yamada2023} of $\log (L_{2-10 \rm \ keV} / \rm erg \ s^{-1}) = 41.90 \pm 0.10$. To fit the spectrum with \ctpo, we used a \texttt{mekal} component with kT $= 0.67_{-0.10}^{+0.12}$ keV, and estimated the LOS obscuration to be $\rm N_{H,LOS} = 48_{-15}^{+25} \times 10^{22} \ cm^{-2}$, which agrees with the \citet{Yamada2023} estimate of $\rm N_{H,LOS} = 50 \pm 10 \times 10^{22} \ cm^{-2}$. We found an intrinsic luminosity estimate of $\log (L_{2-10 \rm \ keV} / \rm erg \ s^{-1}) = 42.00_{-0.15}^{+0.21}$, which agrees with the other estimates within the uncertainties.

\subsection*{NGC0424}

    NGC0424 is a known Compton-thick AGN with a reflection-dominated X-ray spectrum \citep[see][]{Tanimoto2022, Akylas2024} that was observed by \textit{NuSTAR} in 2020 and \textit{Chandra} in 2019. There is a significant difference in flux between the two different observations, so the average luminosity is reported in Tables \ref{tab:physical_results} and \ref{tab:carrot_results}. To fit the spectrum with \myt, we required an additional \texttt{mekal} component with kT $= 0.66\pm 0.05$ keV, and scattering with $f_{\text{scatt}} = 1.51_{-0.63}^{+1.23} \%$. We estimated the column densities to be $\rm N_{H,LOS} = 2.04^{+0.34}_{-0.29} \times 10^{24} \ cm^{-2}$ and $\rm N_{H,eq} = 0.21_{-0.05}^{+0.08} \times 10^{24} \ cm^{-2}$, which are lower than the estimates from \citet{Zhao2021} of $\rm N_{H,LOS} = 3.24 \pm 0.15 \times 10^{24} \ cm^{-2}$ and $\rm N_{H,tor} = 0.65_{-0.10}^{+0.08} \times 10^{24} \ cm^{-2}$, found using the \texttt{Borus} model. These slight differences are likely due to the different model geometries, as \texttt{Borus} assumes a smooth spherical distribution with conical cutouts at the poles. The \myt \ intrinsic luminosity estimate was $\log (L_{2-10 \rm \ keV} / \rm erg \ s^{-1}) = 43.71_{-0.29}^{+0.28}$ for the \textit{NuSTAR} epoch, and $\log (L_{2-10 \rm \ keV} / \rm erg \ s^{-1}) = 43.56_{-0.29}^{+0.27}$ for the \textit{Chandra} epoch, the average of which agrees with the estimate from \citet{Zhao2021} of $\log (L_{2-10 \rm \ keV} / \rm erg \ s^{-1}) = 43.54 \pm 0.06$. With \ctpo, we used a \texttt{mekal} component for the soft excess with kT $= 0.71_{-0.09}^{+0.11}$ keV, and found $\rm N_{H,eq} = 3.77_{-0.51}^{+1.19} \times 10^{24} \ cm^{-2}$, which agrees with the \citet{Zhao2021} estimate within uncertainties. We found an intrinsic luminosity of $\log (L_{2-10 \rm \ keV} / \rm erg \ s^{-1}) = 43.39_{-0.33}^{+0.30}$ for the \textit{NuSTAR} epoch, and $\log (L_{2-10 \rm \ keV} / \rm erg \ s^{-1}) = 43.28_{-0.35}^{+0.30}$ for the \textit{Chandra} epoch, the average of which agrees with both the previous estimate and the one from \citet{Zhao2021}.

\subsection*{NGC0526A}

    NGC0526A is a known Changing-Look AGN \citep[see][]{Jana2025}, that has most recently been classified as a Seyfert 1.9 based on its optical spectrum \citep{Temple2023}. To analyse this source, we used a 2003 \textit{XMM-Newton} observation. We did not use the MOS1 observation, as it suffered from pile-up. To model the spectrum phenomenologically, we used a \texttt{mekal} component with kT $= 0.25 \pm 0.01$ keV for the soft excess, and two \texttt{zgauss} components for the Fe K$\alpha$ and K$\beta$ emission lines at $6.39 \pm 0.03$ keV and $6.88 \pm 0.07$ keV. With this model we estimated the LOS column density to be $\rm N_{H,LOS} = 1.05 \pm 0.02 \times 10^{22} \ cm^{-2}$, which agrees with the \citet{Jana2025} estimate of $\rm N_{H,LOS} \sim 1-4 \times 10^{22} \ cm^{-2}$. We found an intrinsic luminosity of $\log (L_{2-10 \rm \ keV} / \rm erg \ s^{-1}) = 43.29 \pm 0.01$, which agrees with the value from \citet{Akylas2024}. We didn't require any additional components to fit this spectrum with \ctpo, and found a LOS column density of $\rm N_{H,LOS} = 5.24_{-0.13}^{+0.18} \times 10^{22} \ cm^{-2}$, which is slightly higher than the \citet{Jana2025} estimate, but given that our LOS values are only indicative estimates, this discrepancy is acceptable. We estimated an intrinsic luminosity of $\log (L_{2-10 \rm \ keV} / \rm erg \ s^{-1}) = 43.30 \pm 0.01$, which agrees with our previous and the \citet{Akylas2024} estimates.

\subsection*{NGC1125}

    NGC1125 is a known Compton-thick AGN \citep[see][]{Tanimoto2022} that we analysed using a 2019 \textit{NuSTAR} observation and a 2018 \textit{Chandra} observation. There were no significant flux differences between the observations. The \myt \ fitting required both \texttt{mekal} component (kT $= 0.68_{-0.17}^{+0.13}$ keV) and scattering ($f_{\text{scatt}} = 2.0_{-1.6}^{+4.7} \%$) for the soft excess. We found $\rm N_{H,LOS} = 0.93^{+0.49}_{-0.25} \times 10^{24} \ cm^{-2}$, which agrees with the \citet{Tanimoto2022} estimate of $\rm N_{H,LOS} = 1.16^{+0.92}_{-0.50} \times 10^{24} \ cm^{-2}$ found using \texttt{XClumpy}. The intrinsic luminosity estimated with this model was $\log (L_{2-10 \rm \ keV} / \rm erg \ s^{-1}) = 42.8_{-1.4}^{+0.6}$, which agrees within the uncertainties with the value from \citet{OsorioClavijo2022} of $\log (L_{2-10 \rm \ keV} / \rm erg \ s^{-1}) = 42.53 \pm 0.26$. The large uncertainties in our estimate are likely due to the Compton-thick nature of this source, which makes studying both the obscuration and underlying emission difficult. To fit the spectrum with \ctpo, we did not require any additional components. We estimated the column density to be $\rm N_{H,LOS} = 1.49^{+0.16}_{-0.12} \times 10^{24} \ cm^{-2}$ and the intrinsic luminosity to be $\log (L_{2-10 \rm \ keV} / \rm erg \ s^{-1}) = 42.75_{-0.12}^{+0.23}$, both of which agree with the other estimates.

\subsection*{NGC1194}

    We analysed this Compton-thick AGN \citep{Marchesi2018} using a \textit{NuSTAR} observation from 2020 and \textit{Chandra} observations from 2019/2020. There were no significant flux differences between the observations. To fit this reflection-dominated spectrum with \myt, we used an additional scattering component with $f_{\text{scatt}} = 1.11_{-0.35}^{+1.25} \%$. With this model we found $\rm N_{H,LOS} = 0.80^{+0.11}_{-0.16} \times 10^{24} \ cm^{-2}$ and $\rm N_{H,eq} \geq 4.76 \times 10^{24} \ cm^{-2}$ which agree with the \citet{Zhao2021} estimates of $\rm N_{H,LOS} = 0.98^{+0.12}_{-0.15} \times 10^{24} \ cm^{-2}$ and $\rm N_{H,eq} \geq 2.4 \times 10^{24} \ cm^{-2}$ within the uncertainties. We found an intrinsic luminosity of $\log (L_{2-10 \rm \ keV} / \rm erg \ s^{-1}) = 42.60_{-0.22}^{+0.07}$, which again agrees with the \citet{Zhao2021} estimate of $\log (L_{2-10 \rm \ keV} / \rm erg \ s^{-1}) = 42.71_{-0.09}^{+0.07}$. When fitting the spectrum with \ctpo, we didn't require any additional components. We estimated $\rm N_{H, eq} = 4.32^{+0.58}_{-1.11} \times 10^{24} \ cm^{-2}$, and found an intrinsic luminosity of $\log (L_{2-10 \rm \ keV} / \rm erg \ s^{-1}) = 42.63_{-0.26}^{+0.19}$, both of which agree with the other estimates.

\subsection*{NGC1320}

    This Compton-thick Seyfert 2 galaxy was observed with \textit{NuSTAR} in 2013 and \textit{Chandra} in 2020. There was no significant flux difference between the two observations, despite the temporal difference. This spectrum features an incredibly strong emission line at $\sim 6.4$ keV and prominent soft excess, making it difficult to properly fit the underlying power law. We initially fit the spectrum with \myt, needing an extra \texttt{mekal} component (kT $= 0.98_{-0.22}^{+0.17}$ keV) for the soft excess. We were not able to constrain the obscuration, finding $\rm N_{H,LOS} \geq 3.09 \times 10^{24} \ cm^{-2}$ and $\rm N_{H,eq} \geq 4.59 \times 10^{24} \ cm^{-2}$. We found many different values for the obscuration in literature, namely $\mathrm{N_{H, LOS}} = 5.75^{+1.16}_{-1.86} \times 10^{24} \ \mathrm{cm}^{-2}$ from \citet{OsorioClavijo2022} (using a phenomenological model with \texttt{zphabs} for the obscuration), $\rm N_{H, eq} = 4^{+4}_{-2} \times 10^{24} \ cm^{-2}$ from \citet{Balokovic2014} (found using a ``reflection only'' version of \myt), $\rm N_{H, eq} \geq 29.19 \times 10^{24} \ cm^{-2}$ from \citet{Brightman2015} (found using the \texttt{Torus} model), and $2.09 \times 10^{24} \ \mathrm{cm}^{-2} \leq \mathrm{N_{H, LOS}} \leq 67.61 \times 10^{24} \ \mathrm{cm}^{-2}$ across the different models used by \citet{Boorman2025}. It is clear that this highly obscured and reflection-dominated source is incredibly hard to properly fit with the currently available data, even with physical models. We could only get a lower limit for the intrinsic luminosity of $\log (L_{2-10 \rm \ keV} / \rm erg \ s^{-1}) \geq 42.56$ due to the unconstrained obscurations. This agrees with the estimate of $\log (L_{2-10 \rm \ keV} / \rm erg \ s^{-1}) = 42.79_{-0.09}^{+0.12}$ from \citet{Brightman2015}. When fitting with \ctpo, we had to fix the opening angle and inclination at $\Theta = 60^{\circ}, \ i = 45^{\circ} $ respectively, and we required an additional \texttt{mekal} component with kT $= 1.36_{-0.34}^{+0.71}$ keV. Overall, the fit was completely reflection-dominated, with an incredibly steep photon index $(\Gamma \sim 2.6)$. We estimated the column density to be $\rm N_{H,eq} =2.39^{+1.86}_{-0.58} \times 10^{24} \ cm^{-2}$, which agrees with literature. With this model we estimated an intrinsic luminosity of $\log (L_{2-10 \rm \ keV} / \rm erg \ s^{-1}) = 43.74_{-0.07}^{+0.19}$, which is much higher than the \citet{Brightman2015} estimate (although theirs might also be a lower limit as they could not constrain the obscuration), but agrees with our lower limit.

\subsection*{NGC1365}

    NGC1365 is a complicated source, being a known changing-look AGN that oscillates between being Compton-thin and Compton-thick \citep[see][]{Walton2014, Liu2021}. It was observed simultaneously with \textit{NuSTAR} and \textit{XMM-Newton} four times between 2012 and 2013. During these observations, NGC1365 experienced variable levels of obscuration (but remained in the Compton-thin regime), and also showed evidence of relativistic reflection \citep[see][for further details]{Walton2014}. It is an incredibly complex spectrum, needing many different phenomenological components to be successfully fit. For this analysis, we are using the first pair of observations from July 2012. For the soft excess, we used \texttt{mekal} with kT $= 0.69 \pm 0.01$ keV. We used three \texttt{zgauss} components for the Fe K$\alpha$ emission line at $6.42 \pm 0.01$ keV and two absorption lines at $6.74 \pm 0.01$ keV and $7.01^{+0.02}_{-0.01}$ keV respectively. We think that the absorption lines are due to a high-velocity outflow, as is discussed in \citet{Rivers2015}. \citet{Walton2014} found two additional absorption lines at $\rm E \sim 8 \ keV$; however, we found that adding these components did not make a significant difference to the overall quality of the fit. To model the relativistic reflection, we used \texttt{relxill}, keeping all parameters frozen at the default values except for the inclination $(i=40.29_{-7.56}^{+5.84} \ \rm degrees)$, ionisation parameter $(\log \xi = 3.99_{-0.05}^{+0.04})$, iron abundance $(\rm A_{Fe} \geq 8.20)$, and we fixed the reflection fraction $= -1$. We tied the photon index and normalisation to that of the main power law. To model the obscuration, we required both \texttt{zxipcf} ($\log \xi = 2.41 \pm 0.05$) and \texttt{zpcfabs}. It is hard to compare LOS obscuration as both of these models are partial absorbers, so we assume the equivalent neutral LOS column density of this model is $\mathrm{N_{H, LOS}} \sim 2-3 \times 10^{23} \ \mathrm{cm}^{-2}$. This agrees with the average of values found by \citet{Walton2014} for the first pair of observations, which is $\rm N_{H,LOS} = 23.98_{-2.78}^{+3.33} \times 10^{22} \ cm^{-2}$. We estimated an intrinsic luminosity of $\log (L_{2-10 \rm \ keV} / \rm erg \ s^{-1}) = 41.59_{-0.17}^{+0.09}$, which is lower than the average of the estimates from \citet{Rivers2015} of $\log (L_{2-10 \rm \ keV} / \rm erg \ s^{-1}) = 42.16_{-0.13}^{+0.10}$. This may be explained by the difference in photon indices, which may in turn be due to the different components used to model the reflection (\texttt{RELCONV} $\times$ \texttt{xillver} $+$ \texttt{xillver} vs our singular \texttt{relxill}). To fit the source with \ctpo, we again used \texttt{mekal} with kT $= 0.66 \pm 0.01$ keV for the soft excess, and two \texttt{zgauss} components for absorption lines at $6.72 \pm 0.01$ keV and $7.02 \pm 0.01$ keV. We used an additional \texttt{relxill} component to model the relativistic reflection that cannot be accounted for by \ctpo. It was multiplied by \texttt{zphabs} and \texttt{cabs} components to account for the LOS absorption, with the column density of both components linked to the \ctpo \ $\rm N_{H, eq}$ via Equation \ref{eq:nh_conversion}. Once again, we kept all \texttt{relxill} parameters frozen at the default values except for the inclination ($i=45.01_{-0.99}^{+1.28}$ degrees), ionisation parameter $(\log\xi = 2.49_{-0.07}^{+0.06})$, iron abundance $(\rm A_{Fe} =4.03_{-0.18}^{+0.16})$, and we fixed the reflection fraction $= -1$. We also tied the photon index and normalisation to that of the main \texttt{C2POTorusD} component. We did not link the inclination to that of \ctpo \ as the torus and accretion disk may not lie in the same plane, and thus used the \ctpo \ inclination to tie \texttt{zphabs} and \texttt{cabs} $\rm N_{H, eq}$. The estimated LOS obscuration is $\rm N_{H,LOS} = 19.09_{-0.23}^{+0.21} \times 10^{22} \ cm^{-2}$, which is slightly lower than the \citet{Walton2014} estimate. The intrinsic luminosity we estimated with this model was $\log (L_{2-10 \rm \ keV} / \rm erg \ s^{-1}) = 42.17 \pm 0.01$, which agrees with the average of the \citet{Rivers2015} estimates.

\subsection*{NGC1566}

    NGC1566 is a known changing-look AGN that underwent an outburst in June 2018 and experienced post-outburst flares afterwards \citep{Jana2021}. We used \textit{NuSTAR} and \textit{XMM-Newton} observations from 2019 that were taken 10 days apart, but do not have a significant difference in flux. We initially modelled the spectrum using an unobscured power law, one \texttt{zgauss} component for the Fe K$\alpha$ emission line at $6.39 \pm 0.04$ keV, and \texttt{pexrav} with relative reflection $= -0.31_{-0.12}^{+0.10}$ for the reflection. This resulted in an intrinsic luminosity of $\log (L_{2-10 \rm \ keV} / \rm erg \ s^{-1}) = 41.84 \pm 0.01$, which agrees with the average of the estimates from \citet{Jana2021} of $\log (L_{2-10 \rm \ keV} / \rm erg \ s^{-1}) = 42.58_{-1.07}^{+0.55}$. This average value has such large error bars as it includes estimates from before, during and after the outburst. When fitting with \ctpo \ we did not require any additional parameters, and were only able to find an upper limit for the inclination. However, the combination of the opening angle and inclination upper limit indicates to us that this source is unobscured, a conclusion also reached by \citet{Ricci2017}. We estimated the intrinsic luminosity to be $\log (L_{2-10 \rm \ keV} / \rm erg \ s^{-1}) = 41.82_{-0.01}^{+0.13}$, which agrees with the other two estimates within the uncertainties.

\subsection*{NGC2992}

    NGC2992 is a highly-variable Seyfert 2 AGN \citep{Middei2022} that we analysed using nearly simultaneous \textit{NuSTAR} and \textit{XMM-Newton} observations from 2019. We required both scattering ($f_{\text{scatt}} = 0.94 \pm 0.07 \%$) and a \texttt{mekal} component (kT $= 0.61 \pm 0.02$ keV) to model the soft excess, as well as three \texttt{zgauss} components for the Fe K emission lines at $6.38 \pm 0.01$ keV, $6.68 \pm 0.05$ keV and $6.95_{-0.02}^{+0.03}$ keV. We estimated the intrinsic luminosity to be $\log (L_{2-10 \rm \ keV} / \rm erg \ s^{-1}) = 43.09 \pm 0.01$, which agrees with the estimate of $\log (L_{2-10 \rm \ keV} / \rm erg \ s^{-1}) = 43.03_{-0.09}^{+0.11}$ from \citet{Middei2022} for the same observations. We did not require any additional components when fitting with \ctpo, and estimated the LOS column density to be $\mathrm{N_{H, LOS}} = 0.62 \pm 0.02 \times 10^{22} \ \mathrm{cm}^{-2}$, which is slightly lower than the estimate from \citet{Middei2022} of $\mathrm{N_{H, LOS}} = 0.78 \pm 0.01 \times 10^{22} \ \mathrm{cm}^{-2}$, but as our value is only an indicative estimate, we are satisfied with this result. We found an intrinsic luminosity of $\log (L_{2-10 \rm \ keV} / \rm erg \ s^{-1}) = 43.07 \pm 0.01$, which agrees with the other estimates within the uncertainties.
    
\subsection*{NGC4593}

    NGC4593 is a Seyfert 1 galaxy that was observed simultaneously with \textit{NuSTAR} and \textit{XMM-Newton} in 2014, however we used a longer 2016 \textit{XMM-Newton} observation for this analysis as there was no significant flux variation between the two observations. The spectrum was complex requiring many different phenomenological components to be fit successfully. To model the soft excess, we used \texttt{mekal} with kT $= 0.11 \pm 0.01$ keV. We used two \texttt{zgauss} components to model the Fe K$\alpha$ and K$\beta$ emission lines at $6.41 \pm 0.01$ keV and $7.03_{-0.05}^{+0.04}$ keV, and used \texttt{pexrav} with relative reflection $= -1.24_{-0.13}^{+0.11}$ to model the reflection. We used \texttt{zxipcf} to model the obscuration, finding $\log \xi = 2.14^{+0.01}_{-0.03}$. We estimated the intrinsic luminosity to be $\log (L_{2-10 \rm \ keV} / \rm erg \ s^{-1}) = 42.58 \pm 0.01$, which is slightly higher than the estimate of $\log (L_{2-10 \rm \ keV} / \rm erg \ s^{-1}) = 42.50 \pm 0.01$ from \citet{VictoriaCeballos2023}, but they utilised a different absorber and found a slightly shallower slope ($\sim 1.8$ vs our $\sim 1.9$), which could explain the luminosity difference. To fit the spectrum with \ctpo, we used an extra \texttt{zbbody} with kT $= 0.09 \pm 0.01$ keV for the soft excess, and the estimated opening angle and inclination result in an unobscured LOS for this source. \citet{VictoriaCeballos2023} estimated the LOS obscuration to be $\mathrm{N_{H, LOS}} = 0.39_{-0.30}^{+0.33} \times 10^{22} \ \mathrm{cm}^{-2}$, which disagrees with our result, but it is possible that the LOS intersects with a cloud `above' the torus (at an inclination that doesn't intersect with the torus according to the opening angle), that we cannot account for. We estimated the intrinsic luminosity to be $\log (L_{2-10 \rm \ keV} / \rm erg \ s^{-1}) = 42.56 \pm 0.01$, which agrees with our previous estimate within the uncertainties.

\subsection*{NGC4602}

    This faint Seyfert 2 galaxy was simultaneously observed by \textit{NuSTAR} and \textit{XMM-Newton} in 2021, however the \textit{NuSTAR} observation did not detect any signal above the background noise for this source. Due to the low counts $(\sim 400)$, we utilised C-statistics during fitting. We found the spectrum to be incredibly reflection-dominated, requiring \texttt{pexrav} with relative reflection $= -10.75_{-26.89}^{+7.95}$ in addition to the underlying unobscured power law emission. To model the soft excess, we used \texttt{mekal} with kT $= 0.70_{-0.23}^{+0.19}$ keV. This model produced an intrinsic luminosity of $\log (L_{2-10 \rm \ keV} / \rm erg \ s^{-1}) = 39.70_{-0.11}^{+0.19}$, which agrees with the upper limit derived by \citet{Rush1996} of $\log (L_{2-10 \rm \ keV} / \rm erg \ s^{-1}) \leq 40.99$. We required no additional components when fitting with \ctpo, and were only able to determine an upper limit for the obscuration. We estimated the intrinsic luminosity to be $\log (L_{2-10 \rm \ keV} / \rm erg \ s^{-1}) = 39.84\pm 0.09$, which agrees with \citet{Rush1996} and the previous estimate. As \citet{Rush1996} did not calculate the obscuration, we cannot compare our $\rm N_H$ results.

\subsection*{NGC5135}

    NGC5135 is a known Compton-thick Seyfert 2 \citep[see][]{Johnstone2025}, which we analysed using 2015 \textit{NuSTAR} and 2001 \textit{Chandra} observations. Despite the large temporal difference, there was no significant flux variation between the two observations. Initially, we modelled this source using \myt, including additional scattering ($f_{\text{scatt}} = 1.01_{-0.29}^{+0.64} \%$) and \texttt{mekal} (kT $= 0.75_{-0.04}^{+0.03}$ keV) for the soft excess. We estimated the obscuration to be $\rm N_{H,LOS} = 2.51_{-0.55}^{+1.49} \times 10^{24} \ cm^{-2}$ and $\rm N_{H,eq} = 5.00_{-2.02}^{+4.04} \times 10^{24} \ cm^{-2}$, which are very close to the values estimated by \citet{Yamada2020} using \texttt{XCLUMPY} of $\rm N_{H,LOS} = 6.7_{-2.8}^{+16.6} \times 10^{24} \ cm^{-2}$ and $\rm N_{H,eq} = 9.5_{-2.9}^{+24.0} \times 10^{24} \ cm^{-2}$. We estimated the intrinsic luminosity to be $\log (L_{2-10 \rm \ keV} / \rm erg \ s^{-1}) = 42.85_{-0.27}^{+0.17}$, which is lower than the estimate by \citet{Yamada2020} of $\log (L_{2-10 \rm \ keV} / \rm erg \ s^{-1}) = 43.30_{-0.26}^{+0.42}$. This may be due to the differences in model geometry. To fit the spectrum with \ctpo, we required a \texttt{mekal} component with kT $= 0.75 \pm 0.03$ keV for the soft excess, and were only able to find a lower limit for the average obscuration of $\rm N_{H,eq} \geq 9.78 \times 10^{24} \ cm^{-2}$, which agrees with the \citet{Yamada2020} estimate. We found an intrinsic luminosity of $\log (L_{2-10 \rm \ keV} / \rm erg \ s^{-1}) = 43.72_{-0.16}^{+0.11}$, which is much higher than the previous estimate, but agrees with \citet{Yamada2020} within the uncertainties.

\subsection*{NGC5506}

    NGC5506 is a bright Seyfert 1 galaxy that was observed by \textit{NuSTAR} in 2014 and \textit{XMM-Newton} in 2015. There was no significant flux variation between the two observations, despite it being a known variable source \citep[see][]{Guainazzi2010}. Due to large amounts of pile-up, we discarded the MOS1/2 observations. To model the spectrum phenomenologically, we used scattering ($f_{\text{scatt}} = 0.59 \pm 0.03 \%$), \texttt{mekal} (kT $= 0.78 \pm 0.02$ keV) for the soft excess, two \texttt{zgauss} components for the Fe K$\alpha$ and K$\beta$ emission lines at $6.39 \pm 0.01$ keV and $6.96_{-0.02}^{+0.03}$ keV, and \texttt{pexrav} (relative reflection $= -0.76_{-0.11}^{+0.10}$) for the reflection. We estimated the LOS obscuration to be $\mathrm{N_{H, LOS}} = 3.15 \pm 0.03 \times 10^{22} \ \mathrm{cm}^{-2}$, which agrees with the \citet{Matt2015} estimate of $\mathrm{N_{H, LOS}} = 3.10_{-0.20}^{+0.21} \times 10^{22} \ \mathrm{cm}^{-2}$. This model produced an intrinsic luminosity of $\log (L_{2-10 \rm \ keV} / \rm erg \ s^{-1}) = 42.71 \pm 0.01$, which agrees with the estimate of $\log (L_{2-10 \rm \ keV} / \rm erg \ s^{-1}) \sim 42.72$ from \citet{Matt2015}. When fitting with \ctpo, we used \texttt{zbbody} (kT $= 0.33_{-0.01}^{+0.02}$ keV) to model the soft excess, as we were unable to do so with \texttt{mekal}. Using Equation \ref{eq:nh_conversion} we estimated the LOS obscuration to be $\mathrm{N_{H, LOS}} = 48.08 \pm 0.68 \times 10^{22} \ \mathrm{cm}^{-2}$, which is much higher than the previous and \citet{Matt2015} estimates. However, we again stress that this is an indicative estimate based on the assumption of a smooth torus, and that the LOS may be intersecting fewer clouds than anticipated, which would result in a true LOS obscuration that is closer to the \citet{Matt2015} estimate. We found an intrinsic luminosity of $\log (L_{2-10 \rm \ keV} / \rm erg \ s^{-1}) = 42.88 \pm 0.03$, which is higher than the previous estimates, but agrees with the \citet{Guainazzi2010} estimate of $\log (L_{2-10 \rm \ keV} / \rm erg \ s^{-1}) \simeq 42.93_{-0.15}^{+0.11}$.

\subsection*{NGC5995}

    We analysed this bright Seyfert 2 with \textit{NuSTAR} and \textit{Chandra} observations from 2014 and 2015 respectively. There was a significant flux difference between the two observations ($\sim 30 \%$), so the average intrinsic luminosity has been reported in Tables \ref{tab:phenomenological_results} and \ref{tab:carrot_results}. To phenomenologically model this spectrum, we required one \texttt{zgauss} component for the Fe K$\alpha$ emission line at $6.40 \pm 0.07$ keV, and \texttt{pexrav} with relative reflection $= -0.64_{-0.53}^{+0.39}$ for the reflection. We estimated the LOS column density to be $\mathrm{N_{H, LOS}} = 1.06_{-0.10}^{+0.11} \times 10^{22} \ \mathrm{cm}^{-2}$, which agrees with the \citet{Ricci2017} estimate of $\mathrm{N_{H, LOS}} = 0.93 \pm 0.14 \times 10^{22} \ \mathrm{cm}^{-2}$. We found an intrinsic luminosity of $\log (L_{2-10 \rm \ keV} / \rm erg \ s^{-1}) = 43.38 \pm 0.01$ for the 2014 epoch, and $\log (L_{2-10 \rm \ keV} / \rm erg \ s^{-1}) = 43.49 \pm 0.02$ for the 2015 epoch. The average is slightly higher than the estimate from \citet{Ricci2017} of $\log (L_{2-10 \rm \ keV} / \rm erg \ s^{-1})\sim 43.33$. When fitting with \ctpo, we required no additional components, and found a LOS obscuration estimate of $\mathrm{N_{H, LOS}} = 1.31_{-0.21}^{+0.24} \times 10^{22} \ \mathrm{cm}^{-2}$, which is slightly higher than the \citet{Ricci2017} value, but this is only an indicative estimate. We estimated the intrinsic luminosity to be $\log (L_{2-10 \rm \ keV} / \rm erg \ s^{-1}) = 43.32_{-0.03}^{+0.06}$ for the 2014 epoch, and $\log (L_{2-10 \rm \ keV} / \rm erg \ s^{-1}) = 43.46_{-0.04}^{+0.01}$ for the 2015 epoch. The average agrees with the other estimates within the uncertainties.

\subsection*{NGC6810}

    NGC6810 is a faint Seyfert 2 with only one 2004 \textit{XMM-Newton} observation. Due to relatively high background flux, we were only able to fit the spectrum between $0.5- 4.9$ keV. As such, we struggled to constrain parameters with both the phenomenological and \ctpo \ models, especially since this range is nearly entirely dominated by soft excess. The first model consists of an unobscured power law, and multiple components for the soft excess. Specifically, one \texttt{mekal} component (kT $= 0.58_{-0.03}^{+0.02}$ keV), and two \texttt{zgauss} components for emission lines that could not otherwise by fit at $1.35 \pm 0.02$ keV and $1.84 \pm 0.04$ keV. We think that these emission lines are $\rm Mg\, XI$ and $\rm Si\, XIII$ respectively, but more detailed spectroscopic measurements are needed to confirm this. This model produced an intrinsic luminosity estimate of $\log (L_{2-10 \rm \ keV} / \rm erg \ s^{-1}) = 39.88_{-0.08}^{+0.07}$, which agrees with the estimate of $\log (L_{2-10 \rm \ keV} / \rm erg \ s^{-1}) = 39.9 \pm 0.2$ from \citet{Strickland2007}. When fitting with \ctpo, we again required a \texttt{mekal} component with kT $= 0.54_{-0.03}^{+0.04}$ keV, and could only find upper limits for the obscuration of $\rm N_{H, eq} \leq 0.28 \times 10^{22} \ cm^{-2}$ and $\mathrm{N_{H, LOS}} \leq 0.14 \times 10^{22} \ \mathrm{cm}^{-2}$. This is slightly lower than the LOS obscuration found by \citet{Strickland2007} of $\mathrm{N_{H, LOS}} \sim 0.21 \times 10^{22} \ \mathrm{cm}^{-2}$, but they employed multiple absorbers, including ones with fixed $\mathrm{N_H}$ values. We found an intrinsic luminosity of $\log (L_{2-10 \rm \ keV} / \rm erg \ s^{-1}) = 39.78 \pm 0.06$, which agrees with our original estimate.

\subsection*{NGC6860}

    This Seyfert 1.5 galaxy was analysed using 2024 \textit{NuSTAR} and 2023 \textit{XMM-Newton} observations. There is a significant flux difference between the two observations, with the later \textit{NuSTAR} being $\sim 3$ times fainter. This suggests that this may be a changing-look AGN, and/or that the central engine itself is dimming, as the obscuration seems to remain constant. We have been unable to verify this in the literature, though, as we did not find any papers discussing the change in intrinsic luminosity in the recent \textit{NuSTAR} observation. As with the other variable sources, we have reported the average intrinsic luminosity in Tables \ref{tab:phenomenological_results} and \ref{tab:carrot_results}. To fit the spectrum initially, we used scattering with $f_{\text{scatt}} = 7.02_{-0.93}^{+0.91} \%$ for the soft excess. To model the reflection, \texttt{pexrav} with relative reflection $= -0.67_{-0.44}^{+0.33}$ was used. To model the obscuration, we used both a neutral absorber (\texttt{zphabs}) and partially-ionised partial absorber (\texttt{zxipcf}) with $\log \xi = 2.16^{+0.02}_{-0.04}$. The estimated intrinsic luminosity of this model was $\log (L_{2-10 \rm \ keV} / \rm erg \ s^{-1}) = 42.12 \pm 0.02$ for the \textit{NuSTAR} epoch, and $\log (L_{2-10 \rm \ keV} / \rm erg \ s^{-1}) = 42.50_{-0.01}^{+0.02}$ for the \textit{XMM-Newton} epoch. The average is lower than the estimate of $\log (L_{2-10 \rm \ keV} / \rm erg \ s^{-1}) \sim 42.66$ from \citet{Winter2009}. However, we do expect this, given the much lower flux of the \textit{NuSTAR} observation. With \ctpo, we needed a \texttt{mekal} component with kT $= 0.23 \pm 0.01$ keV for the soft excess, and found a LOS column density of $\mathrm{N_{H, LOS}} = 2.27_{-0.30}^{+0.09} \times 10^{22} \ \mathrm{cm}^{-2}$, which is slightly lower than the \citet{Winter2009} value of $\mathrm{N_{H, LOS}} = 4.5 \pm 1.3 \times 10^{22} \ \mathrm{cm}^{-2}$, but our value is only an indicative estimate. We found an intrinsic luminosity of $\log (L_{2-10 \rm \ keV} / \rm erg \ s^{-1}) = 42.13_{-0.03}^{+0.02}$ or the \textit{NuSTAR} epoch, and $\log (L_{2-10 \rm \ keV} / \rm erg \ s^{-1}) = 42.64 \pm 0.02$ for the \textit{XMM-Newton} epoch. This average agrees with our original estimate, and the \textit{XMM-Newton} epoch estimate is very close to the \citet{Winter2009} value.

\subsection*{NGC6890}

    NGC6890 is known to be a Compton-thick AGN \citep{Chen2025}, for which we have 2018 \textit{NuSTAR} and 2005 \textit{XMM-Newton} observations. There is an extreme difference in flux between the two observations (the \texttt{NuSTAR} observation is $\sim 14$ times brighter than the \texttt{XMM-Newton} one), leading us to suspect that this may also be a variable/changing-look AGN. Despite this, we were able to successfully fit the combined spectrum using \myt. As with the other variable sources, we have reported the average intrinsic luminosity in Tables \ref{tab:physical_results} and \ref{tab:carrot_results}. For the soft excess, we required both \texttt{mekal} with kT $= 0.34_{-0.10}^{+0.24}$ keV, and scattering with $f_{\text{scatt}} = 16.17_{-6.90}^{+10.66} \%$. We estimated the equatorial column density to be $\rm N_{H,eq} = 4.02_{-1.43}^{+4.12} \times 10^{24} \ cm^{-2}$, which is higher than the average of the estimates from \citet{Saade2022} of $\rm N_{H,eq} = 0.56_{-0.48}^{+2.01} \times 10^{24} \ cm^{-2}$, found using \texttt{Borus}. As was found by \citet{Saade2022}, there is roughly an order of magnitude luminosity difference between the two epochs, and the intrinsic luminosity we found was $\log (L_{2-10 \rm \ keV} / \rm erg \ s^{-1}) = 41.81_{-0.11}^{+0.7}$ for the \textit{NuSTAR} epoch, and $\log (L_{2-10 \rm \ keV} / \rm erg \ s^{-1}) = 40.56 \pm 0.16$ for the \textit{XMM-Newton} epoch. These values are much lower than the estimates from \citet{Saade2022} of $\log (L_{2-10 \rm \ keV} / \rm erg \ s^{-1}) = 43.66_{-0.01}^{+0.09}$ for the \textit{NuSTAR} epoch, and $\log (L_{2-10 \rm \ keV} / \rm erg \ s^{-1}) = 42.25_{-0.24}^{+0.89}$ for the \textit{XMM-Newton} epoch. These discrepancies may be due to the different model geometries, as well as the different approaches used to model the changing fluxes \citep[][kept the power law normalisation fixed while allowing the obscuration to vary]{Saade2022}. To fit the soft excess when using \ctpo, we included a \texttt{mekal} component with kT $= 0.26_{-0.04}^{+0.07}$ keV, and found an equatorial column density of $\rm N_{H,eq} = 7.66_{-1.72}^{+1.29} \times 10^{24} \ cm^{-2}$, which is again higher than the \citet{Saade2022} estimates from \texttt{Borus}. We estimated an intrinsic luminosity of $\log (L_{2-10 \rm \ keV} / \rm erg \ s^{-1}) = 43.32_{-0.34}^{+0.26}$ for the \textit{NuSTAR} epoch, and $\log (L_{2-10 \rm \ keV} / \rm erg \ s^{-1}) = 42.23_{-0.39}^{+0.30}$ for the \textit{XMM-Newton} epoch. These are much closer to (and in the case of the \textit{XMM-Newton} epoch, agree with) the \citet{Saade2022} estimates.

\subsection*{NGC7130}

    NGC7130 is a Compton-thick Seyfert 2, which we analysed using 2016 \textit{NuSTAR} and 2001 \textit{Chandra} observations. There was no significant difference between the fluxes of the two epochs, despite the large temporal gap. We used \myt \ to model the spectrum, needing \texttt{mekal} (kT $= 0.67 \pm 0.03$ keV) and scattering ($f_{\text{scatt}} = 2.00_{-0.66}^{+0.83} \%$) to model the soft excess. We estimated the column densities to be $\rm N_{H,LOS} = 1.86_{-0.28}^{+0.39} \times 10^{24} \ cm^{-2}$ and $\rm N_{H,eq} = 5.00_{-1.55}^{+2.68} \times 10^{24} \ cm^{-2}$, which agree with the \citet{Zhao2021} estimates of $\rm N_{H,LOS} \geq 1.45 \times 10^{24} \ cm^{-2}$ and $\rm N_{H,eq} \geq 1.95 \times 10^{24} \ cm^{-2}$. This model resulted in an estimated intrinsic luminosity of $\log (L_{2-10 \rm \ keV} / \rm erg \ s^{-1}) = 42.62_{-0.13}^{+0.10}$, which agrees with the estimate of $\log (L_{2-10 \rm \ keV} / \rm erg \ s^{-1}) \leq 43.59$ from \citet{Zhao2021}. We again used \texttt{mekal} (kT $= 0.68_{-0.03}^{+0.08}$ keV) when modelling with \ctpo, and found $\rm N_{H,LOS} = 2.08_{-0.56}^{+0.76} \times 10^{24} \ cm^{-2}$ and $\rm N_{H,eq} = 4.5_{-1.2}^{+1.6} \times 10^{24} \ cm^{-2}$, which also agree with the \citet{Zhao2021} estimates. We estimated an intrinsic luminosity of $\log (L_{2-10 \rm \ keV} / \rm erg \ s^{-1}) = 43.14_{-0.25}^{+0.18}$, which is higher than the previous estimate but also agrees with the \citet{Zhao2021} upper limit.

\subsection*{NGC7213}

    We analysed this variable Seyfert 1 with \textit{NuSTAR} and \textit{XMM-Newton} observations from 2014 and 2009 respectively. As there was a significant variation in flux between the two observations (the \texttt{NuSTAR} observation is $\sim50 \%$ brighter than the \texttt{XMM-Newton} one), we report the average intrinsic luminosity in Tables \ref{tab:phenomenological_results} and \ref{tab:carrot_results}. The initial model uses a \texttt{mekal} component for the soft excess (kT $= 0.66_{-0.07}^{+0.10}$ keV), three \texttt{zgauss} components for Fe K emission lines at $6.40 \pm 0.01$ keV, $6.68 \pm 0.02$ keV and $7.00 \pm 0.02$ keV, and \texttt{zxipcf} for the absorption with $\log \xi = -1.25^{+0.20}_{-1.75}$. We estimated the intrinsic luminosity to be $\log (L_{2-10 \rm \ keV} / \rm erg \ s^{-1}) = 42.12 \pm 0.01$ for the \textit{NuSTAR} epoch, and $\log (L_{2-10 \rm \ keV} / \rm erg \ s^{-1}) = 41.94 \pm 0.01 $ for the \textit{XMM-Newton} epoch, the average of which agrees with the estimate from \citet{Salvestrini2020} of $\log (L_{2-10 \rm \ keV} / \rm erg \ s^{-1}) = 41.95 \pm 0.07$ within the uncertainties. When using \ctpo, we again utilised \texttt{mekal} with kT $= 0.65 \pm 0.05$ keV for the soft excess. Based on the estimated inclination and torus opening angle, we find this to be an unobscured source, which is corroborated by \citet{Ricci2017}. We found an intrinsic luminosity of $\log (L_{2-10 \rm \ keV} / \rm erg \ s^{-1}) = 42.13 \pm 0.01 $ for the \textit{NuSTAR} epoch, and $\log (L_{2-10 \rm \ keV} / \rm erg \ s^{-1}) = 41.96 \pm 0.01$ for the \textit{XMM-Newton} epoch. The average of these estimates agrees with the phenomenological and \citet{Salvestrini2020} estimates.

\subsection*{NGC7469}

    NGC7469 is a Seyfert 1 that was simultaneously observed by \textit{NuSTAR} and \textit{XMM-Newton} in 2015. It is a bright, `bare' AGN with a complex spectrum \citep{Nandi2023}, requiring many phenomenological components to fit successfully. To model the soft excess, we used two \texttt{mekal} components with kT $= 0.26 \pm 0.01$ keV and kT $\leq 0.09$ keV, and a \texttt{zgauss} component for an emission line at $1.23^{+0.02}_{-0.03}$ keV that we think is the $\rm Mg \ K\alpha$ line, but more detailed spectroscopy is needed to confirm this. We used an additional \texttt{zgauss} component to model the Fe K$\alpha$ emission line at $6.42 \pm 0.01$ keV. To model the reflection, we used \texttt{pexrav} with relative reflection $= -0.58 \pm 0.13$. Our estimated intrinsic luminosity was $\log (L_{2-10 \rm \ keV} / \rm erg \ s^{-1}) = 43.32 \pm 0.01$, which agrees with the estimate from \citet{Nandi2023} of $\log (L_{2-10 \rm \ keV} / \rm erg \ s^{-1}) = 43.14^{+0.59}_{-0.36}$ within the uncertainties. When modelling with \ctpo, we used \texttt{mekal} (kT $= 0.22 \pm 0.01$ keV) to model the soft excess, and found an inclination that results in an unobscured LOS. \citet{Ricci2017} found a near-negligible LOS column density of $\mathrm{N_{H, LOS}} = 3 \pm 1 \times 10^{20} \ \mathrm{cm}^{-2}$, which we consider to be consistent with our result. We estimated the intrinsic luminosity to be $\log (L_{2-10 \rm \ keV} / \rm erg \ s^{-1}) = 43.37 \pm 0.01$, which also agrees with the \citet{Nandi2023} estimate.

\subsection*{NGC7496}

    NGC7496 is a faint Seyfert 2 galaxy, which we were only able to detect within \textit{XMM-Newton} observations from 2022. \textit{NuSTAR} observations were available, but there was no signal detected above the background. Due to the low counts ($\sim 390$), we utilised C-statistics during fitting. The initial model consisted of an unobscured power law, a \texttt{mekal} component (kT $= 0.47 \pm 0.15$ keV), and a \texttt{zgauss} component at $6.74^{+0.15}_{-0.11}$ keV. This model yielded an intrinsic luminosity of $\log (L_{2-10 \rm \ keV} / \rm erg \ s^{-1}) = 39.32^{+0.08}_{-0.09}$, which is in agreement with the upper limit estimate of $\log (L_{2-10 \rm \ keV} / \rm erg \ s^{-1}) \leq 40.04$ from \citet{Rush1996}. Working with \ctpo, we had to freeze the majority of parameters, and only left the column density and photon index free to vary. We estimated an intrinsic luminosity of $\log (L_{2-10 \rm \ keV} / \rm erg \ s^{-1}) = 39.25^{+0.04}_{-0.07}$, which agrees with our previous estimate. \citet{Rush1996} did not calculate the obscuration, so we cannot compare our $\rm N_H$ results.

\subsection*{NGC7603}

    This Seyfert 1 galaxy only had one \textit{XMM-Newton} observation from 2006 available, for which both MOS1/2 instruments were set to `timing mode' and thus could not be used for spectral fitting. Despite this, we were able to fit this bright spectrum with both a phenomenological model and \ctpo. The phenomenological model consisted of a \texttt{mekal} component for the soft excess (kT $= 0.19 \pm 0.02$ keV), \texttt{pexrav} for the reflection (relative reflection $= -5.85_{-0.72}^{+0.71}$), and one \texttt{zgauss} component for the Fe K$\alpha$ emission line at $6.37^{+0.07}_{-0.08}$ keV. We estimated the intrinsic luminosity to be $\log (L_{2-10 \rm \ keV} / \rm erg \ s^{-1}) = 43.44 \pm 0.01$, which is slightly lower than the absorption-corrected luminosity from \citet{Ricci2017} of $\log (L_{2-10 \rm \ keV} / \rm erg \ s^{-1}) \sim 43.65$. When fitting with \ctpo, we once again needed a \texttt{mekal} component for the soft excess, with kT $= 0.18 \pm 0.02$ keV, and estimated the LOS column density to be $\rm N_{H,LOS} = 11.8_{-2.5}^{+3.0} \times 10^{22} \ cm^{-2}$. This disagrees with \citet{Ricci2017}, who found the source to be unobscured. However, as our LOS estimate assumes a smooth dust distribution, it is possible that our LOS instead `peaks through' the clouds, resulting in an unobscured spectrum, which we cannot tell from our model. We found an intrinsic luminosity of $\log (L_{2-10 \rm \ keV} / \rm erg \ s^{-1}) = 43.70^{+0.06}_{-0.05}$, which agrees with the value from \citet{Ricci2017}.

\subsection*{NGC7674}

    NGC7674 is a Seyfert 1.5 galaxy with \textit{NuSTAR} and \textit{XMM-Newton} observations available from 2014 and 2004, respectively. However, this source has previously been identified as Compton-thick with no noticeable variation over $\sim 20$ years \citep[see][]{Gandhi2017}, and we also did not find any noticeable flux variations between the observations. To model this source, we used \myt, with an additional \texttt{mekal} for the soft excess (kT $= 0.74_{-0.09}^{+0.05}$ keV). We struggled to fit this source with the decoupled \myt, much like \citet{Gandhi2017}, and our best fit could not properly constrain many of the parameters. We were able to constrain the LOS obscuration, finding $\mathrm{N_{H, LOS}} = 0.34 \pm 0.09 \times 10^{24} \ \mathrm{cm}^{-2}$, but could only find a lower limit for the equatorial column density of $\rm N_{H, eq}  \geq 4.8 \times 10^{24} \ cm^{-2}$. Similarly, \citet{Gandhi2017} found $\rm N_{H,LOS}\geq 3.2 \times 10^{24} \ cm^{-2}$ and $\rm N_{H,eq} \geq 5.9 \times 10^{24} \ cm^{-2}$, using \myt \ in its `coupled' configuration. Using the \texttt{Torus} model, they estimate the column density to be $\rm N_{H}  \geq 26 \times 10^{24} \ cm^{-2}$ (equatorial and LOS are equal), which surpasses the upper limit of \myt. We thus suspect that \myt \ may not be the best model to use for this source, as its obscuration seemingly lies outside the range of possible values. \citet{Tanimoto2020} used the \texttt{XCLUMPY} model to fit this source, and estimated $\mathrm{N_{H, LOS}} = 0.24_{-0.10}^{+0.22} \times 10^{24} \ \mathrm{cm}^{-2}$ and $\mathrm{N_{H, Eq}} = 10_{-4.1}^{+9.2} \times 10^{24} \ \mathrm{cm}^{-2}$, which our results agree with. We estimated an average intrinsic luminosity of $\log (L_{2-10 \rm \ keV} / \rm erg \ s^{-1}) = 42.71^{+0.04}_{-0.05}$, which is much lower than the result from \citet{Gandhi2017} of $\log (L_{2-10 \rm \ keV} / \rm erg \ s^{-1}) = 43.6^{+0.5}_{-0.6}$. However, as we think \myt \ is underestimating the obscuration, the luminosity is likely to be underestimated as well. When fitting with \ctpo, we again required a \texttt{mekal} component with kT $= 0.68_{-0.10}^{+0.08}$ keV. We estimated the obscuration to be $\mathrm{N_{H, LOS}} = 6.4_{-2.0}^{+1.9} \times 10^{24} \ \mathrm{cm}^{-2}$ and $\rm N_{H, eq} = 16.3_{-5.1}^{+4.9} \times 10^{24} \ cm^{-2}$, which are higher than the previous estimates, as expected. These values agree with the \myt \ lower limits from \citet{Gandhi2017}, but the $\mathrm{N_{H, eq}}$ is lower than their \texttt{Torus} estimate. This may be due to differences in the model geometries. Our $\mathrm{N_{H, eq}}$ values agrees with those from \citet{Tanimoto2020} within the uncertainties, but our $\mathrm{N_{H, LOS}}$ is much higher. This may be due to how the $\mathrm{N_{H, LOS}}$ is found when using \ctpo, which is an estimation based on an assumption of a smooth torus. The random clump placement within the torus may result in a relatively unobscured LOS in comparison to the $\mathrm{N_{H, eq}}$, if the LOS `peeks through' the clumps. As is the case with NGC 5506, this possible scenario cannot be accurately represented by the $\mathrm{N_{H, LOS}}$ estimates presented, and it should be re-emphasised that our $\mathrm{N_{H, LOS}}$ values are merely indicative estimates rather than fitted values. The estimated intrinsic luminosity of $\log (L_{2-10 \rm \ keV} / \rm erg \ s^{-1}) = 43.74^{+0.06}_{-0.05}$ agrees with the \citet{Gandhi2017} estimate.

\subsection*{TOLOLO1238-364}

    Also known as IC3639, we analysed this Compton-thick Seyfert 2 galaxy using a \textit{NuSTAR} observation from 2015 and a \textit{Chandra} observation from 2004. There was a significant variation in flux between the two observation epochs (the \texttt{NuSTAR} observation is $\sim 2.6$ times brighter than the \texttt{Chandra} one), and so we have reported the average intrinsic luminosity in Tables \ref{tab:physical_results} and \ref{tab:carrot_results}. We modelled the spectrum using \myt, using a \texttt{mekal} component with kT $= 0.65_{-0.09}^{+0.15}$ keV and a scattering component with $f_{\text{scatt}} = 0.64_{-0.33}^{+0.63} \%$ for the soft excess. We were only able to get lower limits for the obscuration, finding $\mathrm{N_{H, LOS}} \geq 3.1 \times 10^{24} \ \mathrm{cm}^{-2}$ and $\rm N_{H, eq}  \geq 4.6 \times 10^{24} \ cm^{-2}$. \citet{Boorman2016} used \myt \ in its `coupled' configuration, and were also only able to find lower limits of $\mathrm{N_{H, LOS}} \geq 3.61 \times 10^{24} \ \mathrm{cm}^{-2}$ and $\rm N_{H, eq} \geq 5.9 \times 10^{24} \ cm^{-2}$, which agree with our limits. We estimated lower limits for the intrinsic luminosity of $\log (L_{2-10 \rm \ keV} / \rm erg \ s^{-1}) \geq 42.31$ for the \textit{NuSTAR} epoch, and $\log (L_{2-10 \rm \ keV} / \rm erg \ s^{-1}) \geq 42.13$ for the \textit{Chandra} epoch. These limits agree with the estimate of $\log (L_{2-10 \rm \ keV} / \rm erg \ s^{-1}) = 43.4^{+0.6}_{-1.1}$ from \citet{Boorman2016}. We also used a \texttt{mekal} component when modelling with \ctpo, finding kT $= 0.67_{-0.10}^{+0.16}$ keV. We estimated the intrinsic luminosity to be $\log (L_{2-10 \rm \ keV} / \rm erg \ s^{-1}) = 43.38^{+0.22}_{-0.13}$ for the \textit{NuSTAR} epoch, and $\log (L_{2-10 \rm \ keV} / \rm erg \ s^{-1}) = 43.22_{-0.12}^{+0.36}$ for the \textit{Chandra} epoch. These estimates both agree with the \citet{Boorman2016} estimate within the uncertainties.
    

\bsp	
\label{lastpage}
\end{document}